\documentclass[a4paper,11pt]{article}
\pdfoutput=1
\usepackage{mystyle}
\usepackage{graphicx}
\usepackage{color}
\usepackage{amssymb}
\usepackage{amsmath}
\usepackage{mathabx}
\usepackage{upgreek}
\usepackage{bbm}
\usepackage{multirow}
\usepackage{multicol}
\usepackage{amscd}
\usepackage{ipa}
\usepackage{mfpic}
\usepackage{verbatim}
\usepackage{hyperref}
\usepackage{cancel}
\usepackage{chemarrow}
\usepackage{tikz-cd}
\usepackage{tocloft}
\usepackage{MnSymbol} % for double brackets, but induces conflicts

\definecolor{cream}{rgb}{.97, .95, .88}
\definecolor{darkcream}{rgb}{1., .88, .5}
\definecolor{lightpink}{rgb}{0.98, 0.88, 0.87}
\definecolor{lightwhite}{rgb}{1., 0.98, 0.95}
\definecolor{lightsalmon}{rgb}{1., 0.95, 0.90}
\definecolor{lightviolet}{rgb}{0.9, 0.8, 0.9}
\definecolor{lightgray}{rgb}{.96, .96, .96}  
\definecolor{lgray}{rgb}{.75, .75, .75}
\definecolor{LemonChiffon}{rgb}{0.95, 1., 0.7}
\definecolor{lightolivegreen}{rgb}{0.84, 0.89, 0.25}
\definecolor{lightgreen}{rgb}{.664, 1., .52}
\definecolor{llgreen}{rgb}{.900, .983, .960}
\definecolor{tristle}{rgb}{0.87, 0.67, 0.77} %{0.792, 0.609, 0.698}
\definecolor{pink}{rgb}{0.95, 0.45, 0.75}
\definecolor{magenta}{rgb}{1., 0, 1.}
\definecolor{violet}{rgb}{0.9, 0.20, 0.85}
\definecolor{darkolivegreen}{rgb}{0.55, 0.65, 0.35}
\definecolor{maroon}{rgb}{0.7, 0.26, 0.56}
\definecolor{lightmaroon}{rgb}{0.85, 0.38, 0.58}
\definecolor{darkmaroon}{rgb}{0.604, 0.169, 0.451}
\definecolor{ddarkmaroon}{rgb}{0.2, 0.03125, 0.150}
\definecolor{mediumorchid}{rgb}{0.8, 0.33, 0.83}
\definecolor{mediumorchidd}{rgb}{1., 0.33, 0.63}
\definecolor{darkgreen}{rgb}{0.1, 0.6, 0.13}
\definecolor{lightyellow}{rgb}{1., 1., 0.82}
\definecolor{turquoise}{rgb}{0.042, 0.586, 0.512}
\definecolor{turquoisel}{rgb}{0.66, 0.94, 0.83}
\definecolor{darkturquoise}{rgb}{0.21, 0.55, 0.50}
\definecolor{coral}{rgb}{1., 0.6, 0.21}
\definecolor{lightorange}{rgb}{1., 0.88, 0.75}
\definecolor{orangered}{rgb}{1., 0.5, 0.}
\definecolor{orange}{rgb}{1., 0.65, 0.1}
\definecolor{orangel}{rgb}{1., .85, .3}
\definecolor{darkorange}{rgb}{0.875, 0.4, 0.204}
\definecolor{ddarkorange}{rgb}{.675, .218, .05}
\definecolor{bluesky}{rgb}{0.48, 0.53, 1.}
\definecolor{gold}{rgb}{1., 0.85, 0.25}
\definecolor{goldd}{rgb}{0.95, 0.75, 0.05}
\definecolor{darkviolet}{rgb}{0.54, 0.04, 0.84}
\definecolor{ddarkviolet}{rgb}{.382, .063, .657}
\definecolor{lightblue}{rgb}{0.30, 0.86, 0.89}
\definecolor{LightBlue}{rgb}{0.68, 0.85, 0.9}
\definecolor{lblue}{rgb}{0.78, 0.90, 0.95}
\definecolor{darkblue}{rgb}{.105, .308, .707}
\definecolor{lightmaroon}{rgb}{0.85, 0.38, 0.58}
\definecolor{darkmaroon}{rgb}{0.604, 0.169, 0.451}
\definecolor{darkpink}{rgb}{0.879, 0.020, 0.766}
\definecolor{ddarkpink}{rgb}{0.738, 0.195, 0.406}
\definecolor{grey}{rgb}{0.717, 0.717, 0.717}
\definecolor{lightgrey}{rgb}{0.800, 0.800, 0.800}
\definecolor{brown}{rgb}{0.740, 0.323, 0.182}
\definecolor{redbrown}{rgb}{.575, .158, .05}
\definecolor{darkbrown}{rgb}{0.34, 0.25, 0.05}
\definecolor{orangebrown}{rgb}{0.433, 0.262, 0.06}
\definecolor{pinkl}{rgb}{1., 0.788, 0.918}
\definecolor{salmon}{rgb}{1., 0.66, 0.5}
\definecolor{lightbrown}{rgb}{0.703, 0.508, 0.121}

\def\Journal#1#2#3#4{{#1} {\bf #2} (#3) #4}

\def\etal{{\it et al.}}
\def\Name#1#2 {{ #1 }{#2}}

\def\AIP{\em AIP Conf. Proc.}

\def\ANM{\em Annals Math.}
\def\APH{\em Annals Phys.}

\def\ATM{\em Adv. Theor. Math. Phys}

\def\CMP{\em Commun. Math. Phys.}

\def\CQG{\em Class. Quant.~Grav.}

\def\FOP{\em Found. Phys.}

\def\GRG{\em Gen. Rel. Grav}

\def\IMA{{\em Int. J. Mod. Phys.} \emph{A}}
\def\IMD{{\em Int. J. Mod. Phys.} \emph{D}}

\def\JCA{\em J. Cosmol. Astrop. Phys.}

\def\JGP{\em J. Geom. Phys.}
\def\JHA{\em J. High Energy Astrophys.}
\def\JHE{\em J. High Energy Phys.}
\def\JMM{\em J. Math.Mech.}
\def\JMP{\em J. Math. Phys.}
\def\JPA{\em J. Phys. A: Math. Theor.}
\def\JPC{\em J. Phys. Conf. Ser.}

\def\JPU{\em J. Phys. (USSR)}

\def\LNP{\em Lect. Notes Phys.}

\def\MDU{\em MDPI Universe J.}

\def\MPL{{\em Mod. Phys. Lett.} \emph{A}}
\def\MRA{\em MNRAS}

\def\NAC{\em Nature Commun.}

\def\NAT{\em Nature}

\def\NJP{\em New J. Phys.}

\def\NPB{{\em Nucl.~Phys.}~\emph{B}}

\def\PAP{\em Publ. Astro. Soc. Pac.}

\def\PLA{{\em Phys. Lett.}~\emph{A}}
\def\PLB{{\em Phys. Lett.}~\emph{B}}

\def\PRA{{\em Phys.~Rev.}~\emph{A}}

\def\PRD{{\em Phys.~Rev.}~\emph{D}}

\def\PRE{\em Phys.~Rep.}
\def\PRL{\em Phys. Rev. Lett.}
\def\PRR{\em Phys.~Rev.~Research}
\def\PRV{\em Phys.~Rev.}
\def\PRX{{\em Phys.~Rev.}~\emph{X}}

\def\PTO{\em Phys. Today}

\def\PYA{\em Physica D} 
\def\QNM{\em Quantum}
 
\def\RMP{\em Rev. Mod. Phys.}
\def\RPP{\em Rept. Prog. Phys.}
\def\SCI{\em Science}

\def\SYM{\em Symmetry}

\def\be{\begin{equation}}
\def\ee{\end{equation}}
\def\bea{\begin{eqnarray}}
\def\eea{\end{eqnarray}}
\def\bes{\begin{equation*}}
\def\ees{\end{equation*}}
\def\beas{\begin{eqnarray*}}
\def\eeas{\end{eqnarray*}}
\def\tr{\text{tr}}

\DeclareMathVersion{normal2}
\mathversion{normal2}

\def\mf{\mathsf f}
\def\mF{\mathsf F}
\def\mg{\mathsf g}
\def\cf{\mathtt f}

\def\bm{\mathcal B}

\def\dm{\mathcal D}
\def\fm{\mathcal F}
\def\gm{\mathcal G}
\def\hm{\mathcal H}
\def\lm{\mathcal L}
\def\mm{\mathcal M}

\def\ym{\mathcal Y}
\def\hA{\hat{A}}
\def\hB{\hat{B}}

\def\hH{\hat{H}}
\def\hJ{\hat{J}}
\def\hL{\hat{L}}
\def\hO{\hat{O}}
\def\hP{\hat{P}}
\def\hR{\hat{R}}
\def\hS{\hat{S}}
\def\hT{\hat{T}}
\def\hU{\hat{U}}
\def\hX{\hat{X}}
\def\hY{\hat{Y}}
\def\hZ{\hat{Z}}
\def\hrho{\hat{\rho}}
\def\sD{\cancel{D}}

\def\sqgr{$SU(\infty)$-QGR}
\def\suinf{{SU(\infty)}}
\def\suinfa{{\mathcal{SU}(\infty)}}

\title{\boldmath $\suinf$ Quantum Gravity: Emergence of Gravity in an Infinitely Divisible Quantum Universe}
\author[a,b]{Houri Ziaeepour}
\affiliation[a]{Institut UTINAM, CNRS UMR 6213, Observatoire de Besan\c{c}on, Universit\'e de Franche Compt\'e, 41 bis ave. de l'Observatoire, BP 1615, 25010 Besan\c{c}on, France}
\affiliation[b]{Mullard Space Science Laboratory, University College London, Holmbury St. Mary, GU5 6NT, Dorking, UK}
\emailAdd{houriziaeepour@gmail.com}

\newcounter{propos}

\abstract{
\sqgr~is a foundationally quantum approach to cosmology and gravity. It assumes that the Hilbert space of the Universe as a whole represents the symmetry group $\suinf$, and demonstrates this symmetry for Hilbert spaces of infinite number of subsystems, which randomly emerge and represent arbitrary finite rank {\it internal} symmetries. The aim of present work is in depth study of the foundation and properties of this model. We show that the global $\suinf$ symmetry manifests itself through the entanglement of each subsystem with the rest of the Universe. We demonstrate that the states of subsystems depend on a dimensionful parameter arising due to the breaking of a global $U(1)$ symmetry. A relative dynamics arises when an arbitrary subsystem is selected as a quantum clock with a time parameter. Thus, states of subsystems are characterized by 4 continuous parameters related to their $\suinf$ symmetry and dynamics, plus discrete parameters characterizing their internal symmetries. We demonstrate the irrelevance of the geometry of this parameter space for observables. On the other hand, we use quantum speed limits to show that the perceived classical spacetime and its Lorentzian geometry emerge as the effective (average) path of subsystems in their Hilbert space. In this respect, \sqgr~fundamentally deviates from gauge-gravity duality models. The invariance under reparameterization restricts the action of subsystems dynamics to a Yang-Mills quantum field theory defined on the (3+1)-dimensional parameter space for both $\suinf$ - gravity - and internal symmetries. Consequently, \sqgr~is renormalizable, but predicts a spin-1 mediator for quantum gravity. Nonetheless, it is proved that when quantum gravity effects are not detectable, the dynamics is perceived as the Einstein-Hilbert action. We also briefly discuss \sqgr~specific models for dark energy.


}

\begin{document}
\maketitle
\flushbottom
%\tableofcontents

\section{Introduction} \label{sec:intro}
So far most Quantum GRavity (QGR) candidates have been based on the classical models quantized 
in one way or another. Moreover, the geometry of spacetime, sometimes extended to include the 
matter sector have had a prominent role in the construction of many QGR theories. For instance, 
Kaluza-Klein model~\cite {kaluza,klein} and its successors such as string 
theory~\cite{stringrev,stringrev0} and related models - random 
matrices~\cite{qgrmatrixm,qgrmatrix0} and tensor product models~\cite{qgrtensornet,qgrtensornet0} - 
are all geometry inclined. Loop Quantum Gravity (LQG)~\cite{lqgrev,lqgrev0} and its extension 
to Group Field Theory (GFT)~\cite{qgrgroup,qgrgroup0} are even more geometrical and consider 
matter only as {\it external} fields without direct relationship with spacetime. In general, 
in these models a quantum gravity is obtained through quantization of a D-dimensional space, 
where D can be 2, 3, or higher. Of course, the reason behind this trend is the close 
relationship of classical gravity with the geometry of perceived spacetime in the Einstein 
general relativity. 

In geometrical approaches to QGR a classical model do exists for $\hbar \rightarrow 0$, because 
their departure point is essentially a classical model. However, a Universe in which 
$\hbar \rightarrow 0$ cannot correspond to ours, even as an asymptotic or approximate 
description. The reason is the innumerable experiments which have shown that matter behaves 
quantum mechanically even at macroscopic scales. On the other hand, inconsistency of a 
classical gravity with a quantum matter is well 
documented~\cite{grinconsist,grinconsist0,grinconsist1,houriqgr}\footnote{Nonetheless, 
skepticism still persists, see e.g.~\cite{grclassic,grclassic0}.}. Thus, to construct a 
quantum model of gravity and spacetime, it seems more logical to treat the Universe as an 
intrinsically quantum system from the beginning, rather than starting with a classical theory 
and subsequently quantizing it.

Another point, which is often overlooked by traditional approaches to quantum gravity and 
quantum cosmology is the fact that quantum systems need an environment for 
decoherence~\cite{statesymmbreak} and formation of a system looking classical and random in 
intermediate scales. We remind that at microscopic/UV scales the Universe is fully quantum 
mechanical. Furthermore, due to the expansion of the Universe superhorizon scales, which are at 
present causally disconnected, were previously interacting and might have kept their quantum 
coherence even now~\cite{infcohermode,infcohermode0}. The decoherence of other modes was 
necessarily due to an {\it environment}~\cite{infdecohere,infdecohere0}, presumably provided 
by the rest of the Universe. Thus, in a quantum model of gravity and spacetime it is imperative 
to consider the Universe as an ensemble of interacting, either coherently or causally, quantum 
subsystems~\cite{qbitdef,sysdiv,houriqmsymm}. In addition, a consistent and frame-independent 
definition of quantum dynamics and observables needs a quantum reference 
(observer)~\cite{qmref,qmrefchangedecoher,qmrefdecohere,qmrefsubsys} and a quantum 
clock~\cite{qmtimepage,qmtimedef}, which are not explicitly considered in many QGR proposals. 
These requirements highlight the crucial role of quantum subsystems in a fundamentally quantum 
approach to a Universe with a universal interaction similar to gravity between its parts. 
%Note: paragraphs which may be removed QM clock is discussed here but not QM ref./observer. Note: reference frame - subsys perspective is discussed in the new paper
Of course, a Quantum Field Theory (QFT) by definition includes infinite number of subsystems 
(particles). But, in standard formulation of QFTs the presence of an observer (a reference) 
and a quantum clock is implicit. Moreover, their quantum correlation and effects, for instance 
on the algebra of observables~\cite{qmrefsubsys}, decoherence through a change of reference 
frame~\cite{qmrefchangedecoher}, and dependence of entanglement and superposition on the choice 
of quantum reference frame~\cite{qmrefdecohere} are not explicit.

Quantum approaches to QGR are relatively recent additions to the jungle of QGR candidates. 
In absence of a hint from classical gravity, QFT, or experiments, this class of models 
usually have to use new priors as their departure point. Concepts and conjectures used so 
far include: principle of {\it locality} and geometrization of quantum mechanics - considered 
to be crucial for describing black holes and their puzzles~\cite{qgrhistory,qgrlocalqm}; 
holographic principle
%, which conjectures that maximum possible information capacity of a quantum system is proportional to the area of its null (lightlike) boundary rather than its volume
~\cite{hologprin,hologprin0,hologprin1}; and conjecture of the duality between quantum gravity 
in an asymptotically Anti-de-Sitter (AdS) bulk and a Conformal Field Theory (CFT) on the 
boundary, see e.g.~\cite{adscftrev} and references therein. For a brief review of these 
proposals and their comparison with \sqgr~see~\cite{hourisqgrcomp}.

In~\cite{houriqmsymmgr}, we proposed $\suinf$ Quantum Gravity or in short \sqgr~as a model 
for a quantum Universe. It is based on a few well motivated axioms,  in particular it assumes 
the Universe has infinite number of independent and mutually commuting observables. 
We studied some of the properties of such a quantum Universe as a single system, notably 
we showed that it is static and topological. We also demonstrated that it becomes dynamical 
after its division to subsystems. Specifically, the definition of an operational time as a 
parameter describing variation of the states of subsystems with respect to that of an 
arbitrary subsystem called a {\it clock} - \`a la Page \& Wootters~\cite{qmtimepage} or 
similar procedures~\cite{qmtimedef} - induces a dynamics. Moreover, in~\cite{hourisqgrcomp} 
we showed that gravity emerges as a universal quantum force having the form of a $\suinf$ 
Yang-Mills gauge theory defined on the (3+1)-dimensional parameter space of representations 
of the $\suinf$ symmetry by infinite number of subsystems and the dynamics. Therefore, 
in contrast to other QGR models, \sqgr~explains the origin of observed dimensionality of the 
classical spacetime. It is also shown that at low energies, that is when the quantum details 
of $\suinf$ Yang-Mills is not detectable, the action of the gravity sector of the model looks 
like that of the Einstein-Hilbert gravity. 

A discriminating aspect of \sqgr~is the strict classical spacetime. In fact, there is 
no background spacetime in the model. What is perceived as classical spacetime is an effective 
presentation of the parameters characterizing quantum states of subsystems and their dynamics.  
Moreover, its geometry presents the average path of quantum subsystems in the space of 
parameter. For this reason quantization of spacetime is meaningless. It is useful to remind 
that Einstein's opinion was that without the graviton field $g_{\mu\nu}(x)$, that is in absence 
of a gravity source, the spacetime itself has no physical meaning, see e.g.~\cite{etherhistory} 
(and references therein) for a concise review of the history of physical interpretations of 
spacetime. Einstein motivation for his conclusion was the failure of all experiments to 
detect a medium - {\it the ether} - as the empty space. However, general relativity does not 
provide any explanation for why and how a non-existent entity emerges in presence of matter. 
Therefore, the interpretation of spacetime as being an effective representation of parameters 
describing the quantum state matter provides the missing explanation in general relativity. 
Indeed, as Einstein suggested, in \sqgr~spacetime does not have any physical meaning without 
$\suinf$ Yang-Mills field (gravity) and matter. It is analogous to charge or spin, which 
cannot exist without matter and its interactions. 

To further advance preliminary results reported in~\cite{houriqmsymmgr} 
and~\cite{hourisqgrcomp}, in the present work we will investigate more thoroughly the structure 
and properties of the states of subsystems, the parameter space of representations of their 
$\suinf$ symmetry, and the Lagrangian of \sqgr. We will discuss in more details the 
fundamentally quantum nature of the model and provide an extended discussion about the 
parameter space and its geometry. In particular, we will prove that its geometry is irrelevant 
and can be transformed to that of a flat space by a $\suinf$ gauge transformation. This 
property is crucial for the model, otherwise the number of degrees of freedom of gravity 
sector would be doubled and one would have to introduce new rules, for example Einstein 
equation, for fixing geometry of the parameter space. Another characteristic of \sqgr, which 
was not discussed in the previous works, is the entanglement of subsystems of the Universe, 
induced by $\suinf$ symmetry at both local and Universe-wide levels. We will show that 
symmetries at these two scales are not independent, but intertwined. The importance of 
entanglement for QGR is recently explored in various 
approaches~\cite{qgrentangle,qgrentangle1,qgrentangle2}, specially those inspired by quantum 
information\cite{qmrefdecohere,qmrefsubsys}. In \sqgr~the omnipresent entanglement can be 
considered as quantum analogue of gravitational waves 
memory~\cite{gwmemory,gwmemory0,gwmemory1,gwmemory2}. We also discuss the emergence of local 
symmetries and invariance (or not) of subsystems under discrete symmetries of the parameter 
space. An important subject, which was not addressed in the previous works on \sqgr, is the 
possibility of having a constant term in the action, in other words a cosmological constant. 
We will introduce this topic here and will briefly present several processes specific to 
\sqgr, which may induce a constant term in the effective action at classical limit. 

In Sec. \ref{sec:modelreview} we review the rationale for proposing an intrinsically quantum 
model for gravity, axioms of \sqgr, and properties of its symmetry, algebra, Hilbert space, 
and quantization. Algebraic arguments for the emergence of local orthogonal symmetries and 
division of the Universe to subsystems are studied in Sec. \ref{sec:division}. We also 
discuss the existence of a permanent entanglement between every subsystem and the 
rest of the Universe and its consequences in this section. Algebra and parameter space of 
the subsystems are discussed in Sec. \ref{sec:subsysparam}. For this purpose we introduce 
an abstract quantum clock, associate a parameter to the variation of its state, and call this 
parameter {\it time}. Then, we explain the full parameter space of subsystems and its 
symmetries. We also discuss the geometry of the parameter space. In Sec. \ref{sec:geometry} 
Mandelstam-Tamm uncertainty relation is used to introduce a parameter, which we identify as 
the affine parameter of the perceived classical spacetime. It is related to the quantum state 
of subsystems. Moreover, we prove that the signature of classical metric must be negative. 
This topic was first explored in~\cite{houriqmsymmgr}. The purpose for its reminder here is 
its necessity for the construction of full parameter space and study of its properties.
%Some of these topics have been discussed in~\cite{houriqmsymmgr}. Here we provide more physical insight to their interpretation in conjunction with the global entanglement, coherence and causality. 
Finally, the dynamics and evolution of the Universe in \sqgr~is presented in 
Sec. \ref{sec:qmevol}. In addition, we study the classical limit of gravitation sector 
and propose a few processes which may introduce a cosmological constant in the effective 
Lagrangian. Outline and perspective for future research, notably application of \sqgr~to 
phenomena in which quantum gravity may be important are given in Sec. \ref{sec:outline}. 
Supplementary sections contain additional discussions and details of calculations for the 
subjects discussed in the main text.

\section{A quantum Universe with infinite independent observables} \label{sec:modelreview}
Rationale for a purely quantum approach to the Universe and to quantum gravity are reviewed 
in Sec. \ref{app:rationale}. They justify axioms based on which we construct \sqgr~in this 
section and we develop them further in the rest of this work.

\subsection{Axioms and construction of \sqgr} \label{sec:axioms}
Considering the conceptual background described in Sec. \ref{app:rationale}, we follow a 
purely quantum mechanical approach for constructing the Universe and its content. We begin 
by designing a quantum Universe based on three well motivated assumptions:
\setcounter{enumi}{0}
\renewcommand{\theenumi}{\Roman{enumi}}
\begin{enumerate}
\item Quantum mechanics is valid at all scales and applies to every entity, including the 
Universe as a whole; \label{uniaxiom1}
\item Any quantum system is described by its symmetries and its Hilbert space represents 
them; \label{uniaxiom2}
\item The Universe has infinite number of independent degrees of freedom, reflected in as 
many mutually commuting observables. \label{uniaxiom3} 
\end{enumerate}
The last assumption means that the Hilbert space of the Universe $\hm_U$ is infinite 
dimensional and linear operators acting on it represent the group $SU(\infty)$. We call the 
vector space of these operators $\bm[\hm_U]$. 

The $\suinfa$ algebra of $\bm[\hm_U]$ is defined as the following\footnote{In this work, all 
vector spaces and algebras are defined on complex numbers field $\mathbb{C}$, unless 
explicitly mentioned otherwise.}:
\be
[\hL_a,\hL_b] = \frac{i\hbar}{cM_P} f_{ab}^c \hL_c = i L_P f_{ab}^c \hL_c , \quad \quad 
L_P \equiv \frac{\hbar}{cM_P} \label{statealgebra}
\ee
where operators $\hL_\alpha \in \bm[\hm_U]$ are generators of $\suinfa$ and constants 
$f_{ab}^c \equiv f_{abc}$ with anti-symmetric indices are structure constants of the algebra. 
In Sec. \ref{app:stringcomp} $\suinfa$ is compared with Virasoro algebra.

In the sphere basis~\cite{suninfhoppthesis,suninftorus} (reviewed in~\cite{houriqmsymmgr})  
structure constants are proportional to 3j symbols defined for three pairs of indices 
$(l,m),~(l',m'),~(l'',m'')$, see (\ref{lharminicexp}). Notice that in (\ref{statealgebra}), 
operators $\hL_\alpha$ are normalized such that the r.h.s. explicitly depends on the Planck 
constant $\hbar$ and Planck mass $M_P \equiv \sqrt{\hbar c / G_N}$ or equivalently Planck 
length $L_P \equiv \hbar /cM_P$. This normalization is justified by dynamics and classical 
limit of the model in Sec. \ref{sec:qmevol}. In addition, in Sec. \ref{sec:subsysparam} we 
will show how a dimensionful scale arises in the model and becomes relevant for the above 
algebra. With this normalization if $\hbar \rightarrow 0$, or $M_P \rightarrow \infty$, the 
length scale $L_P \rightarrow 0$, and the algebra becomes Abelian and homomorphic to 
$\bigotimes^{N \rightarrow \infty} U(1)$. In agreement with the definition of $\hbar = 0$ as 
classical limit, this corresponds to the symmetry of configuration space of a classical 
system with infinite number of independent observables. Thus, the quantumness of the 
Universe according to \sqgr~imposes both $\hbar \neq 0$ and $M_P < \infty$. In other words 
gravity and quantumness are inseparable. 
%Note (from wiki: structure constants): Jacobi identity: [T^a, [T^b, T^c]] + [T^b, [T^c, T^a]] + [T^c, [T^a, T^b]] = 0 => F_ad^e f_bc^d + f_bd^e f_ca^d + f_cd^e f_ab^d = 0 

\subsection{Symmetry and Hilbert space of the Universe}  \label{sec:symm}
It is proved that in the limit of $N \rightarrow \infty$ the algebra 
$\mathcal{SU}(N \rightarrow \infty)$ is homomorphic to the algebra of Area-preserving 
Diffeomorphism $ADiff(D_2)$ of 2D compact Riemann surfaces 
$D_2$~\cite{suninfhoppthesis,suninfym,suninftorus,suninfrep,suninfrep0}\footnote{The 
relationship between $SU(N \rightarrow \infty)$ and $ADiff(D_2)$ group is subtle and depends 
on whether $N$ is considered to be rational or irrational. The latter cases correspond to 
infinite number of mutually non-homomorphic algebras~\cite{suninfrep}. Consequently, some 
of them should not be homomorphic to $ADiff(D_2)$~\cite{suninfrep}. Nonetheless, in a 
follow up of the present work~\cite{hourisqgrym} we demonstrate that their limit is unique 
and corresponds to $ADiff(D_2)$. Thus, in this work (\ref{diffeohomo}) indicates this 
unique limit.}. Thus:
\be
\bm[\hm_U] \cong SU(\infty) \cong ADiff(D_2) \label{diffeohomo}
\ee
where through this work the symbol $\cong$ means homomorphic. Although all representations of 
$SU(\infty)$ are locally homomorphic, they can have different global topology and thereby, 
are globally non-equivalent. Hence, the class of equivalent representations of $\suinf$ 
is isomorphic to the class of topologically equivalent $D_2$ surfaces, and we can associate to 
each representation of $\suinf$ a compact Riemann surface that its area preserving 
diffeomorphisms are homomorphic to application of a member of $\suinf$. As a shorthand we call 
such a surface the {\it diffeo-surface}. In \sqgr~topologically non-equivalent representations 
of $\suinf$ and their diffeo-surfaces correspond to globally different universes. Topology of 
the Universe and its Hilbert space would be important for nonlocal phenomena, for example when 
entanglement or black hole singularity~\cite{qgrentangle} makes part of the Hilbert space or 
equivalently its parameter space inaccessible, see Sec. \ref{sec:size} for more details. 

Due to (\ref{diffeohomo}), members of $\suinfa$ algebra are parameterized by two angular 
coordinates $(\theta, \phi)$ of the diffeo-surface, and generators $\hL_a$ of the algebra can 
be expanded with respect to spherical harmonic functions~\cite{suninfhoppthesis,suninftorus}: 
%Note: \hP = i\hbar \partial/\partial x is used with good metric. -i\hbar... bad metric (wiki "Momentum operator") => [X, P] = -i\hbar
\bea
\hL_{lm} (\theta, \phi) & = & i\hbar ~ \biggl (\frac{\partial Y_{lm}}{\partial \cos \theta} 
\frac{\partial}{\partial \phi} - 
\frac{\partial Y_{lm}}{\partial \phi} \frac{\partial}{\partial \cos \theta} \biggr ) = 
i\hbar ~ \sqrt{|g^{(2)}|} \epsilon^{\mu\nu} (\partial_\mu Y_{lm}) \partial_\nu, \nonumber \\
&& {\mu,~\nu \in \{\theta,\phi \}}, \quad l\geqslant 0,~ -l \leqslant m \leqslant l 
\label{lharminicexp} \\
\hL_{lm} Y_{l'm'} & = & -i\hbar \{Y_{lm},~Y_{l'm'}\} = -i \hbar f ^{l"m"}_{lm,l'm'} Y_{l"m"} 
\label{lapp} \\
\{\mf,~\mg\} & \equiv & \frac{\partial \mf}{\partial \cos \theta} 
\frac{\partial \mg}{\partial \phi} - \frac{\partial \mf}{\partial \phi} 
\frac{\partial \mg}{\partial \cos \theta}~~, \quad \forall ~ \mf,~\mg \label{fgbrac}
\eea
For the sake of simplicity from now on we will absorb the numerical factor 
$\frac{\hbar}{cM_P}$ in (\ref{statealgebra}) in the normalization of operators, unless when 
its presence is important for the discussion. The symbol $\{~,~\}$ in (\ref{fgbrac}) is the 
Poisson bracket with respect to local coordinates of the diffeo-surface and $g^{(2)}$ is the 
determinant of 2D metric of the diffeo-surface. Equalities in (\ref{lapp}) show that spherical 
harmonic functions have the same algebra as $\suinfa$ under Poisson bracket. This property 
will be useful for understanding various properties of \sqgr.

Cartan decomposition of $SU(\infty)$ allows to write $\hL_{lm}$ as a tensor product of Pauli 
matrices~\cite{suninfhoppthesis,cartandecomp}: 
\bea
\suinf & \cong & \bigotimes^\infty SU(2)  \label{suinfsu2decomp} \\ 
\hL_{lm} & = & \mathcal {R} \sum_{\substack{i_\alpha = {1,~2,~3}, \\ \alpha = {1, \cdots, l}}} 
a^{(m)}_{i_1, \cdots i_l} \sigma_{i_1} \cdots \sigma_{i_l}, \quad l\geqslant 0,~ -l 
\leqslant m \leqslant l \label {llmdef}
\eea
where $\sigma_{i_\alpha}$'s are Pauli matrices~\cite{suninfhoppthesis} and $\mathcal {R}$ is a 
normalization constant. Coefficients $a^{(m)}$ are determined from expansion of spherical 
harmonic functions with respect to spherical description of Cartesian 
coordinates~\cite{suninfhoppthesis}. We have omitted tensor product symbols between the 
factors, because one can also consider $\sigma$'s as $N \rightarrow \infty$ representation 
of Pauli matrices~\cite{suninfhoppthesis}. In this case, their products would be the 
ordinary matrix product. As the fundamental representation of $SU(2)$ corresponds to $l=1/2$, 
this relation shows that decomposition (\ref{llmdef}) with half-integer $(l, m)$ is also a 
representation of $\suinf$. 

The importance of $SU(2) \cong SO(3)$ symmetry in QGR models is discussed 
in~\cite{hourisqgrcomp}. The obvious reason is the fact that it is the symmetry of the 
Euclidean $\mathbb{R}^{(3)}$ space - the classical physical space. However, in contrast to 
other QGR proposals, here the decomposition (\ref{suinfsu2decomp}) does not have any special 
physical interpretation, except that $SU(2)$ is the smallest non-Abelian member of $SU$ 
groups. In supplementary section \ref{app:suinf} we show that $\suinf$ can be expressed 
as tensor product of any $SU(K), ~K < \infty$. 

%Although $\hL_{lm}$ operators are indexed by two integer, we continue to use single letters for indices when there is no need for their explicit description. 
It is possible - at least formally - to define generators of $\suinfa$ only with respect to 
continuous $(\theta, \phi)$ variables by summing over indices $(l,m)$ (See Appendix D 
of~\cite{houriqmsymmgr} for more details):
\bea
[\hL (\theta_1,\phi_1), \hL (\theta_2,\phi_2)] & = & \int d\Omega_3 ~ 
\cf((\theta_1,\phi_1),(\theta_2,\phi_2); (\theta_3,\phi_3)) ~ \hL (\theta_3,\phi_3) 
\label{xcommute} \\
\cf((\theta,\phi),(\theta',\phi'); (\theta",\phi")) & \equiv & 
\sum_{lm,l'm',l"m"} Y^*_{lm}(\theta,\phi) 
Y^*_{l'm'}(\theta',\phi') Y_{l"m"}(\theta",\phi'') f ^{l"m"}_{lm,l'm'} \label{fthetaphidef} \\
\hL (\theta, \phi) & \equiv & \sum_{l,m} Y^*_{lm} \hL_{lm}, \quad \quad d\Omega \equiv 
d\phi ~ d(\cos\theta) \sqrt{|g^{(2)}|} \label{xdef}
\eea
Eigen vectors of $\hL_{lm}$ and $\hL (\theta, \phi)$ are discussed in Appendix E.1 
of~\cite{houriqmsymmgr}. Although the number of generators of $\suinf$ group $\hL_{lm}$ or 
equivalently $Y_{lm}$ that generates diffeomorphism of diffeo-surfaces at the vicinity of a 
point $(\theta,\phi)$ is countable, the total number of generators is innumerable. The reason 
is that spherical harmonic functions $Y_{lm} (\theta,\phi)$ at different points are linearly 
independent. This is an interesting property, because it shows that the Lagrangian for the 
whole Universe (\ref{twodlagrang}) and that of its subsystems (\ref{lagranges}) are inherently 
nonlocal. Moreover, parameters of \sqgr~related to $\suinf$ and dynamics (see, Sec. 
\ref{sec:qmevol}), and their effective values interpreted as the classical spacetime 
(see Sec. \ref{sec:geometry}) are continuous. This is in contrast to QGR models with 
symplectic geometry, which never become truly continuous. Despite explicit continuity of 
operators in (\ref{xcommute}), description of algebra, Cartan decomposition of $SU(\infty)$ 
to $SU(2)$, and Poisson bracket representations of $\suinf$ are more suitable for practical 
applications. 

From (\ref{lharminicexp}) we conclude that vectors of $\hm_U$ are differentiable complex 
valued functions of $(\theta, \phi)$. In particular, $\theta$ and $\phi$ can be themselves 
considered as vectors of the Hilbert space $\hm_U$, They present an orthogonal basis and any 
state of the Hilbert space can be decomposed to orthogonal spherical harmonic functions 
$Y_{lm}$. The relationship between structure constants of $\suinfa$ algebra and Poisson 
brackets of $Y_{lm}$ in (\ref{lapp}) guarantees the consistency of the whole structure. 
Moreover, from (\ref{fthetaphidef}) we conclude that structure constants of the algebra 
(\ref{xcommute}) are generated by 3-point Green functions. Therefore, not only \sqgr~is 
inherently nonlocal but also non-Gaussian.

From now on the pair $(l,m)$ of indices is assumed to satisfy the constraint 
(\ref{lharminicexp}), even if it is not explicitly mentioned. Moreover, if there is no risk 
of confusion, we simplify the notation by using e.g. $a \equiv (l,m)$ as index of 
generators, i.e. $\hL_a \equiv \hL_{lm}$.

\subsection{Quantization} \label{sec:quant} 
In absence of a background spacetime the canonical quantization does not apply to \sqgr. 
Nonetheless, non-commuting algebra of $\suinf$ can replace canonical quantization. The
supplementary section \ref{app:quant} provides physical and mathematical arguments for this 
claim and examples of systems considered to be quantum due to their non-commutative algebra. 
%Specifically, when the $\suinfa$ algebra is written with respect to dimensionful operators proportional to $\hbar$, the Lie algebra commutation relation (\ref{statealgebra}) plays the role of a quantization relation. 
On the other hand, as the algebra (\ref{statealgebra}) is infinite dimensional, there should 
be also a commutation relation similar to the canonical quantization relations - Heisenberg 
uncertainties - between generators of $\suinfa$ algebra $\hL_a$ and their duals conjugate 
$\hJ_b$\footnote{We remind that dual space $V^*$ of a vector space $V$ is the space of maps 
$v^*: V \rightarrow \mathbb{F}$, where $\mathbb{F}$ is the field on which the vector space 
$V$ is defined. A map $v^* \in V^*$ defines a scalar product 
$\langle v^*, v \rangle = v^*(v) \in F, ~ \forall ~ v \in V$.}:
\be
[\hJ_a,\hL_b] = -i \hbar \delta_{ab} \mathbbm{1}, \quad \quad \hJ_a \in \bm[\hm_U^*] 
\label{lquantize}
\ee
where dual conjugate operators $\hJ_a$'s act on the dual Hilbert space $\hm_U^*$ and belong 
to the space of (bounded) linear operators $\bm[\hm_U^*] \cong \bm[\hm_U]$. In particular, 
considering representation (\ref{lharminicexp}) for the generators of $\suinfa$, it is clear 
that operators $\hJ_a$'s are complex valued functions of the parameters $(\theta, \phi)$. 
Consequently, spherical harmonic functions $Y_{lm}$, which according to (\ref{lapp}) have the 
same algebra as $\hL_{lm}$, can be also considered as generators of $\suinfa$ of the dual 
Hilbert space $\hm_U^*$. In supplementary section \ref{app:llmconj} we obtain the expression 
for family of $\hJ_a$ functions that satisfy (\ref{lquantize}). In supplementary 
\ref{app:positionop} we obtain the expression for conjugates of parameters $(\theta, \phi)$, 
that is $i\hbar \partial / \partial (\cos \theta)$ and $i\hbar \partial /\partial \phi$, 
respectively, with respect to generators $\hL_{lm}$ of representation (\ref{lharminicexp}). 
The result shows that their expressions are degenerate, in the sense that they can be 
calculated using any $\hL_{lm}$. Consequently, they carry much less information about the 
quantum Universe constructed here than $\hL_{lm}$'s. In fact giving this degeneracy, it is 
possible to consider $i\hbar \partial / \partial (\cos \theta)$ and 
$i\hbar \partial /\partial \phi$ as being a superposition of $\hL_{lm}$ operators. This 
degeneracy is not a surprise because the subspaces independently generated by 
$i\hbar \partial / \partial (\cos \theta)$ and $i\hbar \partial /\partial \phi$ are merely 
Abelian subspaces of $\suinf$.  
 
%This interpretation can be used to test \sqgr, for instance in experiments sensitive to $\hL_{lm}$ with lowest $(l,m)$. We leave details of this test proposal to a dedicated work on the experimental signatures of this model.
%Note: commutation between a vector and its dual make sense. If \phi is conjugate of v, it means \phi(v) \in C. [\phi, v] v' \equiv \phi(v) v' - v \phi(v'). For square matrix rep. of operators, the dual can be also represented as a square matrix. Therefore, the usual definition of commutation makes sense.
%Note: from wiki Baker–Campbell–Hausdorff formula e^X e^Y = e^Z , Z = X + Y + 1/2 [X,Y] + 1/12 [X, [X,Y]] + ..... For [X,Y] = cte I higher terms are zero.

%Considering the differential form of the dual of $\theta$ and $\phi$, and the fact that $\hm_U \cong \hm_U^*$, operators $\hJ_a$'s are functions of $(\theta, \phi)$. 

As the dual space $\bm[\hm_U^*] \cong \suinf$, it has a representations similar to 
(\ref{lharminicexp}) with respect to two parameters $(p_\theta, p_\phi)$, which are analogous 
to momentum components. The existence of such expansions is a direct consequence of the 
Stone-von Neumann theorem. It states that two sets of operators satisfying standard 
commutation relations are related by a unitary transformation~\cite{qmmathbook}. Specifically, 
these operators must satisfy exponentiated Weyl quantization relation~\cite{weylquant} 
(see e.g.~\cite{weylquantiz} for a recent review) defined as: 
\be
e^{i x_i \hX_i} e^{i p_i \hP_i} = e^{i \hbar \mathbf{x.p}} e^{i p_i \hP_i} e^{i x_i \hX_i}  \label{expocommut}
\ee
where $\mathbf{x}$ and $\mathbf{p}$ are C-number vectors in $d$-dimensional space and 
$0 \leqslant i \leqslant d-1$, and $\hX_i$ and $\hP_i$ are the corresponding dual conjugate 
operators. Using Baker–Campbell–Hausdorff formula: 
$e^{\hX} e^{\hY} = e^{\hZ},~ \hZ = \hX + \hY + 1/2 [\hX,\hY] + 1/12 [\hX, [\hX,\hY]] + \cdots$, 
it is straightforward to show that the canonical commutation relation leads to 
(\ref{expocommut}). Nonetheless, this reformulation of the standard commutation relation is 
for ensuring that the 
domain of the unbounded operators $\hX_i$ and $\hP_i$ cover each other~\cite{qmmathbook}. 
In \sqgr~the number of independent operators $d \rightarrow \infty$. If we use representation 
(\ref{lharminicexp}) of the generators of $\suinfa$, which depend on numerable indices 
$(l,m)$, the exponentiated form of (\ref{lquantize}) becomes:
\be
e^{i L_{lm}(\theta, \phi) \hL_{lm}} e^{iJ_{lm}(\theta, \phi) \hJ_{lm}} = e^{i \hbar L_{lm} J_{lm} (\theta, \phi)} 
e^{i J_{lm}(\theta, \phi) \hJ_{lm}} e^{i L_{lm}(\theta, \phi) \hL_{lm}}. \label{expocommutlm}
\ee
In contrast to Fourier transform of functions or Hilbert space vectors defined on 
$\mathbb{R}^d$ background spacetime, non-Abelian nature of $\suinfa$ algebra means that 
operators $\hL_a$ cannot be considered as analogous to position vector operators 
$\vec{\hX}$ in $\mathbb{R}^d,~d \rightarrow \infty$ space. Hence, \sqgr~is not a quantized 
2D geometry, or as it is the case in string theory, the QFT of a finite number of fields 
on a 2D surface~\cite{hourisqgrcomp}. 

These results complete the demonstration that in \sqgr~the algebra (\ref{statealgebra}) is 
equivalent to the canonical quantization of unbounded operators, specially those operators 
that at low energies are interpreted as the classical spacetime. This property of \sqgr~is 
distinguishable from other QGR models based on non-Abelian algebras, such as those with 
non-commutative geometry~\cite{noncummut,noncummut0,qgrnoncommut,qgrmatrixnoncommut,noncommutstring,qgrgaugedual,qgrmatrix} and can be used to test and discriminate this model. 

\section{Division of the Universe to subsystems} \label{sec:division}
According to a corollary of Symmetry Description of Quantum Mechanics (SDQM) an indivisible 
Universe is trivial~\cite{houriqmsymm,houriqmsymmgr}. Indeed, even in the whole Universe 
model described in the previous section, we can recognize at least two categories of objects: 
pure states which are vectors of the Hilbert $\hm_U$, and linear operators which act on them 
and constitute the space $\bm[\hm_U] \cong \suinf$. On the other hand, in the context of 
a physical model, states and operators are more than mere {\it mathematical objects}, and 
have {\it physical} existence. Therefore, division of the Universe to entities/subsystems 
is inevitable. The issue, however, is how to discriminate these entities from each others. 
Specifically, considering the properties of $\suinf$ group reviewed in supplementary section 
\ref{app:suinf}, homomorphism of one and tensor product of many copies of $\suinf$ according 
to equation (\ref{suinfmult}) means that they cannot be distinguished from each others. 
Therefore, additional structures are necessary for discriminating them\footnote{On the 
other hand, as a quantum system with $\suinf$ symmetry is homomorphic to many copy of 
itself, it would be meaningful to associate a probability to states of the Universe.}. 

In this section we show that such structure can arise due to quantum fluctuations - in other 
words random $\suinf$ self-interaction. In the following subsections we describe how the 
emergence of finite rank symmetries divide the Universe to subsystems. We also discuss how 
the global $\suinf$ symmetry sticks together these subsystems and entangle them.

\subsection{Infinitely divisible Universe}  \label{sec:infdivis}
Necessary conditions for dividing a quantum system to subsystems are described 
in~\cite{sysdiv} and reviewed in supplementary section \ref{app:subsyscriteria}. In the 
case of \sqgr, properties of $\suinf$ symmetry of the Hilbert space $\hm_U$ and possibility 
of its decomposition to any finite rank $SU$ group (\ref{suinfsukdecomp}) imply that the 
Universe can be divided to infinite number of subsystems, each having its own symmetry 
group $G_i$, where index $i$ is used to discriminate subsystems. 

The complementarity condition (\ref{subcomplement}) of subsystem definition imposes the 
following constraint on the symmetries represented by subsystems: 
\be
\bigotimes_i G_i \cong \suinf  \label{gidecomp} 
\ee
Division of the Hilbert space $\hm_U$ to those of subsystems $\hm_i$ representing $G_i$'s is 
discussed in supplementary section \ref{app:subsyssymm}. Considering the properties of 
$\suinf$ tensor product by itself and by other groups (\ref{sunnp} - \ref{bmsuinftensor}), 
subsystems symmetries $G_i$'s can also contain infinite rank factors. 

% Note (from https://mathoverflow.net/questions/27853/infinite-dimensional-unitary-representations-of-su2-for-non-half-integer-j): According to the Peter-Weyl Theorem, all irreducible Hilbert space representations of a compact group (e.g. SU(2)) are finite dimensional. Thus, any infinite dimensional Hilbert space representation will be reducible. Note however, that this applies to finite rank Lie groups. 
% Writing \suinf as tensor product of finit rank SU and the fact that the number of factors can be irrational means that for any rational number of factors it remains infinite many, which can be still considered to be homomorphic to \suinf. This is a consequence of Hoppe's theorem and my demonstration. Therefore, if the number of subsystems of the Universe are innumerable, symmetry of subsystems is always G_i \times \suinf where G_i is finite or rational rank.

\subsubsection{Emergence of internal symmetries and locality in \sqgr}  \label{sec:symmorig}
In~\cite{houriqmsymmgr} we explicitly demonstrated that a symmetry invariant Lagrangian-like 
functional for the whole \sqgr~Universe can be written (we discuss this functional in details 
in Sec. \ref{sec:qmevol}). After application of variational principle to this functional, one 
obtains a trivial vacuum, which is nonetheless unstable. This instability provides the 
necessary conditions for division of the Universe to subsystems, as described in 
supplementary section \ref{app:subsyscriteria}. To demonstrate the emergence of clustering 
and decomposition of the Hilbert space $\hm_U$ and $\bm[\hm_U]$ to approximately orthogonal 
subspaces, we use an operational approach rather than dynamical, because at this stage the 
model is still static. 

A state $|\psi_U\rangle \in \hm_U$ can be expanded with respect to spherical harmonic 
functions:
\bea
|\psi_U\rangle & = & \int d\Omega ~ \psi_U (\theta, \phi) ~ |\theta, \phi \rangle \nonumber \\
& = & \int d\Omega \sum_{\substack{l \geqslant 0 ,\\ -l \leqslant m \leqslant l}} 
\psi_U (l, m, \theta, \phi) ~ |\theta, \phi \rangle  \label{univstate} \\  
\psi_U (\theta, \phi) & \equiv & \sum_{\substack{l \geqslant 0 ,\\ -l \leqslant m \leqslant l}} 
\psi_U (l, m, \theta, \phi) \equiv \sum_{\substack{l \geqslant 0 ,\\ -l \leqslant m \leqslant l}} 
\psi_U^{lm} Y_{lm} (\theta, \phi)  \label{stateexpand}
\eea
The state $|\psi_U\rangle$ is pure because by construction no degree of freedom is traced 
out. Therefore, the density matrix of the Universe as a whole can be written as 
$\hrho_U = |\psi_U\rangle \langle \psi_U|$.
%Note: The following paragraph is not true. \theta and \phi dependence in Y_{lm} can be separated. This means that bais |\theta, \phi\rangle can be separated. 
%It is important to emphasize that in contrast to quantum mechanical description of systems in a spacetime background, the basis $|\theta, \phi\rangle$ cannot be decomposed to $|\theta\rangle \otimes |\phi\rangle$, in the same way that $SU(2)$ eigen vectors $|l,m\rangle$ cannot be decomposed to $|l\rangle \otimes |m\rangle$. In both cases eigen states are characterized by two inseparable parameters. 
Using the last equality in (\ref{stateexpand}), we define another basis for $\hm_U$:
\be
|\psi_U\rangle = \int d\Omega \sum_{\substack{l \geqslant 0 ,\\ -l \leqslant m \leqslant l}} \psi_U^{lm} 
|\ym_{lm} (\theta, \phi)\rangle, \quad \quad |\ym_{lm} (\theta, \phi)\rangle \equiv 
Y_{lm} (\theta, \phi) |\theta, \phi \rangle  \label{ystateexpan}
\ee
The interest of this basis is that amplitudes $\psi_U^{lm}$ do not depend on the continuous 
parameters $(\theta, \phi)$. In this basis a state with constant amplitude 
$\psi_U^{lm} = \mf = cte., ~ \forall ~ (l,m)$ is completely coherent. Such state is unique and 
we call it $|\psi^{cc}\rangle$. Unless a strict fine-tuning - preparation - is applied to the 
system, the uniqueness of $|\psi^{cc}\rangle$ means that the probability of its random 
occurrence is much smaller than partially incoherent states, which in general are less 
homogeneous, in the sense that amplitudes of their components $\psi_U^{lm}$ differ from each 
others. Thus, these states have an intrinsic clustering. Specifically, consider the 
application of $\hL_{l_1m_1}$ on $|\psi^{cc}\rangle$. Using (\ref{lapp}) we find:
\bea
\hL_{l_1m_1}(\theta, \phi) |\psi^{cc} \rangle & = & \mf \sum_{\substack{l \geqslant 0 ,\\ 
-l \leqslant m \leqslant l}} \hL_{l_1m_1} (\theta, \phi) Y_{lm} (\theta, \phi) 
|\theta, \phi \rangle  \nonumber \\
& = & -i \hbar \mf \sum_{\substack{l \geqslant 0 , -l \leqslant m \leqslant l \\ 
l' \geqslant 0 , -l' \leqslant m' \leqslant l'}} f ^{l'm'}_{l_1 m_1,lm} Y_{l'm'} (\theta, \phi) 
|\theta, \phi \rangle  \nonumber \\
& = & -i \hbar \mf \sum_{\substack{l \geqslant 0 , -l \leqslant m \leqslant l \\ 
l' \geqslant 0 , -l' \leqslant m' \leqslant l'}} f ^{l'm'}_{l_1 m_1,lm} 
|\ym_{l'm'} (\theta, \phi) \rangle  \equiv |\mg_{ l_1 m_1}(\theta, \phi) \rangle  
\label{opstateexpand}
\eea
where in the second line we have used equation (\ref{lapp}). Structure constants 
$f ^{l'm'}_{l_1 m_1,lm}$ are proportional to 3j symbols and depend on the indices 
$(l,m), (l',m'), (l'_1,m'_1)$. Thus, in general the new state 
$|\mg_{l_1 m_1}(\theta, \phi)\rangle$ is not any more a completely coherent state and some of 
the components in $|\ym_{l'm'}\rangle$ basis have larger (or smaller) amplitudes than others. 
Consequently, the density matrix $\hrho_U$ becomes more structured and is approximately 
decomposed to blocks. Moreover, this grouping increases with successive application of 
$\hL_a$'s. Inversely, modification of $|\psi_U\rangle$ affects the distribution of 
$\hL_{a}$'s, because $\hrho \in \bm[\hm_U]$ and can be expanded with respect to $\hL_{a}$'s. 
Consequently, its modification alters the probability of interaction and further changes. 
These processes are analogous to absorption of gauge bosons, which are generators of symmetry 
algebra, by matter fields - states - and re-emission of both in a Compton-like interaction. 
In addition, non-commutative algebra of $\hL_{a} \in \suinfa$ implies that they can also 
source themselves, in the same way as in non-Abelian Yang-Mills models such as QCD. 

A priori if the probability of operation/interaction with $\hL_{l_1m_1}$ is the same for all 
$(l_1, m_1)$, the equilibrium can be reestablished. However, due to the randomness of 
operations and infinite number of {\it colors} of $\hL_a$'s, the process of clustering, 
that is the formation of blocks inside $\hrho_U$, is quasi-irreversible and probability 
of returning to the completely coherent state $|\psi^{cc} \rangle$ is negligibly small. 
Clustered subspaces of $\hrho_U$ are by definition (approximately) orthogonal to each 
others and (approximately) satisfy criteria (\ref{subcommut}-\ref{subcomplement}) for 
division of a quantum system to subsystems. 

Finally, after many times action of $\hL_a$'s, the density $\hrho_U$ becomes approximately 
restricted to a {\it blocked} subspace of $\bm[\hm_U]$:
\be
\bm[\hm_U] \supset \bigoplus_i \bm[\hm_i] \ownsbar \hrho_U  \label{hilbertsum}
\ee
where the symbol $\ownsbar$ means {\it approximately belong to}. Sub-Hilbert spaces $\hm_i$ 
can be approximately considered as disconnected subsystems, that is 
$\hm_i \cap \hm_j \approx \emptyset$, for $i \neq j$. In other words, operators 
$\bm[\hm_i] \subseteq \bm[\hm_U]$, which dominantly affect $\hm_i \subset \hm_U$, 
approximately commute with operators in $\bm[\hm_j] \subseteq \bm[\hm_U]$, which dominantly 
affect $\hm_j \subseteq \hm_U$, for $j \neq i$. Thus, they approximately satisfy the condition 
(\ref{subcommut}) for dividing a quantum system to subsystems, and for all practical purpose 
the Hilbert space to which the state of the Universe belongs can be approximated by a tensor 
product:
\bea
&& \hm_U \leadsto \bigoplus_i \hm_i \leadsto \bigotimes_i \hm_i  \\  \label{hilbertmulti}
&& \bm[\hm_U] \leadsto \bigoplus_i \bm[\hm_i] \leadsto \bigotimes_i \bm[\hm_i] \label{hilbertsum}
\eea
where, the symbol $\leadsto$ means {\it approximately lead to}. It is crucial to emphasize 
that the decomposition of density matrix of the Universe $\hrho_U$ is always an approximation, 
otherwise each block could be considered as a separate universe. Moreover, one has to assume 
that as the pointer/basis states in one subspace $\hm_i$ have little coherent mixing 
(superposition) with other subspaces, they acquire a distinguishable physical reality - 
existence - of their own, relative to other subspaces. 

Giving the continuous nature of parameters $(\theta, \phi)$ of the Hilbert space $\hm_U$ and 
$\bm [\hm_U]$, subspaces $\hm_i$ and operators $\bm[\hm_i]$ are also approximately local in 
the parameter space. This property fulfills the second condition - criterion (\ref{sublocal}) 
in supplementary section \ref{app:subsyscriteria} - for division to subsystems. Moreover, the 
complementarity condition (\ref{subcomplement}) is trivially satisfied, because any operator 
$\hO \in \bm [\hm_U]$ is either $\hO \in \otimes_i \bm[\hm_i] \subseteq \bm [\hm_U]$ or 
belongs to a new set of operators $\{A'\}$ such that $\otimes_i \bm[\hm_i] \bot \{A'\}$ and 
$\otimes_i \bm[\hm_i] \otimes \{A'\} \cong \bm [\hm_U]$. Thus, 
$\{A'\} \subseteq \bm[\hm'], ~ \hm' \subseteq \hm_U$ and $\bm[\hm'] \subseteq \bm[\hm_U]$. 

Each set of operators $\{A_i\} \subseteq \bm[\hm_i]$ represents a symmetry group $G_i$ with 
a rank at most equal to the number of linearly independent members of $\{A_i\}$ set. This 
number is necessarily finite, otherwise $\{A_i\} \cong \bm [\hm_U] \cong \suinf$. In this 
case  clustering in $\hrho_U$ was not enough to break the Universe  approximately to 
subsystems. Such case brings back the state of the Universe to the initial state considered 
for this analysis. The argument about subsystems and locality arising from a global symmetry 
can be also inverted. Assume many (infinite) number of indistinguishable quantum systems, 
each representing a finite rank symmetry group $G$. Their Hilbert space represents 
$G \times G \times \ldots \cong \suinf$. Thus, $\suinf$ symmetry in \sqgr~can be considered 
to be due to a bath of large (infinite) number of indistinguishable (abstract) 
decoherence-free quantum subsystems~\cite{qndecohersubsys} with finite rank symmetries. 
In~\cite{hourisqgrym} we use an alternative definition of compositeness of a quantum system to 
obtain these results.

The first decomposition in (\ref{hilbertsum}) is similar to the expansion of a vector space 
with respect to a basis. At this stage of clustering the Universe as a whole is similar to a 
strongly coupled quantum system, where despite presence of some discernible features grouped 
in subspaces $\hm_i$, they cannot be isolated and studied separately. Consequently, unitarity, 
Born rule, and global $U(1)$ phase symmetry of quantum states apply only to the ensemble of 
$\hm_i$. However, when these subspaces become sufficiently isolated, such that $\hm_U$ can be 
approximately considered as their tensor product, unitarity, Born rule, and $U(1)$ phase 
symmetry can be approximately applied to each $\hm_i$ individually. 

\subsection{Symmetries and entanglement of subsystems}  \label{sec:subsyssymm}
Due to the assumed global $\suinf$ symmetry, the fulfillment of conditions 
(\ref{subcommut}-\ref{subcomplement}) for division to subsystems and decomposition 
(\ref{hilbertsum}) of $\hm_U$ to tensor product of $\hm_i$'s are only approximately true. 
Indeed, conservation of $\suinf$ means that subspaces $\hm_i$ are not projected to themselves 
under application of $\hL_a \in \bm[\hm_U]$ with non-zero components outside the $\bm[\hm_i]$
blocks. Therefore, the division of the Universe to subsystems is an approximation 
and no strictly isolated subsystem exists. In the language of quantum information, maximum 
quantum complexities of the Universe and its subsystems is infinite, because preparation 
of their states needs infinite number of qubits and infinite number of binary-gate operations. 
Therefore, to make these properties explicit for each subsystem, we add a $\suinf$ factor 
to the full symmetry $G'_i$ of subsystem $i$ with finite rank local symmetry $G_i$, 
represented by its block in the density matrix of the Universe, that 
is\footnote{In~\cite{hourisqgrym} we prove analytically that (\ref{gdecomp}) is indeed full 
symmetry of purified states of subsystems.}: 
\be
G'_i = G_i \times \suinf \label{gdecomp}
\ee
Considering the properties of tensor products containg $\suinf$ in 
(\ref{suinfgi} - \ref{nginf}), such decomposition remains consistent with the global $\suinf$ 
symmetry. Without loss of generality we can consider the same {\it local/internal} symmetry 
$G$ for all subsystems. Specifically, $G$ can be chosen to be a tensor product containing all 
inequivalent $G_i$'s. Then, some subsystems may represent some of the factor groups in $G$ 
trivially. Indeed, symmetry group of the Standard Model (SM) of elementary particles has such 
a product structure. It is also notable that SM includes the two lowest rank SU groups which 
cannot be further decomposed to non-Abelian groups. 

\subsubsection{Global entanglement}  \label{sec:entangle}
The global $\suinf$ symmetry has a crucial role in keeping together the Universe as a 
quantum system. We discuss this property in the following proposal: 

\refstepcounter{propos}
{\bf Proposition \thepropos \label{propentang}:} {\it In \sqgr~every subsystem is entangled to 
the rest of the Universe:}

{\bf Proof:} This is a consequence of the invariance under the global $\suinf$. To demonstrate 
this claim, consider a division of the Universe to two subsystems with Hilbert spaces $\hm_A$ 
and $\hm_B$ representing symmetry groups $G_A$ and $G_B$, respectively. The mutual information 
of subsystem $A$ and the rest of the Universe $B$ is 
$I(A:B) \equiv S(\hrho_A) + S(\hrho_B) - S(\hrho_{AB})$, where the function $S \geqslant 0$ is 
the von-Neumann entropy of each system in isolation. For two independent system $A$ and $B$, 
the entropy of their ensemble $S(\hrho_{AB}) = S(\hrho_A) + S(\hrho_B)$ and $I(A:B) = 0$. 
In \sqgr~Universe, the state of the whole Universe is by definition pure, because its state 
does not depend on an external system and no observable is traced out. Thus, 
$S(\hrho_{AB}) = 0$. On the other hand, $\hrho_A$ and $\hrho_B$ are obtained by tracing out the 
other subsystem\footnote{In~\cite{hourisqgrym} we obtain explicit expression of these states}. 
Therefore, in general these states are not pure and $S(\hrho_A) \geqslant 0$ and 
$S(\hrho_B) \geqslant 0$. Thus, $I(A:B) \geqslant 0$ and there is no isolated subsystem in 
the \sqgr~Universe. They are all entangled to the rest of the Universe. Variation of state of 
each subsystem leads to a global adjustment to restore the purity and symmetry of their 
ensemble. This is the quantum analogous of gravitational memory in the classical 
spacetime~\cite{gwmemory,gwmemory0}. In~\cite{hourisqgrym} we quantify this entanglement. 

The omnipresent entanglement of subsystems shown here is not a surprise. Irrespective of the 
quantum model constructed for the Universe and more generally for an isolated composite 
quantum systems, there are abundant evidence that its subsystems must be coherently 
correlated. This entanglement manifest itself in various way in composite isolated systems. 
For example, change of quantum reference frame may lead to decoherence and/or 
superselection~\cite{qmrefchangedecoher}. In particular, limitations on the transfer of 
information about the reference frame is equivalent to quantum 
decoherence~\cite{qmref,qmrefdecohere}. Moreover, entanglement and superposition of states 
become perspective dependent~\cite{qmrefdecohere,qmframeperspect}. We should also remind that 
according to the foundation of quantum mechanics properties of an isolated system is 
observable only by its components, because any interaction by an external system violates 
the assumed isolation. In particular, a reference frame used for measurements would be one 
of its components. The emergence of reference-dependent perspectives is equivalent to gauge 
fixing in QFT and general relativity~\cite{qmframeperspect,qmframeperspect0}. The common 
aspect of these phenomena is the reduction of quantum coherence. This property and its 
relationship with symmetry breaking are further discussed in supplementary section 
\ref{app:subsysentangle}. Other consequence of the global entanglement and its observability 
are discussed in supplementary sections \ref{app:entanglecons} and \ref{app:entangleobs}, 
respectively. Perspective dependence in \sqgr~is discussed in~\cite{hourisqgrym}.

\section {Parameter space and algebra of subsystems} \label {sec:subsysparam}
The division of the Universe to subsystems makes it possible to define a reference quantum 
observer and a quantum clock. In this section we study Hilbert spaces of subsystems and 
parameters that characterize their states. We also describe the relation between these 
parameters and what we perceive as a classical spacetime.

\subsection{Emergence of an area/length scale as a measurable} \label{sec:size}
The area of diffeo-surface $\mm$ of a $\suinf$ representation is irrelevant when only one 
representation is considered. In fact, area preserving diffeo-surfaces represent 
$\suinf \times \mathbb{R} \cong \suinf \times U(1) = U(\infty)$, where $\mathbb{R}$ or its 
compactified version $U(1)$ represents scaling symmetry $\mathtt{\Lambda}$ of the 
diffeo-surface. It can be chosen, for instance, to be the area of diffeo-surface or its 
square-root - a size scale. Moreover, diffeomorphism of a compact Riemann manifold $\mm$ can 
be decomposed to an area preserving diffeomorphism and a global scaling, that is:
\be 
Diff (\mm) \cong ADiff (\mm) \times \mathtt{\Lambda} (\mm) \cong \suinf \times U(1) 
\label{diffeodecomp}
\ee
The second homomorphism in (\ref{diffeodecomp}) shows that the scaling symmetry of 
diffeo-surfaces can be identified with the irrelevant global $U(1)$ phase of the 
corresponding Hilbert. On the other hand, in~\cite{hourisqgrym} we demonstrate that when the 
Universe is divided to subsystems, each representing $\suinf$, the $\otimes_i U(1)$ symmetry 
of tensor product of their Hilbert spaces and diffeo-surfaces breaks becomes to a single 
global $U(1)$. The physical consequence of this symmetry breaking is that scaling of 
diffeo-surfaces (or equivalently Hilbert spaces) of subsystems become comparable, hence a 
relative observable. Of course, breaking of scaling symmetry does not mean that area of 
diffeo-surface become fixed, or equivalently Hilbert spaces of subsystems lose their 
projective nature. Rather, in addition to the diffeo-surface parameters $(\theta, \phi)$, 
their states depend on a new parameter $r$ determined with respect to arbitrary phase of 
the state of a reference subsystem or equivalently with respect to arbitrary area of its 
diffeo-surface. 

\subsection{Quantum clocks and emergence of dynamics in \sqgr} \label{sec:clock}
Relative time in quantum mechanics and general properties of quantum clocks, specially in the 
context of \sqgr~are reviewed in supplementary section \ref{app:reldyn}. Overlooking the 
complexity of realistic quantum clocks and issues such as their finite size, which induces 
uncertainties~\cite{qmclock} and decoherence~\cite{qmclock0}, a quantum clock and a time 
parameter satisfying properties described in supplementary section \ref{app:timeprop} can be 
defined by an operator $\hO_c \in \bm [\hm_C]$, which its application on the clock subsystem 
leads to the variation of its state $\hrho_c$. If the clock is tomographically complete, its 
state $\hrho_c$ and its variation $\delta\hrho_c$ under application of $\hO_c$ can be 
determined. Then, the time parameter or more precisely its variation $\delta t$ can be defined 
as a monotonic scalar function of measurements. For example, time variation may be defined 
as being proportional to the amplitude of state variation 
$\delta t ~\propto~ |\delta \psi_c| \equiv |\psi'_c - \psi_c|$ or proportional to the fidelity 
function: %Note: this fidelity function is also called Bures fidelity.
\be
\delta t ~\propto~ F (\hrho_c , \hrho'_c) \equiv \biggl (\tr \sqrt{\sqrt{\hrho_c} ~ \hrho'_c 
\sqrt{\hrho_c}} \biggr)^2, \quad \quad  \hrho'_c \equiv \hrho_c + \delta \hrho_c\label{timefidelity}
\ee
If the clock is not tomographically complete, expectation value or outcome of POVM measurements 
of an observable and its variation can be used as $\delta t$. Alternatively, the time parameter 
$t$ can be defined by parameterizing the average trajectory of the clock state $\hrho_c$ in its 
Hilbert space. The trajectory may be measured using an observable $\hO_c$ of the clock. This 
approach is closer to classical definition of time and $\delta t$ can be defined as 
$\delta t ~ \propto ~ |\tr~(\hrho'_c \hO_c) - \tr~(\hrho_c \hO_c)|$. Obviously, variation 
range of the clock state $\hrho_c$ and observable $\hO_c$ are usually limited. Nonetheless, 
book-keeping of variations extends the range of $t$ to infinity~\cite{qmclockevent}. We remind 
that due to the proposition \ref{propentang}, there is no isolated clock and decomposition of 
the clock observable $\hat{\tt{O}}_c \approx \hO_c \times \mathbbm{1}$ where $\mathbbm{1}$ 
presents the operation on the rest of the Universe is always an approximation. On the other 
hand, the omnipresence of entanglement between clock and the rest of the Universe induces a 
permanent arrow of time. 

\subsubsection{Hamiltonian associated to a clock}  \label{sec:hamiltonian}
Due to the Proposition \ref{propentang} there is no isolated subsystem in the Universe. 
Therefore, if we want to concentrate on the state one subsystem, we must trace out those of 
others and the result would be a mixed quantum state. In~\cite{hourisqgrym} we calculate 
density matrix of such state and its purified form $\hrho_s$. Both mixed and purified density 
matrices of a subsystem depend on the time parameter $t$ according to an arbitrary clock. 
Assuming projective measurements of $t$, we define the operator $\hH_s (t) \in \bm [\hm_s]$ 
called Hamiltonian of subsystem $s$ such that:
\be
\frac{\partial\hrho_s (t)}{\partial t} = -\frac{i}{\hbar} [\hH_s, \hrho_s]  \label{hamiltonian}
\ee
A notable difference between the definition of Hamiltonian here and in quantum models 
originated from a classical formulation is that $\hm_s$ is associated to the time parameter 
$t$, not the other way around. If $t$ is redefined, $\hm_s$ should be also redefined such 
that (\ref{hamiltonian}) remain valid. On the other hand, in the next section we argue that 
time and one of 3 other parameters characterizing states of $\hm_s$ can be considered as 
parameters of $\suinf$ symmetry of the subsystem. We explicitly prove this 
in~\cite{hourisqgrym}. In supplementary sections \ref{app:llmconj} and \ref{app:positionop} 
we find the expression of these parameters and their conjugates as functions of $\suinf$ 
generators. 

Giving the fact that in \sqgr~the whole Universe is static~\cite{houriqmsymmgr}:
\be
\biggl (\sum_{\{s\}} \hH_s \biggr) |\psi_U \rangle = \sum_{\{s\}} \hH_s |\psi_s \rangle = 0, 
\quad \quad \hrho_s \equiv |\psi_s \rangle \langle \psi_s|, \quad \quad 
|\psi_U\rangle \equiv \bigotimes_{\{s\}} |\psi_s\rangle  \label{hamiltonconst}
\ee
where the ensemble ${\{s\}}$ of subsystems includes also the clock. The Hamiltonian $\hH_s$ 
applies non-trivially only on $|\psi_s\rangle$. Implicit in our notation is the equivalence 
of all states $|\psi_U\rangle$ up to a $\suinf$ symmetry transformation. Moreover, it is 
meaningless to interpret the sum over Hamiltonian operators in the l.h.s. of 
(\ref{hamiltonconst}) as the Hamiltonian of the Universe, because by definition and 
construction the Universe is static - there is no clock outside the Universe. Thus, in 
contrast to the ADM formalism~\cite{admgr} and other canonical quantization of gravity, this 
property is not directly related to diffeomorphism invariance of the parameter space and 
dynamics.

\subsection {Quantum states of subsystems}  \label{sec:qmstatesub}
At this point we can conclude that in general the set $\alpha$ of parameters describing 
representation of {\it internal} symmetry $G$ of subsystems and 4 continuous quantities 
$(t, r, \theta, \phi) \equiv x$ related to representations of $\suinf$ by their Hilbert spaces 
and relative variation of states fully characterize their quantum states and relative 
dynamics. Therefore, purified state of a subsystem $|\psi_s \rangle$ can be decomposed 
to~\cite{hourisqgrym}:
\be
|\psi_s \rangle = \sum_{t, r, \theta, \phi, \alpha} \psi^\alpha_s (t, r, \theta, \phi) ~ 
|t, r, \theta, \phi \rangle \otimes |\alpha \rangle  \label{subsysstate}
\ee
where $|t, r, \theta, \phi \rangle \otimes |\alpha \rangle$ is an eigen basis for the Hilbert 
space $\hm_s$\footnote{In mixed states parameter $(t, r, \theta, \phi)$ appear as 
{\it external}, i.e. their quantum origin is lost. In this case, 
$\psi^\alpha_s (t, r, \theta, \phi) |\alpha \rangle$ can be interpreted as a quantum field 
defined on a classical background with coordinates $(t, r, \theta, \phi)$~\cite{hourisqgrym}, 
and summation over them can be considered as statistical averaging.}. The wave function 
$\psi^\alpha_s (t, r, \theta, \phi)$ is not in general factorizable, because parameters 
$(r, \theta, \phi)$ characterize diffeo-surface of $\suinf$ of subsystems and their relative 
area. Although in principle time parameter $t$ is only related to the clock, its 
entanglement with the rest of the Universe and constraint (\ref{hamiltonconst}) indirectly 
relates $t$ to other parameters and the 4 parameters are effectively inseparable. Of course, 
the state of a subsystem may depend on other continuous parameters. The best example is 
temperature in the class of thermal states. However, in contrast to $x$, such parameters are 
not indispensable.

Considering the dimension and interpretation of $x$, it is natural to relate it to what is 
perceived as classical spacetime. We give further precision about exact nature of this 
relation and classical limit of \sqgr~in the following sections. An important consequence 
of this observation is that as a parameter space, the classical spacetime should not be 
quantized. Thus, in this respect \sqgr~is completely opposite of other QGR proposals, which 
usually quantize the unobservable background spacetime and/or its geometry\footnote{This 
trend in construction of QGR models is considered by some authors as re-emergence of the 
infamous concept of ether as a physical entity~\cite{qgrether,etherhistory}.}. 

The parameter space $\Xi = \{ x \}$ have several homomorphic descriptions:
\be
D_2 \times \mathbb{R}^1 \times \mathbb{R}^1 \cong D_2 \times U(1) \times U(1) \cong 
\mathbb{R}^{(1+3)}  \label{paramspace}
\ee
The middle homomorphism in (\ref{paramspace}) is useful for description of $\suinfa$ algebras 
of subsystems, because it is compact. Hence, its closed 2D subspaces are compact and can be 
straightforwardly consider as diffeo-surface. Also notice that in (\ref{paramspace}) we used 
$\mathbb{R}^{(1+3)}$ rather than $\mathbb{R}^4$. We justify this point in 
Sec. \ref{sec:geometry}. Nonetheless, in both cases the 4D parameter space is simply 
connected - except for transformation of parameters under discrete symmetries C, P and T, see 
Sec. \ref{sec:poincare} - even if $D_2$ alone have a non-trivial topology. Considering the 
inseparability of the components of $x \in \Xi$, we can use a local Cartesian for $\Xi$ and 
indicate components of $x$ more democratically as $x^0, x^1, x^2, x^3$.

It is useful to remind that if $\Xi$ was a physical entity rather than a property, 
compactification of some of its coordinates could have drastic consequences, for instance 
end of everything in a finite time due to singularities~\cite{compacttime}; violation of 
Copernican principle~\cite{comosclosed}; and various cosmological effects impacting the 
inflation and the Cosmic Microwave Background (CMB)~\cite{comosclosedcmb,comosclosedcmb0}, 
and the Large Scale Structure (LSS)~\cite{lssclose,lssclose0}. By contrast, compactification 
of the parameter space $\Xi$ corresponds to a redefinition of parameters, under which physics 
should be invariant. Hence, it would not induce any new physical observable. In any case, 
giving the emergence of locality in many-body systems due to the limited speed of quantum 
information transfer (see Sec. \ref{sec:geometry} for details) and the growth of 
inhomogeneities~\cite{qmmbl}, there would be negligible impact on the physical process due to 
the compactification (or not) of the parameter space. 

The collective parameters $\alpha$ which characterize representations of the finite rank 
symmetry $G$ are usually discrete. In this work we do not further discuss $G$ symmetry and 
its representations by subsystems. The particle physics motivated by \sqgr~is an important 
subject and needs a dedicated investigation, which we leave to future works.

\subsubsection{The $\mathbf{\suinfa}$ algebras of subsystems and their commutation relations} 
\label {sec:subsysalgebra}
Due to the inseparability of the 4 continuous components of the parameter vector $x \in\Xi$ 
every pair $(\eta, \zeta)$ of components can be used to characterize $\suinf$ symmetry of the 
Hilbert space a subsystem and its algebra as defined in (\ref{lharminicexp}-\ref{lapp}) and 
(\ref{xcommute}-\ref{xdef}). More precisely, the diffeo-surface of $\suinf$ symmetry at $x$ is a 
2D compact surface with induced coordinates $(\eta, \zeta)$. Without loss of generality we can 
parameterize $\Xi$such that at each point $x$ two of coordinates $x^\mu, ~ \mu = 0, \cdots, 3$ 
be equal to $(\eta,\zeta)$, see~\cite{hourisqgrym} for details, in particular, it is shown that 
different choices correspond are equivalent up to a $\suinf$ transformation which does not affect 
observables. The arbitrariness of the embedding of the diffeo-surface in the parameter space 
means that the division of the Universe into subsystems is not rigid. Moreover, the compactness 
of diffeo-surfaces of $\suinf$ representations can be overlooked because the definition of 
generators $\hL_{lm}$ in (\ref{lharminicexp}) would be still usable. This would be equivalent 
of compactifying diffeo-surface, and thereby $\Xi$, at infinity. Indeed, spherical harmonics 
are special cases of the generalized hypergeometric functions~\cite{inttablebook} and the 
associated Legendre polynomial $P_{lm}$ in (\ref{sphereharmdef}) is related to the Gauss 
hypergeometric functions $_2F_1$:
\be
P_{lm} (\eta) = (-1)^m \frac{\Gamma (l + m +1) (1 - \eta^2)^{m/2}}
{2^m \Gamma(l - m +1) m!}~ _2F_1 (m-l, m+l+1; m+1; (1-\eta)/2)  \label{legendrehypergeo} 
\ee
with $l+m \geqslant 0$ and $\mathcal{R}e (\eta) \geqslant -1$. Therefore, spherical harmonics 
and their Poisson algebra,, and thereby $\suinfa$, can be analytically extended outside the 
range of compact angular parameters $(\theta, \phi)$, except for singular points at $\pm 1$ 
and $\infty$\footnote{We discuss the case of what is called the torus basis for $\suinf$ 
generators in~\cite{hourisqgrym}. They have a Fourier transform expression in a 2D flat space.}. 
%Note: Relation between P_{lm} and 2F1 = F page 968 formula 8.751 in integral book
%Although ranges of $r$ and $t$ are usually chosen to be the non-compact $\mathbb{R}$, the homomorphism $\mathbb{R} \cong U(1)$ means that by identifying points at infinity, that is $x^i \rightarrow \pm\infty ~ i = 0, \cdots, 3$ in Cartesian coordinates, the parameter space can be made compact such that the surface passing through every pair of coordinates become a 2D compact (pseudo)-Riemann surface and its $ADiff$ a representation of $\suinf$~\cite{suninfrep}. 

Expansion and commutation relations of one of these parameters, which we call $\eta$, and its 
conjugate $i\hbar \partial / \partial \eta$ with respect to $\hL_{lm}$ generators are described 
in supplementary section \ref{app:positionop}. They apply to all parameters of 
$\suinf$ symmetry of subsystems, irrespective of how they are defined\footnote{The limiting 
points such as $|\eta| \rightarrow \infty$ may need special treatment, in particular if 
diffeo-surfaces have non-trivial topology.}. Thus, one can construct a $\suinfa$ algebra from 
any pair of the 4 parameters of the Hilbert space of subsystems:
\be 
[\hL_{l,m} (\eta,\zeta), \hL_{l',m'}(\eta,\zeta)] = f_{lm,l'm'}^{l"m"} \hL_{l'',m''} (\eta,\zeta) 
\label{algebra4d}
\ee 
under the condition that an appropriate associated Legendre polynomial is used in 
(\ref{lharminicexp}). Crucially, giving the similarity of commutation relations between 
parameters and their conjugates to the standard quantum mechanics, the usual 
uncertainty relations of quantum mechanics~\cite{qmheisenuncert,qmspeed} can be applied 
to the time parameter $t$, irrespective of how a quantum clock is defined.

\subsection {Parameter transformation} \label{sec:paramgeo}
The expression (\ref{subsysstate}) of pure state $|\psi_s\rangle$ of a subsystem shows that 
the Hilbert space $\hm_s$ can be decomposed to $\hm_s = \hm_{s_\infty} \times \hm_{s_G}$, where 
$\hm_{s_G}$ represents $G$ symmetry\footnote{See ~\cite{hourisqgrym} for more details about 
structure and properties of the states of subsystems in \sqgr.}. The choice of a basis for 
each component is arbitrary. Nonetheless in this section we concentrate on the $\suinf$ 
symmetry and dynamics of subsystems and only discuss transformation of subsystems states 
under basis transformation of $\hm_{s_\infty}$, which corresponds to tracing out the dependence 
of $|\psi_s\rangle$ on $\alpha$\footnote{A possible effect of this action is further 
entanglement of quantum states of subsystems and enforcement of inseparability of remaining 
parameters.}. 

A basis transformation of the Hilbert space $\hm_{s_\infty}$ can be written as:
\be
|x' \rangle = \sum_{\Xi} \hU |x \rangle, \quad \quad \hU \in \bm [\hm_{s_\infty}]  
\label{basetrans}
\ee
As every pair of components $x^i,~i=0, \cdots, 3$ of the parameter $x \in \Xi$ can be used for 
characterizing $\suinf$, the transformation (\ref{basetrans}) can be also interpreted as 
$\suinf$ transformation. The relative nature of parameters means that comparison of multiple 
subsystems is meaningful only if the basis transformation is applied to all of their Hilbert 
spaces. 

In Sec. \ref{sec:symm} we showed that states of $\hm_s$, and thereby $\hm_{s_\infty}$, are complex 
functions on the parameter space $\Xi$. Thus, the basis transformation (\ref{basetrans}) of 
$\hm_{s_\infty}$ can be equally interpreted as a diffeomorphism of the parameter space $\Xi$: 
\be
x = (t, r, \theta, \phi) \rightarrow (t' (x), r' (x), \theta' (x), \phi' (x)), \quad \quad x, ~x' \in \Xi  \label{paramdiffeo}
\ee

\subsubsection{Poincar\'e symmetry}  \label{sec:poincare}
In addition to diffeomorphism of the parameter space $\Xi$, states of subsystems must 
be in non-trivial representations of its global Lorentz and translation transformation, which 
together constitute the Poincar\'e group. A reminding of the relationship between 
diffeomorphism and reparameterization can be found in supplementary section 
\ref{app:reparam}. Observations show that both the Standard Model~\cite{smlorentz} and 
classical gravity~\cite{grlorentz} are invariant under Lorentz group, which is isomorphic to 
$SO(3,1)$. However, in contrast to $\suinf$ and more generally $SU(N) ~ \forall N$, which 
are simply connected, $SO(3,1)$ is doubly connected under discrete parity $P$, conjugation 
$C$, and time reversal $T$ transformations. Considering the assumed $\suinf \cong SL(2,R)$ 
symmetry of subsystems, it is clear that quantum superposition of subsystems must play a 
role in the preservation of $C,~P,~T$ symmetries. For instance, parity conserving fermions 
of SM are in $(1/2,0) \oplus (0,1/2)$ representation of $SO(3,1)$. Thus, each of the 
components representing a connected part of $SO(3,1)$ can be embedded in one $\suinf$. 
However, only the superposition of disconnected sectors preserves $C,~P,~T$. This property 
is used by leptogenesis models to explain matter-antimatter asymmetry using the observed 
parity and CP violation in the leptonic sector of the Standard Model (SM), see 
e,g~\cite{leptobaryo} for a review. Although in the SM the quantum number$B-L$, where $B$ 
and $L$ are baryonic and leptonic number, respectively, is conserved, at present we do not 
know whether these symmetries are conserved in the dark sector. In the framework of \sqgr, 
the fact that the symmetry of the whole Universe as a quantum system is $\suinf$, which is 
simply connected, means that the dark sector should conserve $CPT$. If the visible and dark 
sectors interact non-gravitationally, $CPT$ may break - but not necessarily - in each sector. 
Nonetheless, it must be conserved by their ensemble.

\subsection{Curvature of the parameter space} \label{sec:paramcurve}
The invariance of physical processes under a transformation similar to (\ref{paramdiffeo}) 
means that in general the 4D parameter space $\Xi$ is curved and has a nontrivial metric 
$g_{\mu\nu}$. The invariance under reparameterization (\ref{paramdiffeo}) makes 4 of the 
10 components of $g_{\mu\nu}$ arbitrary. However, the remaining 6 parameters are 
undetermined. In classical Einstein gravity $g_{\mu\nu}$ is related through the Einstein 
equation to the distribution of matter - more precisely to the 6 independent components of 
the energy-momentum tensor $T_{\mu\nu}$. This implies that for a given $T_{\mu\nu}$ - or 
rather its frame independent trace $T$ - the diffeomorphism of the spacetime preserves 
its scalar curvature\footnote{In general relativity for a given distribution of matter, for 
example vacuum, there is an equivalence class of metrics which lead to the same curvature. 
They are related by 4 constraints that present reparameterization (diffeomorphism) of the 
spacetime. The 6 d.o.f of $g_{\mu\nu}$ determine whether the scalar curvature changes by this 
transformation. The remaining 4 d.o.f. are associated to redefinition of coordinates without 
changing observables~\cite{curvatureinvar}.}. Therefore, the question arises how the metric 
$g_{\mu\nu}$ is determined in \sqgr~and whether a curvature preserving condition exists. Here 
we show that the curvature of parameter space $\Xi$ is irrelevant for the physics. A more 
detailed demonstration will be presented in~\cite{hourisqgrym}.

\refstepcounter{propos}
{\bf Proposition \thepropos \label{propparamcurve}:} {\it The curvature of the parameter 
space $\Xi$ of subsystems can be made trivial by a $\suinf$ gauge transformation, under 
which the Universe and its subsystems are invariant.}

{\bf Proof:} Consider the compact version of $\Xi$ defined in (\ref{paramspace}). At any point 
$x \in \Xi$ consider the orthogonal basis $e_i \in T\Xi_x, ~ i=0, \cdots, 3$, where $T\Xi_x$ 
is the tangent space of $\Xi$ at $x$. Ricci curvature tensor and scalar curvature can be 
determined using sectional curvatures of the 2D subspace of $T\Xi_x$ expanded by pairs 
$(e_i, e_j),~ i \neq j, ~ i, j=0, \cdots, 3$ and vis-versa, see supplementary section 
\ref{app:curvature} for coordinate independent definition of curvatures and relations  
between them. Consider the 2D surface 
$\Sigma_{ij} \subset \Xi$ generated by parallel transport of $(e_i, e_j),~ i \neq j$. Such 
surfaces are compact. The compact surface $\Sigma_{ij}, i, j \in \{0, \cdots, 3\}, i \neq j$ 
can be considered as diffeo-surfaces of $\suinf$ symmetry of subsystems. Thus, its $ADiff$ 
deformation is equivalent to application of $\suinf$ symmetry group, which does not change 
observables of the subsystems. Crucially, these $ADiff$ do not need to preserve sectional 
curvatures at $x$. Thus, Ricci curvature tensor is not in general preserved. Nonetheless, 
this does not have any impact on quantum observables. A general diffeomorphism can be 
decomposed as in (\ref{diffeodecomp}). Because area of diffeo-surfaces is irrelevant for 
the homomorphism of their $ADiff$ with the representations of $\suinf$, scaling of all 
$\Sigma_{ij}$'s and thereby $\Xi$ is similarly irrelevant for quantum observables of 
subsystems. Therefore, geometry of the parameter space $\Xi$ is irrelevant for the dynamics 
of the Universe and its quantum subsystems. In particular, in the non-compact version of 
the parameters or when areas of $\Sigma_{ij}$ are scaled to infinity, $\Xi$ can be considered 
as a flat space. 

The Proposition \ref{propparamcurve} has a crucial contribution in: construction of dynamics; 
what we may call classical limit of \sqgr~despite the fact that it is fundamentally quantum; its 
relationship with the Einstein gravity; and physical interpretation of what we perceive as the 
curvature of spacetime in presence of matter and its fluctuations observed as gravitational 
waves. For instance, an immediate conclusion from this proposition is that the parameter 
space $\Xi$ of $\suinf$ symmetry and dynamics of subsystems cannot be itself the classical 
spacetime. Nonetheless, in Sec. \ref{sec:geometry} we show that they are conceptually related. 
From now on we indicate the arbitrary metric of the 4D parameter space 
$\Xi$ as $\upeta_{\mu\nu}$ (or its 2-dimensional analogue for 2D parameter space of each 
diffeo-surface) to distinguish it from the effective classical metric $g_{\mu\nu}$, which we 
obtain in the next section. Finally, the Proposition \ref{propparamcurve}, which demonstrates 
embedding of the geometric connection in the $\suinf$ gauge connection - see 
Sec. \ref{sec:qmevol} - provides a natural explanation for gravity/gauge duality 
conjecture~\cite{qgrgaugesep,qgrgaugesep0,qgrgaugesep1,gaugestringcorr,stringgauge0,stringgauge1}. 

\section{Effective classical geometry from quantum speed limits}  \label{sec:geometry}
%{Classical curved space and its relationship with quantum states of subsystems} 
In the previous section we showed the arbitrariness of the geometry of parameter space $\Xi$ 
of $\suinf$ representations by subsystems and their relative dynamics. Therefore, despite 
its similarity to the observed classical spacetime, $\Xi$ cannot be identified with it. 
Otherwise, \sqgr~were already ruled out, because effects of the curvature of spacetime on 
observable are observed in many gravitational phenomena such as: deformation of x-ray 
spectral lines emitted from accretion disk of black holes~\cite{bhfeline}, mirage images of 
black holes~\cite{bhmirage}, gravitational waves emitted by compact objects, etc. In this 
section we discuss the emergence of the perceived classical curved spacetime in~\sqgr from 
quantum properties of subsystems.

\subsection{Emergence of classical spacetime}  \label{sec:fullpath}
In the supplementary section \ref{app:qslgeom} we briefly remind the historical role of the 
limited speed of light in the development of general relativity, and review its analogous in 
quantum physics, called Quantum Speed Limit (QSL) in Sec. \ref{app:qmspeed}. 
In~\cite{houriqmsymmgr} we used QSL to define a quantity analogous to the infinitesimal 
affine separation of a classical metric. It presents the path of the quantum state of 
subsystems in their Hilbert and can be parameterized with (3+1) parameters defining an 
effective {\it position} for the state in its the parameter space $\Xi$. We interpreted the 
space of these (3+1)D vectors as the perceived classical spacetime and demonstrated that its 
metric has a negative signature\footnote{According to the Proposition \ref{propparamcurve} the 
parameter space $\Xi$ does not need to have a specific signature. Indeed, in some 
circumstances one may need to extend the parameter space, in particular the time parameter, 
to complex plane. For example, thermal QFT can be treated as a zero temperature QFT with 
complex time. Some QFT models use a field as time parameter and signature of the metric may 
change~\cite{qftmetricsignature}. Nonetheless, for being consistent with the average geometry, 
which is an observable, it is convenient to use a Lorentzian metric for $\Xi$ by default and 
treat other cases as exceptions.}. In the supplementary section \ref{app:qslgeom} we review 
these results and extend them to the case where the state of subsystems - excluding the clock 
and the reference - is mixed and evolves as an open quantum system. These results can be 
summarized in the following metrics
\be
ds^2 = \Lambda \mF[\hrho,\delta\hrho] \equiv g_{\mu\nu}(x) ~ dx^\mu dx^\nu, \quad \quad 
\mu, \nu = 0, \cdots, 3, \quad x \in \Xi \label{tracemetricalpha}
\ee
where $\Lambda$ is a dimensionful constant to adjust the units and $\mF[\hrho,\delta\hrho]$ 
is a function of the state $\hrho$ defined in (\ref{densitynog}) and its infinitesimal 
evolution evolution $\delta \hrho$. It can be interpreted as a distance between $\hrho$ and 
$\hrho + \delta \hrho$, see Sec. \ref{app:qmspeed}. The supplementary section 
\ref{app:rhoexpad} describes the expansion of state $\hrho$ and its variation with respect to 
$\hL_{lm}$ generators of $\suinf$. 

The function $\mF[\hrho,\delta\hrho]$ depends on the type of the evolution considered for 
subsystems, and in the case of non-unitary evolution on the metric an distance function used 
for quantifying difference between two states, see Sec. \ref{app:qmspeed}. According to 
QSL's (\ref{aqdefsqrt}) and (\ref{qslrelpurity}) the function $\mF[\hrho,\delta\hrho]$ has the 
following form, respectively:
\be 
\begin{cases}
\mF[\hrho,\delta\hrho] = \tr (\sqrt{\delta\hrho} \sqrt{\delta\hrho}^\dagger), & 
\text{For unitary evolution of pure or mixed state of subsystems;} \\  %Note:Tthe r.h.s. is canonical distillation of \delta \hrho
\mF[\hrho,\delta\hrho] = (\tr(\hrho \delta \hrho))^2, & \text{For Markovian evolution of 
the state of subsystems as an open system.}
\end{cases} \label {distcase}
\ee
In the geometric interpretation of 
QSL's~\cite{qmspeedgeom,qmspeedgeomopen,qmspeedgeomgen0,qmspeedgeomlocal} 
the function $\mF[\hrho,\delta\hrho]$ is the geodesic distance between $\hrho$ and 
$\hrho + \delta \hrho$ for the metric defined in (\ref{ineqgen}). The geodesic path connecting states in the Hilbert 
space can be pushed back to the parameter space $\Xi$. In this way we can define an 
{\it proper distance} for the effective geometry (\ref{tracemetricalpha}):
\bea
dl^2_{12} & \equiv & \Lambda^2 \mF[\hrho,\delta\hrho] |_{t = cte}  \label{ddistace} \\
l_{12} & = & \min \int_{\gamma_{12}} dl_{12}, ~~ \forall {\gamma_{12}}  \label{distancedef}
\eea 
where $\gamma_{12}$ is a path in the Hilbert space connecting states $\hrho_1$ and $\hrho_2$. 
By construction the r.h.s. of (\ref{ddistace}) and (\ref{distancedef}) are independent of the 
Hilbert space basis and reparameterization. The reason for considering all the paths between 
two points is the fact that for non-unitary evolution there are multiple QSL's and metrics, 
which may not be optimal or achievable, see Sec. \ref{app:qmspeed} for more details.

In analogy with exchange of virtual particles with negative amplitude of 4-momentum 
$q^2 < 0$ in t-channel, states with $ds^2 < 0$ can be called virtual. Indeed, for pure and 
reachable states $ds^2 \geq 0$. Therefore, $ds^2 < 0$ presents evolution of $\hrho$ to a 
not causally distinguishable state $\hrho + d\hrho$, see Sec. \ref{sec:causal} for the 
definition of causality. 

\begin{figure}
\begin{center}
\includegraphics[width=8cm]{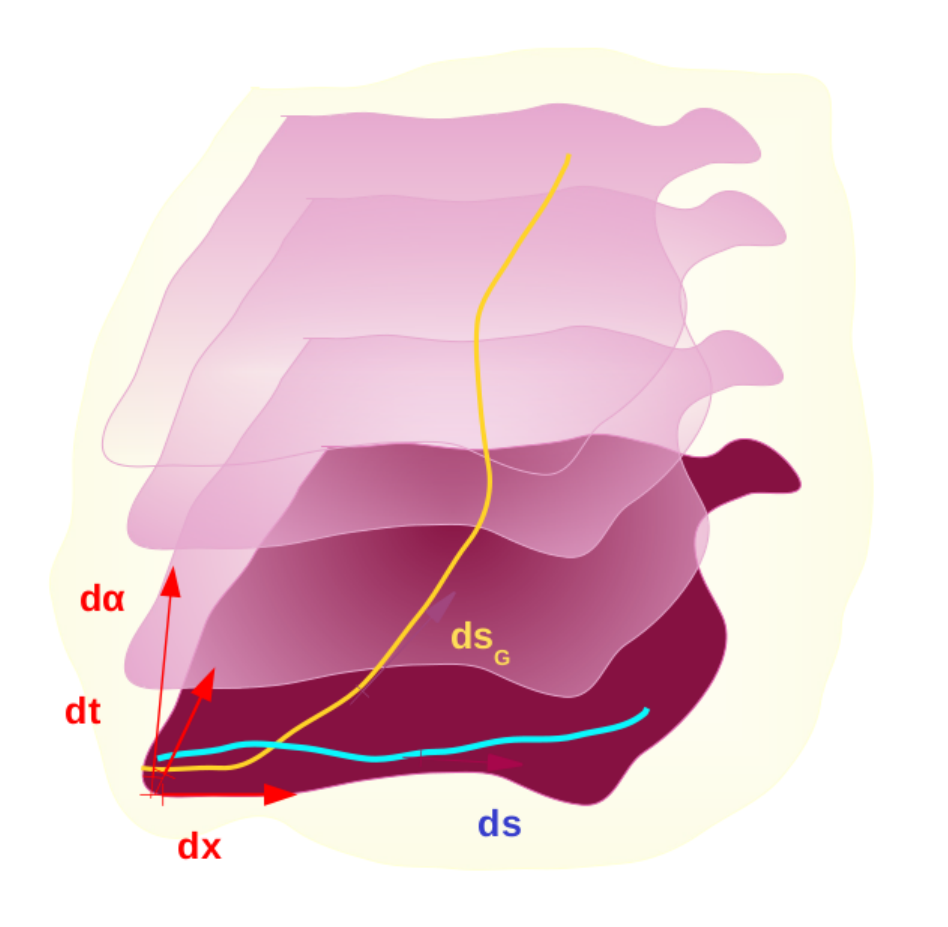} 
\end{center}
\vspace{-1cm}
\caption{A schematic presentation of \sqgr~parameter space. The lowest (darker) surface 
depicts the parameter space of $\suinf$ symmetry - the spacetime - of subsystems. The path 
inside this surface is the average - classical - path of the quantum states of subsystems in 
the parameter space. Orthogonal to this surface depicts the collective parameter $\alpha$ of 
the internal symmetry $G$. Surfaces at discrete values of $\alpha$ remind that as a 
finite-rank Lie group, representations of $G$ are in general characterized by discrete 
quantities - quantum numbers. The background presents parameter space of representations 
of $\suinf \times G$ as a continuum. The path crossing surfaces is the average path of 
subsystems in the full parameter space. \label{fig:paramgeometry}} \end{figure}

For obtaining (\ref{tracemetricalpha}) we traced out the contribution of internal symmetries 
in the density matrices to obtain an effective metric analogous to that of the general 
relativity. A more general metric includes the contribution of collective parameter $\alpha$ 
defined in (\ref{subsysstate}):
\be
ds_G'^2 \equiv \Lambda \mF[\hrho_G,\delta\hrho_G] = g_{\mu\nu} (x', \{\alpha\}) ~ 
dx'^\mu dx'^\nu + g_{ij} (x', \{\alpha\}) ~ d\alpha^i d\alpha^j, \quad \quad 
\mu, \nu = 0, \cdots, 3, \quad i,j \in \{1, \cdots, N\} \label{newmetricalpha} 
\ee
where for the sake of simplicity of notation we have kept the same symbol for the effective 
metric before and after tracing out the contribution of internal symmetries. The index $G$ 
is used wherever the contribution of $G$ symmetry is not traced out. We assumed the the set 
of internal symmetry parameters $\{\alpha\}$ includes $N$ parameters. They characterize 
representations of $G$ by quantum state of subsystems. In (\ref{newmetricalpha}) they are 
treated as continuous such that the metric (\ref{newmetricalpha}) have the form of a 
$(1+3)+N$-dimensional pseudo-Riemannian geometry with $ds_G^2$ as its affine parameter. 
However, as $\suinf$ and $G$ are by definition orthogonal and independent $g_{\mu i} = 0$ 
for any parameterization of the two sectors. Moreover, as by definition $G$ has a finite 
rank, its representations usually have discrete spectra\footnote{This is the case for all 
the observed internal symmetries of elementary particles and many body quantum systems. 
For this reason we only consider linear representations of $G$.}. Therefore, $g_{ij} \neq 0$ 
only for discrete values of $\{\alpha\}$. Fig. \ref{fig:paramgeometry} schematically presents 
the emergent effective geometry according to (\ref{tracemetricalpha}) and 
(\ref{newmetricalpha}), and their relationship\footnote{One might claim that the relation 
between classical affine parameter and quantum state of the subsystems is not new, because 
in semi-classical general relativity geometric part of the Einstein equation is related to 
the expectation value of energy-momentum tensor. However, the latter is a classical quantity 
and it is well known that its quantum counterpart is not well defined and its zero mode 
diverges - at least naively~\cite{devacuum0,devacuum1}. By contrast, density matrix is 
an intrinsically quantum and well defined quantity.}. The effective geometry and its 
relationship with the quantum state of subsystems of the Universe have various consequences. 
They are described in supplementary section \ref{app:effgeometry}.

Despite the crucial role of (\ref{tracemetricalpha}) and (\ref{newmetricalpha}) for 
understanding the nature of classical spacetime and its relationship with the quantum realm 
and contents of the Universe, they have little use for understanding evolution of subsystems, 
because they do not indicate the super-operator of state evolution. This is similar to 
the general relativity. Although a suitable form for the metric may be chosen based on the 
symmetries of matter distribution - for instance a spherically symmetric metric for a 
non-rotating spherically symmetric distribution of matter - a dynamical equation such as 
Einstein equation is needed to relate spacetime geometry to the evolution of matter. 
The next section will describe dynamics of \sqgr. 

\section{\bf Evolution}  \label{sec:qmevol}
%So far we constructed and studied the Hilbert space of \sqgr~as a model for the Universe and its subsystems without considering the dynamics. 
In this section along with the review of dynamics of \sqgr~first reported in~\cite{houriqmsymmgr} 
and with more details in~\cite{hourisqgrcomp}, we clarify some of the points which were not 
discussed in those works. 

Experience from QFT has shown that we can use variational principle at all perturbative orders 
to describe dynamics of many-body quantum systems. In this approach, the Lagrangian 
(or Hamiltonian) is usually constructed based on the classical limit of the QFT and is subsequently 
quantized, for instance using canonical quantization of fields (degrees of freedom) or through 
path integral formalism. In the case of \sqgr, however, the model is constructed as a quantum system, 
and does not have a classical analogue. In fact, as we discussed in Sec. \ref{sec:axioms}, for 
$\hbar \rightarrow 0$ the algebra (\ref{statealgebra}) becomes Abelian and the model trivial. 
Therefore, the construction of a Lagrangian for \sqgr~must be based on symmetries of its Hilbert 
space, that is $\hm_U$ for the whole Universe and $\bigotimes_s \hm_s$ when the Universe is 
presented by the ensemble of its subsystems. In the first case the Lagrangian must a $\suinf$ 
invariant functional, describing all possible excursion of states in the Hilbert space $\hm_U$. 
For the ensemble of subsystems the Lagrangian functional must be, in addition, invariant under the 
finite rank $G$ symmetry.

\subsection{The whole Universe} \label{sec:evolwhole}
 According to a description of quantum mechanics axioms based on symmetry as a foundational 
concept~\cite{houriqmsymm}, a single closed quantum system without structure is static and 
trivial. Thus, a genuinely isolated quantum system such as the whole Universe in~\sqgr is 
necessarily static. Nonetheless, a static Lagrangian - a functional operators $\hO \in \bm[\hm_U]$ 
invariant under application of $\suinf$ and under redefinition of parameters characterizing the 
its representation by $\hm_U$ - can be defined and variational principle can be applied to obtain 
the equilibrium state of the Universe. 
%In \sqgr~the whole Universe represents $\suinf$ % symmetry, where its 
%\footnote{The group $\suinf$ has also representations depending only on discrete parameters~\cite{suninftorus,suninfrep}. However, they are dense in $\mathbb {R^(2)}$. This is the reason for their homomorphism with Virasoro algebra without central charge discussed in Sec. \ref{app:stringcomp}.}, 
%\subsubsection{Formal description of dynamics} \label{sec:evolwholeform}  %#### not referred
%In absence of a background spacetime in \sqgr~the Lagrangian functional must be defined on the Hilbert space and must be invariant under $\suinf$. 

For $SU(N), ~\forall N$ traces of multiple hermitian generators, that is 
$\tr (T^{a_1} T^{a_2} \cdots)$ are invariant under group conjugate operation 
$g^{-1} g_1 g = g^\dagger g_1 g, \forall ~ g, g_1 \in SU(N)$. Moreover, these traces depend only on 
the normalization of generators $\tr (T^a T^b) \equiv 1/2 \delta_{ab}$, anti-symmetric structure 
constants $f_{abc}$, and symmetric anti-commutation constants 
$d_{abc} \equiv 2 ~ \tr (\{T^a, T^b\} T^c)$. For $\suinf$ structure constants in sphere basis are 
defined in (\ref{statealgebra}). Other coefficients are calculated in~\cite{suninfhoppthesis}. 
Therefore, these traces can be used to construct a symmetry invariant Lagrangian. In analogy with 
QFT Lagrangians, it is enough to consider the lowest order traces of generators. Higher orders 
can be considered as higher order quantum corrections and obtained from a path integral.  

Representations $\suinf$, for example (\ref{lharminicexp})\footnote{See~\cite{hourisqgrym} for a 
review of other representations.}, depend on two continuous 
parameters~\cite{suninftorus,suninfrep} and variational method can be straightforwardly applied 
to the Lagrangian functional. Moreover, fluctuations of the Lagrangian and fields - parameter 
dependent coefficients of trace terms - around their equilibrium values can be interpreted as 
quantum fluctuations of the ground state. In the operational view employed in 
Sec. \ref{sec:division}, these fluctuations correspond to random application of the members 
of $\bm[\hm_U] \cong \suinf$ to the quantum state of the Universe. 

Under above considerations, the lowest order $\suinf$-invariant functional Lagrangian of the 
Universe can be written as:
\bea
\lm_U & = & \int d^2\Omega  \biggl [\frac{1}{2} \sum_{a,~b} \tr (L^*_a(\theta,\phi) 
L_b (\theta, \phi) \hL_a \hL_b ) + \frac {1}{2} \sum_a \biggl (L_a (\theta, \phi) 
\tr (\hL_a \hrho_U) + C.C \biggr ) \biggr ] \nonumber \\ 
&& \label{twodlagrang} \\
d^2\Omega & \equiv & \sqrt{|\upeta^{(2)}|} d(\cos \theta) d\phi \label{twodvolume}
%Note: \upeta is used for 4D parameter space \Xi not for 2D diffeo-surface.  
\eea
where $a \equiv (l,m)$ and C-number amplitudes $L_a$ determine the contribution of $\suinf$ 
generators $\hL^a$ to the Lagrangian. The dimensionless $\suinf$ coupling constant is included 
in the amplitudes. The density matrix $\hrho_U$ is the quantum state of the Universe as a 
whole. The scalar function $\upeta^{(2)}$ is the determinant of metric 
$\upeta^{(2)}_{\mu\nu},~ \mu, \nu \in {\theta, \phi}$ of the 2D diffeo-surface $\mm$ of the 
Universe. By definition the whole Universe is in a pure state, because there is nothing 
outside and entangle to it, which could have been possibly traced out. Therefore, its density 
matrix can be written as $\hrho_U = |\Psi_U \rangle\langle \Psi_U|$. We should also remind 
that the integration over angular coordinates of $\mm$ is part of the tracing 
operation, because generators (\ref{lharminicexp}) of the $\suinf$ symmetry are defined at 
each point of the diffeo-surface. This reflects the fact that diffeomorphism of the 
diffeo-surface in the neighborhood of each point is independent from deformation around other 
points up to continuity condition. To be consistent with Lagrangians used in QFT we assume that 
$\lm_U \in \mathbb{R}$. Using representation (\ref{lharminicexp}) for $\hL_{lm}$ it is 
straightforward to show that $\hL^\dagger_{lm} = (-1)^{(m+1)} \hL_{l,-m}$ (see supplementary section 
\ref{app:positionop}). This means that (\ref{lharminicexp}) is an anti-hermitian representation. 
For this reason there is no $\tr (\hL^\dagger_a \hL_b)$ term in (\ref{twodlagrang}) and phases 
of amplitudes $L_a$ in the first term of (\ref{twodlagrang}) are irrelevant. In addition, 
in the second term $\hrho_U$ is hermitian. Therefore, $L_a (\theta, \phi)$ must be real, 
otherwise $\lm_U$ becomes complex. This conclusion is not limited to the representation 
(\ref{lharminicexp}) and applies to all (anti-)hermitian representations of $\suinf$. 

In~\cite{houriqmsymmgr} it is explicitly shown that, as expected, applying variational 
principle with respect to amplitudes $L_a$ and components of $\hrho_U$ leads to a trivial 
vacuum state as the equilibrium solution. This result is consistent with the corollary of 
SDQM stating that an isolated indivisible quantum system is trivial~\cite{houriqmsymm}. 
%\subsubsection{A Yang-Mills Universe}  \label{sec:wholeym}
However, the action $\lm_U$ is a formal description. In particular, it does not clarify how the 
amplitudes $L_a$'s change under global Lorentz and Poincar\'e transformations of coordinates 
$(\theta, \phi)$ of the diffeo-surface $\mm$ and application of $\suinf$ group, which up to a 
global scaling of the area is homomorphic to local reparameterization of $\mm$ - we explicitly 
demonstrate this property in~\cite{hourisqgrym}\footnote{Ay density matrix $\hrho$ is a projector 
operator in an specific basis. Therefore, the second term in the Lagrangian is non-zero only for 
diagonal generators, that is N-dimensional Cartan subspace for $SU(N), N \rightarrow \infty$. 
Notice that $\tr \hrho = 1$. Therefore, its expension to generators of $SU(N)$ must include 
$\mathbbm{I}$. By contrast, the first term does not include unity operator of the symmetry group. 
It is trivially topological - when it is written in (\ref{yminvar}) form - because it does not 
depend on the connection. The second term is not topological for a model written on an 
independent background manifold. But, as we show in~\cite{hourisqgrym}, in~\sqgr the parameter 
space $\Xi$ is related to $\suinf$ symmetry and variation of the geometry of $Xi$ can be 
neutralized by a $\suinf$ gauge transformation.}.
%Note: the above footnote is added later and does exist in v4 of arXiv.

The Lagrangian $\lm_U$ must remain invariant under these transformations. 
%To this end and for getting a better insight into the analytical form of the amplitudes, we analyze the integrand of (\ref{twodlagrang}) for its transformation under aforementioned symmetries. 
To obtain invariant expression of the integrand in (\ref{twodlagrang}) we notice that the 
surface element $d^2\Omega$ is invariant under reparameterization of angular coordinates. 
%Thus, each term in the integrand of (\ref{twodlagrang}) must be reparameterization invariant. 
The amplitudes $L_a$ must be invariant under translation 
$\theta \rightarrow \theta + \theta_0,~\phi \rightarrow \phi + \phi_0$ for arbitrary constant 
shift of the coordinates origin by $\theta_0$ and $\phi_0$. This condition would be satisfied 
if $L_a$'s is a differential operator with respect to coordinates $(\theta, \phi)$. This 
operator must be also invariant under reparameterization of the diffeo-surface and $\suinf$ 
transformation. The only analytical expressions with these properties are 2-forms in of the 
diffeo-surface manifold $\mm$, which must be also singlet of $\suinf$. Thus, 
$L_a \hL_a \equiv F^{\mu\nu}_a \hL_a, ~\mu, \nu \in {\cos \theta, \phi}$ and the Lagrangian 
functional (\ref{twodlagrang}) takes the form of a 2D Yang-Mills Lagrangian for $\suinf$ 
symmetry:
\bea
\lm_U & = & \int d^2\Omega ~ \biggl [ ~\frac{1}{2} ~ \tr (F^{\mu\nu} F_{\mu\nu}) + 
\frac {1}{2} \tr (\sD \hrho_U) \biggr ], \quad \quad \mu, \nu \in {\theta, \phi} 
\label{yminvar} \\
F_{\mu\nu} & \equiv & F_{\mu\nu}^a \hL^a \equiv [D_\mu, D_\nu], \quad 
D_\mu = (\partial_\mu - \Gamma_\mu) \mathbbm{1} + \sum_a i \lambda A_\mu^a \hL^a, 
\quad a \equiv (l,m) \label{yminvardef} \\
F_{\mu\nu}^a F^{\mu\nu}_a & = & L^*_a L^a, \quad \forall a. \label{ltof} 
\eea
where $D_\mu$ is 2D covariant and $\suinf$ gauge invariant derivative with an appropriate 2D 
connection $\Gamma_\mu$, see e.g~\cite{cosmobook,spinor2d0}. The $dimensionless \lambda$ is 
a coupling constant, which can be included in $A_\mu^a$. Nonetheless, here we show it 
explicitly to keep the formulation similar to the standard Yang-Mills theories. The symbol 
$\mathbbm{1}$ presents unit operator of the $\suinfa$ algebra. In analogy with QFT the 
coordinate dependent vector $A_\mu^a$ can be called the $\suinf$ gauge field. Nonetheless, we 
emphasize that in contrast to a QFT on an spacetime, the pair $(\theta, \phi)$ parameterizes 
the representation of $\suinf$ symmetry, see~\cite{hourisqgrym} for more details. The exact 
expression of the differential operator $\sD$ depends on the representation of 2D Euclidean 
group by the state $|\psi_U\rangle$. 
%where $\hrho_U = |\psi_U\rangle \langle \psi_U |$. Such decomposition is possible because the whole Universe is necessarily in a pure quantum state. 
For a scalar-type state $\sD = \overleftarrow {D}_\mu \overrightarrow {D}^\mu = \overleftarrow {\partial}_\mu \overrightarrow {\partial}^\mu$. For a 2D spinor state it can be written using 
zweibein vectors:
\bea
&& \sD = i \sigma^0 \sigma^i z_i^\mu \overleftrightarrow {D}_\mu \quad , \quad   
D_\mu = (\partial_\mu - \Gamma_\mu) \mathbbm{1} + \sum_a i \lambda A_\mu^a \hL^a, \label{spinorder2d} \\
&& \Gamma_\mu \equiv \frac{1}{8} \omega_{\mu ij} [\sigma^i, \sigma^j] \equiv 
\frac{1}{4} \omega_{\mu ij} \sigma^{ij} \quad , \quad [\omega_\mu]^i_j = 
z^i_\nu \biggl(\partial_\mu (z^\nu_j) + z^\sigma_j \Gamma^\nu_{\mu \sigma}\biggr), \label{spinorconnect2d}\\
&& \sigma^{ij} \equiv \frac{i}{2} [\sigma^i, \sigma^j] \quad , \quad 
\eta^{(2)}_{ij}z^i_\mu z^j_\nu = \upeta^{(2)}_{\mu\nu} \quad , \quad \upeta^{(2)}_{\mu\nu}z^\mu_i z^\nu_j = 
\eta^{(2)}_{ij}. \label {zeibein}
\eea
%Note: \sigma_0 is necessary because <\psi| is \hrho does not include it. It is the 2D analogue of Dirac \gamma_0.
where $\sigma^i ~ i= \{1,2\}$ are two of the $(N\rightarrow\infty)$-dimensional representation 
of Pauli matrices~\cite{suninfhoppthesis}, and $\sigma^0$ is the third Pauli matrix. The 
tensors $z_i^\mu$'s are zweibeins (analogous to vierbein (tetrads) for 2D 
geometry)~\cite{spinor2d,spinor2d0}. The 2-dimensional symmetric tensors $\upeta^{(2)}_{\mu\nu}$ 
and $\eta^{(2)}_{ij}$ are arbitrary the metric of parameter space $\Xi = \mm$ of the Hilbert 
space $\hm_U$ and the 2D flat Euclidean metric, respectively.

Despite the presence of a connection term in (\ref{yminvardef}) it does not affect 
$F^{\mu\nu}$. In addition, as a function defined on the 2D diffeo-surface $\Gamma_\mu$ can be 
decomposed to spherical harmonic functions, which are in turn generators of $\suinf$ algebra, 
without loss of generality it can be included in $A_\mu^a$ term in (\ref{yminvardef}). This 
is an explicit demonstration of the Proposal \ref{propparamcurve} applied to the Universe 
as a whole. Consequently, the choice of a special geometry for the diffeo-surface of the 
Universe is equivalent to fixing $\suinf$ gauge and has no effect on measurable quantities, 
namely, $F_{\mu\nu}^a$ - up to an unobservable rescaling of the area of the diffeo-surface. 
We should also remind that in 2D spaces the 2-form $F^a_{\mu\nu}$ has only one independent 
non-zero component. Therefore, the number of degrees of freedom in the two sides of 
(\ref{ltof}) is the same.

\subsubsection{The topological $\mathbf{\suinf}$ Universe}  \label{sec:wholetopol}
Independence of the Lagrangian (\ref{yminvar}) from geometry of diffeo-surface $\mm$ and 
its parameterization means that the $\suinf$ Yang-Mills model of the whole Universe is 
topological. Topology of 2D compact Riemann surfaces are classified by the Euler 
characteristics: 
\be
\int d^2\Omega ~ \mathcal{R}^{(2)} = 4\pi ~ \chi (\mm), \quad\quad \chi (\mm) = 
2 - \gm (\mm)  \label{eulercharct}
\ee
where $\mathcal{R}^{(2)}$ is the scalar curvature of the 2D compact manifold $\mm$. The Euler 
characteristic $\chi (\mm) = 2 - \gm (\mm)$ depends only on the genus $\gm$ of $\mm$. The 
topological nature of the Lagrangian (\ref{yminvar}), in particular the pure gauge term, 
implies that without loss of generality its integrand must be proportional to 
$\mathcal{R}^{(2)}$:
\be
\tr (F^{\mu\nu} F_{\mu\nu}) ~ \propto ~ \mathcal{R}^{(2)}   \label{gaugecurve}
\ee
This relation becomes an equality by changing the arbitrary normalization of fields in 
(\ref{spinorconnect2d}). Thus, (\ref{gaugecurve}) establishes a relationship between geometry 
of the parameter space of $\suinf$ representation and amplitude of the $\suinf$ gauge field. 
On the other hand, at lowest quantum order variational principle relates the gauge field to 
the quantum state of the Universe $\hrho_U$\footnote{We remind that the Lagrangian 
(\ref{yminvar}) is written based on the quantum properties of the model. Therefore, field 
equations obtained from variational principle are not classical, and present lowest order 
quantum dynamics.}, see~\cite{hourisqgrym} for the demonstration. Therefore, there is no 
arbitrary or unrelated quantity in the model. 

%Note: the next paragraphs were not discussed before.
Equation (\ref{gaugecurve}) explicitly demonstrates the relationship between the Yang-Mills 
Lagrangian (\ref{yminvar}) and what we perceive as gravity through the curvature of classical 
spacetime. Indeed, the number of degrees of freedom of pure Yang-Mills fields in dimension 
$D$ is $D(D-1)/2$, which is the same as the number of independent components of the Ricci 
curvature tensor $R_{\mu\nu}$ after taking into account $D$ coordinate reparameterization 
constraints - or equivalently $D$ energy-momentum conservation constraints in classical 
Einstein theory of gravity. Interestingly, this relation exits for any $D$. In any case, 
in (\ref{yminvar}) none of these quantities are dynamical and due to the $\suinf$ symmetry 
all the states are physically equivalent. Nonetheless, the Lagrangian distinguishes between 
non-equivalent representations of $\suinf$, that is different topology of diffeo-surface 
$\mm$, see appendices of~\cite{hourisqgrym} for more details.

In a geometrical view, if the local details of the Universe are not distinguishable, only 
its global - topological - properties can characterize its states. In \sqgr~the only 
characterizing global property is the topology of the diffeo-surface, which determines 
non-homomorphic representations of the $\suinf$ symmetry~\cite{suninfrep,suninfrep0} of the 
Universe. In other words, the only possible difference between whole universes is the 
representation of $\suinf$ by their Hilbert spaces. Thus, in \sqgr the genus of the 
diffeo-surface is the only necessary initial condition for distinguishing our Universe. 
Alternatively, our Universe may be in a superposition of independent representations of 
$\suinf$, that is in superposition of Lagrangians of type (\ref{yminvar}) with different 
values for the integral in (\ref{eulercharct}). We give a short assessment of consequences of 
coherent mixing of global typologies in Sec. \ref{sec:cosmoconst} and leave its detail 
investigation to a future work.

\subsection{Evolution of subsystems}  \label{sec:evolsub}
%Moreover, to avoid inconsistencies associated with null Hamiltonian in models invariant under reparameterization~\cite{admgr}, time and evolution must be treated as relative concepts~\cite{qmtimepage}. 
As we discussed in sections \ref{sec:subsysparam} and \ref{sec:geometry}, when the Universe 
is divided to subsystems and a reference subsystem and a quantum clock are chosen, every 
pair of their 4 continuous parameter space can be considered as parameters of a $\suinf$ 
symmetry, under which physical properties of subsystems must be invariant. The presence of a 
clock means that in contrast to the whole Universe, the ensemble of subsystems can be indeed 
treated as a dynamical quantum system and in analogy with QFT a dynamical, that is time 
dependent Lagrangian can be associated to it. For this purpose we proceed in the same way as 
we did for the whole Universe, and consider the lowest order $\suinf \times G$ invariant 
functional as the Lagrangian:
\bea
\lm_{U_s} & = & \frac{1}{4\pi L_P^4} \int d^4 x \sqrt {|\upeta^{(4)}|} ~ \biggl [~ 
\frac {1}{4} ~ \biggl ( \sum_{l,m,l',m'} \tr (L^*_{lm} (x) L_{l'm'}(x) \hL_{lm} \hL_{l'm'}) + 
\nonumber \\
&& \sum_{a,b} \tr (T^*_a (x) T_b (x) \hT_a \hT_b)\biggr ) + 
\sum_{l,m,a} \tr (L_{lm}(x) T_a (x) \hL_{lm} \otimes \hT_a) + \nonumber \\ 
&& \sum_{lm} L_{lm} ~ \tr (\hL_{lm} \otimes {\mathbbm 1}_G ~ \hrho_s (x)) + 
\frac{1}{2} \sum_a T_a ~ \tr ({\mathbbm 1}_{SU(\infty)} \otimes \hT_a ~ \hrho_s (x)) \biggr ]. 
\label{lagranges}
\eea
where $T^a$'s are generators of the finite rank internal symmetry $G$ of subsystems, and  
$L_P \equiv \sqrt{\hbar G_N / c^2 }$ is the Planck length. In agreement with the 
Proposition~\ref{propparamcurve}, $\upeta_{\mu\nu}, ~\mu, \nu \in \{0, \cdots, 3\}$ 
is an arbitrary metric for the 4-dimensional parameter space $\Xi$, here presented in Cartesian 
coordinates, and $\upeta^{(4)}$ is its determinant. In~\cite{hourisqgrym} we show that the 
contribution of metric can be neutralized by a $\suinf$ transformation. The C-number amplitudes 
$L_{lm}$ and $T_a$ are normalized to be dimensionless. The first two terms present 
self-interaction of $\suinf$ and $G$ symmetries, respectively. If the generators of symmetry 
groups are chosen to be traceless, the third term in (\ref{lagranges}) would be zero. The forth 
and fifth terms present interactions of subsystems due to $\suinf$ and $G$ symmetries, 
respectively. 

Following the same line of arguments given for (\ref{twodlagrang}) about the invariance of 
parameter space under coordinates transformation i.e. Poincar\'e symmetry, invariance under 
both $\suinf$ and $G$ transformations, we conclude that the action $\lm_{U_s}$ must have the 
structure of a Yang-Mills model:
\bea
\lm_{U_s} & = & \int d^4x \sqrt{|\upeta|} ~ \biggl [\frac{1}{16\pi L_P^2} 
\tr (F^{\mu\nu} F_{\mu\nu}) + \frac{1}{4} \tr (G^{\mu\nu} G_{\mu\nu}) + 
\frac {1}{2} \sum_s \tr (\sD \hrho_s) \biggr ], \quad \mu, \nu \in \{0, 1, 2, 3 \}_{\upeta} 
\nonumber \\ && \label{yminvarsub} \\
F_{\mu\nu} & \equiv & F_{\mu\nu}^{lm} \hL^{lm} \equiv [D_\mu, D_\nu], \quad D_\mu = 
(\partial_\mu - \Gamma_\mu) \mathbbm{1} - i \lambda A_\mu, \quad A_\mu, \equiv 
\sum_{lm} A_\mu^{lm} \hL^{lm}, \quad L_P^2 F_{\mu\nu}^{lm} F^{\mu\nu}_{lm} = L^*_{lm} L^{lm}. \nonumber \\
&& \label{yminvardefsuinf} \\
G_{\mu\nu} & \equiv & G_{\mu\nu}^a \hT^a \equiv [D'_\mu, D'_\nu], \quad D'_\mu = 
(\partial_\mu - \Gamma_\mu - i \lambda A_\mu) \mathbbm{1}_G - i \lambda_G B_\mu, \quad B_\mu \equiv 
\sum_a B_\mu^a \hT^a, \quad L_P^4 G_{\mu\nu}^a G^{\mu\nu}_a = T^*_a T^a.  \nonumber \\
&& \label{yminvardefg}
\eea
As explained earlier, $\sD$ depends on the representation of Lorentz symmetry by subsystems. 
For instance, for Dirac spinors it is:
\bea
&& \sD \equiv \gamma^0 \gamma^i e_i^\mu \overleftrightarrow {D}_\mu \quad , \quad 
D_\mu = (\partial_\mu - \Gamma_\mu) \mathbbm{1} - \sum_{lm} i \lambda_s A_\mu^{lm} \hL^{lm} - 
\sum_a i \lambda_G B_\mu^a \hT^a, \label{spinorder4d} \\ 
&& \Gamma_\mu \equiv \frac{1}{8} \Omega_{\mu ij} [\gamma^i, \gamma^j] \equiv 
\frac{1}{4} \Omega_{\mu ij} \Sigma^{ij} \quad , \quad [\Omega_\mu]^i_j = 
e^i_\nu (\partial_\mu (e^\nu_j) + e^\sigma_j \Gamma^\nu_{\mu \sigma}) \label{spinorconnect4d} \\
&& \Sigma^{ij} \equiv \frac{i}{2} [\gamma^i, \gamma^j] \quad , \quad 
\eta_{ij}e^i_\mu e^j_\nu = \upeta_{\mu\nu} \quad , \quad 
\upeta_{\mu\nu}e^\mu_i e^\nu_j = \eta_{ij} \quad , \quad i,j \in \{0, 1, 2, 3 \}_{\eta}. 
\label {vierein}
\eea
where $\gamma^i$ are Dirac matrices, and other quantities in (\ref{spinorder4d}-\ref{vierein}) 
are analogous to those of (\ref{spinorder2d} - \ref{zeibein}), but in the 4-dimensional $\Xi$.  

The Lagrangian (\ref{yminvarsub}) is still somehow formal. In particular, the gauge term 
$\sum_{lm} i \lambda_s A_\mu^{lm} \hL^{lm}$ in the covariant derivative needs to be expanded, 
because as we explained in Sec. \ref{sec:symm}, generators $\hL^{lm}$ of $\suinf$ symmetry 
depend only on two continuous parameters of the diffeo-surface $\mm \subset \Xi$. Thus, what 
is called {\it internal space} in~\cite{suninfym}, which first studied $\suinf$ Yang-Mills 
models, is not an independent space, but embedded in $Xi$. This property of the Lagrangian 
$\lm_{U_s}$ is crucial for its applications and is studied in more detail in~\cite{hourisqgrym}.

\section{Some properties of the \sqgr~Universe}  \label{sec:subsyslagrange}
The only observable of the global \sqgr~Universe is its topology, which its measurement is 
impossible for an observer who only has access to a limited local part of the Universe. By 
contrast, properties of the Lagrangian (\ref{yminvarsub}) and their consequences are a priori 
observable for local observers/subsystems. In this section we present a mostly qualitative 
assessment of the properties important for the consistency of the model with what we know 
about classical gravity and for future tests of QGR.

\subsection {Universality of gravitational interaction}  \label{sec:universality}
As expected there is no cross term between pure gauge field terms of $\suinf$ and $G$, because 
by construction they are orthogonal and the full symmetry of subsystems is their tensor 
product, and interaction of gauge fields is simply similar to QFT in curved spacetime. However, 
according to the Proposition \ref{propparamcurve} the metric $\upeta_{\mu\nu}$ of the parameter 
space is arbitrary. Therefore, the question arises how in do the $G$ gauge fields - including 
those of the Standard Model - interact with gravity ? 

To answer this question, we remind that even in Einstein gravity the difference between flat 
and curved spaces arises only when derivatives of fields are involved~\cite{cosmobook}. 
specifically, the definition of the strength of Yang-Mills fields given in (\ref{yminvardefsuinf}) 
shows that non-zero connection of a curved space does not affect it, and becomes relevant 
only in the field equation. Therefore, at Lagrangian level, only the metric convoys the curvature 
of spacetime (here the parameter space) to gauge fields. Therefore, in this respect there 
is no difference between \sqgr~and Einstein gravity. Nonetheless, in various ways an 
interaction between $F_{\mu\nu}$ and $G^{\mu\nu}$ arises in \sqgr. 

First of all, although $\tr (\hL_{lm} \hT_a) = 0$, higher order cross terms are not null. The 
lowest gauge invariant term not included in $\lm_{U_s}$ is 
$\tr (F^{\mu\nu} F_{\mu\nu})\tr (G^{\mu\nu} G_{\mu\nu})$, which corresponds to a vertex with 4 pads. 
Nonetheless, it does not generate mass, neither for $G$ nor for $\suinf$, because it depends on the field 
strength rather than gauge fields $A_{lm}^\mu$ and $B_{lm}^\mu$\footnote{$A^\mu_{lm}$ and $B^\mu_a$ 
present the decomposition of the wave function of $\suinf$ and $G$ gauge bosons with respect 
to generators. It is also why they appear in pair in the Lagrangian, exactly similar to 
$\hrho$ when it is pure, or when it is written in its canonical purification form 
$\hrho = \sqrt{\hrho} \sqrt{\hrho}$.}. Even without explicitly introducing such higher order 
interaction terms they will arise in the path integral constructed from the Lagrangian 
(\ref{yminvarsub}) by interaction of the density matrix of subsystems with both $\suinf$ 
and $G$ bosons, and thereby in the effective action of the model. In addition, in the field equation 
obtained from the Lagrangian (\ref{yminvarsub}) both $\hrho_s$ and $G_{\mu\nu}$ contribute to 
the source term for the gravity field strength $F_{\mu\nu}$. 

We should remind that even in classical physics both Einstein gravity and Yang-Mills 
action can be derived in an arbitrary background geometry from the assumption of consistent 
self-interactions~\cite{grymselfint}. Another evidence for indirect interaction of gravity 
and gauge bosons in general relativity is the fact that scalar curvature of a universe 
containing only radiation - gauge bosons - vanishes everywhere, because the trace of 
energy-momentum tensor is zero. This means that sectional curvatures must be null or cancel 
each others everywhere, see (\ref{riccicurvsec}). Thus, the spacetime manifold of such 
universe is Ricci-flat and behaves like classical vacuum, despite the presence of energy. 
In conclusion, the $\suinf$ Yang-Mills field - the quantum gravity - couples to all fields, 
including to itself in the same and universal manner. 

Finally, giving the relation between the Pontryagin topological class, average metric 
$g_{\mu\nu}$ defined in (\ref{tracemetricalpha}), and variation of the density matrix of 
subsystems $\hrho_s$, it would be useful to use $g_{\mu\nu}$ rather than an arbitrary metric 
for calculation of physical observables such as Green's functions. Of course, in this case 
all the constituents of the model become intertwined and self-interactions make evolution 
equations highly nonlinear. Nonetheless, in analogy with background field approaches in QFT, 
such as Kadanoff-Beim method which includes condensates, that is non-zero 1-point Green's 
functions, considering $g_{\mu\nu}$ as a background field might lead to a better 
approximation. Moreover, the choice of the metric $\upeta_{\mu\nu}$ can be embedded in  
fixing the $\suinf$ gauge, see~\cite{hourisqgrym} for more details.

\subsection{Possibility of gravitational P, C, and CP violation} \label{sec:cpviol}
%As we described in Sec. \ref{sec:wholetopol}, in the 2D whole Universe the $\suinf$ Yang-Mills/gravity field strength is selfdual, that is $\tilde{F}^{\mu\nu} = F_{\mu\nu}$.
The Lagrangian of the whole Universe $\lm_U $ conserves parity and CP. Intuitively these 
properties make sense, because the parameter space of the $\suinf$ symmetry is not directional. 
On the other hand, it implies that even after the division of the Universe to subsystems, parity 
and CP must be globally preserved. Nonetheless, a priori subsystems may locally break these 
discrete symmetries. There are various reasons for such possibility.

Yang-Mills theories have topologically non-trivial vacuum structure. Because this subject is 
also relevant for the crucial issue of cosmological constant, we discuss it in some details 
in Sec. \ref{sec:cosmoconst}. Here we only remind that in the Standard Model, the 
possibility of non-trivial vacuum in both electroweak and QCD interactions leads to what is 
called {\it strong CP problem}, see e.g.~\cite{qftbook,qcdaxionrev,instantonrev,instantonrev0} 
for review and more details. This problem is closely related to the issue of axial $U(1)$ 
anomaly. In both cases the problem can be summarized in so called a {\it $\Theta$-term} in 
the action:
\be
\lm_\Theta \equiv \frac{\Theta g^2}{32\pi^2} \int d^4x ~ \sqrt{\upeta} ~ 
\tilde{\fm}^{\mu\nu} \fm_{\mu\nu} \quad , \quad \tilde{\fm}^{\mu\nu} \equiv 
\frac{1}{2} \epsilon^{\mu\nu\rho\sigma} \fm_{\mu\nu} \label{thetaact}
\ee
where $\fm^{\mu\nu}$ is gauge field strength and $g$ is its coupling. The integrand in this 
action is a total derivative and the integral is a topological invariant - the Pontryagin 
topological class. However, the dual field $\tilde{\fm}^{\mu\nu}$ has a parity opposite to 
$\fm^{\mu\nu}$. Therefore, addition of (\ref{thetaact}) to the Yang-Mills action violates 
party conservation. Moreover, as conjugation symmetry C is preserved by both actions, CP 
symmetry and thereby time-reversal symmetry T are also broken, if CPT is conserved. Although 
parity symmetry is violated in electroweak interaction, there is stringent constraint on the 
parameter $\Theta$ in QCD.

In \sqgr~the quantum mediator of gravity is a Yang-Mills gauge field. Therefore, complexities 
related to topological structure of vacua, axial $U(1)$ anomaly which arises when fermions 
can be considered massless, and violation of P, T, and CP due to a $\Theta$-term similar to 
(\ref{thetaact}) may arise in the model. In particular, as the energy scale in which quantum 
gravity effects become important - presumably the Planck mass scale $M_P$ - is much larger than 
the mass of all SM particles, they can be considered as massless and axial $U(1)$ anomaly would 
be more serious. Nonetheless, there are various solutions for this issue. For instance, at 
Planck scale strong non-perturbative gravitational interactions can increase the effective 
mass of particles and prevent the anomaly. Indeed, in the strong gravity regime of the 
early Universe these anomalies might have seeded baryons and leptons asymmetries through 
sphaleron-type phenomena, or contributed to formation of these asymmetries, see 
e.g.~\cite{leptosphaleron,leptosphaleron0} for a review. We leave investigation of such 
effects to future works.

In what concerns parity violation in gravity sector, at present there is no observation or 
constraint even for classical gravity. Nonetheless, future generations of gravitational 
wave detectors should be able to test it~\cite{grparityviol,grparityviol0,grparityviol1}. 
Assuming no violation of P or CP in gravitational sector, a possible solution for violation 
of discrete symmetries by gravitational $\Theta-$term in \sqgr~is introducing a pseudo-scalar 
{\it gravitational axion} field $\phi_g$, analogous to a Peccei-Quinn $U(1)$ symmetry and 
its associated axion proposed to solve the QCD strong CP problem. It replaces the constant 
$\Theta$ in an action similar to (\ref{thetaact}). Such gravitational axion must be in close 
relation with the gravity sector. However, it does not seem that in \sqgr~a fundamental 
scalar $\phi_g$ related to $\suinf$ symmetry exists. The only natural $U(1)$ symmetry in the 
model, which emerges after the division of the Universe to subsystems, namely the relative 
size of the diffeo-surfaces of subsystems, is used to introduce an area (or length) 
parameter in the model. On the other hand, in Sec. \ref{sec:cosmoconst} we argue that several 
phenomena may act as an {\it effective axion} at low energies. 

In summary, many of observed and/or predicted complexities of non-Abelian Yang-Mills may 
arise in \sqgr~and must be thoroughly investigated in future works.

\subsection{Classical limit}  \label{sec:classlim}
In quantum mechanics and QFT the classical limit of a model is defined as when 
$\hbar \rightarrow 0$. However, as we discussed in Sec. \ref{sec:axioms}, in \sqgr~such 
assumption makes the algebra (\ref{statealgebra}) Abelian and the symmetry group 
$\bigotimes_1^\infty U(1)$, unless at the same time $M_P \rightarrow \infty$, meaning no 
gravity. In this case, due to the absence of a background spacetime in \sqgr, the structure 
of the model will collapse and become meaningless. Therefore, rather than considering 
$\hbar \rightarrow 0$ limit, we consider the situation where the observer is sensitive to 
classical gravity and can measure the effective metric (\ref{tracemetricalpha}) and its 
associated curvatures, but is unable to detect the underlying quantum $\suinf$ Yang-Mills 
structure. An example of a similar situation in materials physics is magnetism, which was 
known for thousands of years as a classical phenomenon before the discovery of its close 
relationship with quantum properties of matter, in particular spin of elementary particles, 
which is a purely quantum concept. 

%Specifically, for an observer who cannot detect and discriminate the infinite number of $(l,m)$ {\it colors} of the $\suinf$ symmetry and their corresponding quantum fields $F_{lm}^{\mu\nu}$, the only way to become aware of their presence is through the affine parameter $ds$ in (\ref{tracemetricalpha}) - as Einstein suggested - and the effective path of the subsystems in their Hilbert space, perceived through an effective metric $g_{\mu\nu}$ association to an effective spacetime. In this case, the observer {\it sees} the first term of the Lagrangian (\ref{yminvarsub}) as one block - a scalar function of the effective spacetime. 
For the case of 2D parameter space of the whole Universe, we already demonstrated the 
relationship between the $\suinf$ Yang-Mills action and 2D scalar curvature. More generally, 
the lack of sensitivity to quantum structure of gravity means that the purely $\suinf$ gauge 
fields term in the effective Lagrangian appears as a scalar function of effective parameters - 
the classical spacetime. In~\cite{hourisqgrcomp} we used the relation between 4D scalar 
curvature and sectional curvatures, along with the fact that 2D compact subspaces of the 
parameter space $\Xi$ can be considered as diffeo-surfaces representing $\suinf$ of 
subsystems, to relate $\suinf$ Yang-Mills term in (\ref{yminvarsub}) to 4D scalar curvature. 
More generally, it is demonstrated that for (pseudo)-Riemannian spaces with dimension 
$D \geq 3$, there is always a metric for every scalar function such that the scalar 
curvature obtained from that metric is equal to the function~\cite{curvaturfunc} (Chapter 4, 
theorem 4.35). Therefore, the first term of (\ref{yminvarsub}) can be always considered to 
be equal to a scalar curvature:
\be
\int d^4x \sqrt{|\upeta|} ~\tr (F^{\mu\nu} F_{\mu\nu}) ~ \autorightarrow{classical}{limit} ~ 
{\large \propto} \int d^4x \sqrt{|g|} R^{(4)}  \label{r4}
\ee
Similarly, the contribution of $A_\mu^{lm}$ in covariant derivatives can be added to the 
parameter space connection (see explicit demonstration in~\cite{hourisqgrym}), and the 
resulting Lagrangian becomes similar to a $G$ symmetry Yang-Mills model in a curved spacetime 
with Einstein-Hilbert gravity action:
\be
\lm_{U_s} \rightarrow \lm_{cl.gr} = \int d^4x \sqrt{|g|} \biggl [\kappa R^{(4)} + \frac{1}{4} 
\tr (G^{\mu\nu} G_{\mu\nu}) + \frac {1}{2} \sum_s \tr (\sD \hrho_s) \biggr ]. \label{classicgr}
\ee
where the dimensionful constant $\kappa \propto M_P^2$ is necessary for making the Lagrangian 
function dimensionless. It is also clear that other possible scalars obtained from curvature 
tensor, e.g. $R^{\mu\nu} R_{\mu\nu}$ are of higher quantum order and at least $\hbar^2/M_P^2$ 
times smaller than $R^{(4)} \equiv g_{\mu\nu} R^{\mu\nu}$. Therefore, a fundamental spin-1 
mediator for quantum gravity is not in contradiction with the observed spin-2 classical 
graviton and its Einstein-Hilbert action. 

The relationship between a spin-2 gravity model and an underlying spin-1 gauge theory is 
not specific to the classical limit of \sqgr. As mentioned earlier the gauge-gravity duality 
conjecture is investigated by several approaches to QGR, see 
e.g.~\cite{qgrgaugesep,qgrgaugesep0,qgrgaugesep1,gaugestringcorr,stringgauge0,stringgauge1}. 
Moreover, what is called double copy approach, that is the description of gravity and/or QGR 
using two spin-1 gauge models, is used for calculating scattering amplitudes in particle 
physics, QGR, and gravitational waves emitted by merging compact objects, 
see~\cite{ymgwsimul,ymgwsimul0} and references therein. Interestingly, the latter 
application employs gauge/Yang-Mills QFT's with large number of {\it colors}. This can be 
considered as a numerical approximation for the $\suinf$ gauge term in (\ref{lagranges}). 

It is remarkable that we began the construction of \sqgr~from abstract assumptions about 
symmetry and quantum mechanics, but rediscovered exactly Einstein gravity at low energies - 
when gravity seems classical. For instance, if we add a term proportional to 
$F^{\mu\nu} \tilde{F}_{\mu\nu}$ to (\ref{yminvarsub}), in general it cannot be written as a 
function of $R^{(4)}$ for the same metric as that of $F^{\mu\nu} F_{\mu\nu}$. Of course, as 
mentioned in Sec. \ref{sec:cpviol}, such term is topological. But, $\Theta$ in 
(\ref{thetaact}) can be considered as a pseudo-scalar {\it gravitational axion}. 
%It would be an independent pseudo-scalar function, which break parity unless its amplitude is controlled with yet another pseudo-scalar - a gravitational axion - to be consistent with the lack of parity violation in Einstein gravity. 
In this case the classical limit will look like a bimetric 
gravity~\cite{bimetricgr,bimetricgr0,bimetricgr1} gravity where the dual term can be 
considered as scalar curvature of another metric. The second metric is conjectured to become 
significant at high energies. Such models are related to massive graviton 
models~\cite{bimetricmassivegr} and lead to the deviation of the speed of gravitational waves 
propagation from the speed of light~\cite{grspeedvar}. So far such deviation is not detected 
in observations~\cite{grlorentz}.

\subsubsection{Cosmological constant} \label{sec:cosmoconst}
The Lagrangian $\lm_{U_s}$ does not include an explicit constant term, which can be interpreted 
as a cosmological constant. In one hand, as the nature of the observed dark energy is still 
unknown, the need for a cosmological constant in \sqgr~Lagrangian is not certain. On the other 
hand, although a cosmological constant is still consistent with the observed dark energy, 
the growing evidence for what is called the {\it Hubble tension} and other anomalies of the 
standard $\Lambda$CDM, see~\cite{hubbletension} for a review, may point to more elaborate 
models~\cite{hubbletensionmod,hubbletensionmod0,hubbletensionmodrev} for dark energy and/or 
dark matter. In the literature many models are suggested as alternative to a cosmological 
constant, and most of them are relevant in the framework of \sqgr, except those proposing a 
modification of classical Einstein gravity. Nonetheless, in supplementary section 
\ref{app:sqgrcc} we briefly discuss several processes specific to \sqgr~which can potentially 
lead to an effective cosmological constant.

\section{Outlines and future perspectives} \label{sec:outline}
In this work we investigated foundation and principle properties of the 
quantum gravity model dubbed \sqgr. After demonstrating that the model is truly quantum, we 
showed that what is perceived as the classical spacetime can be interpreted as effective 
manifestation of parameters characterizing the quantum state of subsystems of the Universe. 
We related the observed curved metric of the classical spacetime to the quantum state of the 
Universe and demonstrated that it presents the average path of quantum subsystems in the 
parameter space of their Hilbert space. We introduced a relative quantum dynamics and showed 
that its gravitational sector has the structure of a $\suinf$ Yang-Mills quantum field 
theory. Therefore, according to \sqgr~the quantum mediator of gravity is a vector field like 
other fundamental interactions. Nonetheless, we demonstrated that in the classical limit - 
defined as when the quantum structure of purely gravitational part of the Lagrangian and 
is not discernible for the observer - the resulting model is exactly Einstein gravity. We 
also briefly discussed potential explanations for the observed dark energy specific to \sqgr. 

Future works should investigate in detail these proposals as well as application of the \sqgr~
to other puzzling problems in which quantum gravity may play an essential role. Examples 
include: microscopic structure of black holes and apparent loss of information behind their 
horizon; singularities, both cosmological and in black holes; the role of quantum gravity 
in the early Universe and inflation; predictions for observables and discriminating signals 
of these phenomena; methods for testing \sqgr~in laboratory. Regarding this last topic, it 
is important to insist that geometrical test of gravity would not be able to find the 
signature of a vector gravity. Only the detection of intrinsically quantum processes might 
be able to distinguish between spin-1 and spin-2 force mediators. 

Regarding the construction of the model, the approach used here and in the previous works is 
in a QFT perspective. It would be interesting - and probably necessary - to study the model 
in the framework of quantum information theory with subsystems as open quantum systems. 
In particular, quantum reference frame of different subsystems, their comparison, and their 
perspective-dependence should be investigated. Understanding of these topics are crucial 
for studying the microphysics of black holes and for designing test experiments for \sqgr. 

A subject that we did not address in detail here is the emergence of internal symmetry 
$G$. Although the potential for clustering and symmetry breaking is ubiquitous, it is not 
clear why and how the nature prefers some symmetries to others. Standard Model symmetries are 
smallest non-Abelian symmetries and cannot be decomposed to yet smaller non-Abelian groups. 
Is this property related to the decomposition of $\suinf$ symmetry of subsystems ? In 
principle this hypothesis can be investigated in the high energy colliders. Specifically, 
one might search for the emergence of symmetries similar to what is observed in condensed 
matters in the distribution of partons and leptons at very high energies.

\acknowledgments{
The author acknowledges the support of the Institut Henri Poincar\'e (UAR 839 CNRS-Sorbonne 
Universit\'e), and LabEx CARMIN (ANR-10-LABX-59-01).}

\appendix
%\renewcommand\thesection{S.\arabic{section}~}
%\renewcommand\thesubsection{S.\arabic{section}.\arabic{subsection}~}
%\cftsetindents{section}{2em}{4em}
%\cftsetindents{subsection}{3em}{5em}

\section*{Supplementary sections}  \label{supple}
\section{Rationale for \sqgr} \label{app:rationale}
Standard quantization methods fail to provide an inclusive formulation, including not only 
a consistent and problem-free quantum gravity, but also matter and its non-gravitational 
interactions. This conundrum motivates a radically different approach to quantum gravity. 
Details of rationale behind ideas and choices of axioms for \sqgr~are discussed 
in~\cite{houriqmsymmgr}. Here we remind those needed for the present work.

\paragraph*{Symmetry as a foundational concept in quantum mechanics} %\label{app:qmsymm}
Symmetry is one of the mathematically and physically fundamental concepts, which may reveal 
the way forward. Indeed, axioms of quantum mechanics~\cite{qmsymmfound,qmsymmfound0} can be 
reformulated such that they include and emphasize the role of symmetries in its 
foundation~\cite{houriqmsymm} (reviewed in~\cite{houriqmsymmgr}. We call this approach 
{\it Symmetry Description of Quantum Mechanics} (SDQM). In contrast to quantum mechanics 
\`a la Dirac~\cite{qmdirac} and von Neumann~\cite{qmvonneumann}, in which Hilbert spaces of 
quantum systems are a priori abstract Banach spaces, in SDQM they represent symmetries of the 
corresponding quantum systems, related to their coherence - the coherence or state 
symmetry\footnote{In quantum information literature what we call {\it coherence symmetry} is 
confusingly called {\it asymmetry}, emphasizing on the off-diagonal components of density 
matrix when a system is not completely incoherent. For this reason, in~\cite{houriqmsymm} 
the expression {\it state symmetry} was used.}. The result of a complete breaking of 
coherence symmetry is a projective measurement~\cite{statesymmbreak}. On the other hand, 
it can be quantified and used as a resource~\cite{qminfocohere,qmspeedcohere}. In the SDQM 
framework quantum systems and their dynamics can be modeled without resorting to classical 
concepts. This is a crucial requirement for what is called {\it Quantum First} approaches 
to QGR~\cite{qgrlocalqm1}. Thus, in \sqgr~the SDQM is the rationale behind postulating a 
specific symmetry, namely $\suinf$ for the Hilbert space of the Universe. 

\paragraph*{Universe as a composite quantum system} %\label{app:univcomp} 
According to a corollary of SDQM division of a quantum Universe to subsystems is crucial for 
its non-triviality and consistency. It is important to remind that many concepts in quantum 
mechanics rely on the ability to define and distinguish subsystems, for example: entanglement, 
definition of a quantum clock and a quantum mechanically consistent dynamics, 
system-environment discrimination and decoherence, and locality of interactions. In what 
concerns modeling a quantum Universe, its division to subsystems is crucial for the existence 
of a universal interaction similar to gravity\footnote{Considering categorical description of 
quantum mechanics, see e.g.~\cite{qmcategory}, the Universe is the category of categories. 
It incorporates all {\it physical} objects - including itself - and {\it physical} operations 
(functors) applied to them. The purpose of adjective {\it physical} here is to emphasize that 
these entities are not abstract mathematical concepts, but have detectable existence.}. 
Moreover, subsystems are indispensable for a consistent implementation of the concept of 
{\it locality}, which is considered by some QGR 
proposals~\cite{qgrhistory,qgrlocalqm,qgrlocalqm0,qgrentangle,qgrentangle1} to be crucial for 
accommodating black holes and explaining the paradox of information 
loss~\cite{hawkingrad,bhentropy0}. Models that consider the classical spacetime as a 
quantizable physical entity, usually lack a clear description for its division to quantum 
subsystems.

\paragraph*{Infinite Universe} %\label{app:infiniteuniv}
When quantum nature of gravity is ignored all processes, from physics of the early Universe 
and inflation to chemistry, condensed matter, and life can be described by fundamental or 
effective Quantum Field Theories (QFTs), which has infinite number of degrees of freedom. 
Even a Lorentz invariant vacuum can be described as a superposition with negligibly small 
amplitude of all possible states~\cite{hourivacuum}, including Glauber coherent 
states~\cite{coherglauber} and condensates of particles. These observations motivate the 
assumption of infinite number of subsystems for the Universe. The Hilbert space of such a 
quantum system is necessarily infinite dimensional and should have infinite number of 
mutually commuting observables. 

\paragraph*{Nature of the spacetime} %\label{app:spacetime}
Physical observables with continuous spectra are usually related to the spacetime. In 
particular, in QFTs states of the Hilbert space and particles (subsystems) in the Fock space 
are {\it labeled/indexed} by the spacetime or its Fourier conjugate - the momentum (but not 
both). This raises the question whether the spacetime by itself, that is without matter, has 
any physical meaning, or rather what we call matter and spacetime are inseparable properties 
of the same physical reality. General relativity does not clarify the nature of spacetime and 
its difference with matter, and as discussed in the Introduction, Einstein considered 
spacetime as a quantity that parameterizes the spin-2 metric and matter fields. But, this 
interpretation does not clarify the origin of this parameter space and why its geometry is 
determined by a physical field~\footnote{In popular approaches to QGR, specially in string 
theory, spacetime and matter are treated similarly as perturbation modes (fields) of 2D 
world-sheet of strings. The Loop Quantum Gravity (LQG) also treats space as an standalone 
physical entity quantized by triangulation or treated as a spin foam. See~\cite{hourisqgrcomp} 
for an extended discussion of these models and their comparison with \sqgr.}
Irrespective of interpretations and approaches, the undeniable fact is that without matter 
(including a cosmological constant) and with trivial boundary conditions, solution of the 
Einstein equation is a constant metric, which can be also zero - corresponding to no 
spacetime at all. In addition, the choice of a null metric rather than a flat space with 
undetermined volume seems preferable, because there is no experimental evidence for the 
existence of an empty flat Minkowski spacetime at any scale. Indeed, under some assumptions, 
in particular holographic principle~\cite{hologprin,hologprin0,hologprin1}, Einstein equation 
can be obtained and interpreted as an equation of state~\cite{greos}. Therefore, we conclude 
that spacetime and matter are inseparable, and classical perception of spacetime is emergent, 
rather than being fundamental.

\section {Properties of $\suinf$ tensor products} \label{app:suinf}
Using decomposition of $SU(N)$ group:
\be
SU(N) \supseteq SU(N-K) \times SU(K) \label{sundecomp}
\ee
and the fact that factors in its r.h.s are a normal subgroup of $SU(N)$, it is 
straightforward to see that the repetition of (\ref{sundecomp}) with $K = 2$ leads to:
\be
SU(N)\biggl |_{N \rightarrow \infty} \cong \bigotimes^\infty SU(2)  \label{suninfsu2} 
\ee
This is a direct demonstration of (\ref{suinfsu2decomp}) and independent of the decomposition 
of associated generators in (\ref{llmdef}). Using (\ref{suninfsu2}) and the fact that 
$n \times \infty = \infty$ we conclude that tensor products of $\suinf$ are homomorphic to 
itself:
\be
(\suinf)^n \cong \suinf \label{suinfmult}
\ee

It is then straightforward to conclude that:
\bea
\suinf & \supset & \bigotimes^M SU(K), \quad \quad M < \infty, \quad \forall ~ K < \infty,  
\label{subsuk} \\
\suinf & \cong & \bigotimes^\infty SU(K), \quad \quad \forall ~ K < \infty, 
\label{suinfsukdecomp}
\eea
where (\ref{subsuk}) can be concluded from repetition of (\ref{sundecomp}) and the 
assumption that ${SU(N - MK)|_{N, M \rightarrow \infty}}$ is represented trivially. As each 
$SU(K)$ factor in turn has a Cartan decomposition, we can generalize (\ref{suinfsukdecomp}) to: 
\be
\suinf \cong \bigotimes_{i=1}^\infty G_i \label{suinfgi}
\ee
where $G_i$ is a finite rank Lie group. Consider one of $G_i$ factors in (\ref{suinfgi}). We 
call it $G$. Then, the symmetry of the whole Universe\footnote{When there is no risk of 
confusion we identify $\bm[\hm]$ with the symmetry group it represents.} can be written as 
$\suinf \cong G \times G_\infty$, where $G_\infty$ is an infinite rank group. For any $G$ we can 
find finite $N \leqslant N'$ such that $SU(N) \subseteq G \subseteq SU(N')$. Using 
(\ref{sundecomp}) and self-similarity of $\suinf$~\cite{suninfrep0}, we conclude the following 
relations:
\bea
&& SU(N') \times G_\infty \supseteq [G \times G_\infty \cong \suinf] \supseteq SU(N) \times 
G_\infty \label{sunnp}\\
&& SU(N') \times G_\infty \supseteq SU(N'-N) \times (SU(N) \times G_\infty) \cong 
SU(N'-N) \times \suinf \cong \suinf \nonumber \\
\label{ginf}
\eea
Thus, we conclude that $G_\infty \cong \suinf$ and $G \otimes \suinf \cong \suinf$. 
In~\cite{hourisqgrym} we demonstrate this relation explicitly. If $n$ subsystems with $G$ 
symmetry are distinguished from the rest of the Universe, $\hm_U$ can be decomposed to:
\be
G^n \times \suinf \cong G^n \times \suinf^n \cong (G \times \suinf)^n  \label{gsuinftensor} \\
\ee
For $n \rightarrow \infty$:
\bea
&& (G \times \suinf)^{n \rightarrow \infty} \cong \suinf \label{nginf} \\
&& \bm[\hm_U] = \bigotimes_i \bm[\hm_i] \cong \bigotimes_i \biggl(\bm[\hm_{G_i}] \otimes 
\bm[\hm_\suinf] \biggr) \cong \biggl (\bigotimes_i \bm[\hm_{G_i}] \biggr) \otimes 
\bm[\hm_\suinf]) \nonumber \\
\label{bmsuinftensor}
\eea

\subsection{$\suinfa$, Virasoro algebra and their roles in QGR}  \label{app:stringcomp}
It is useful to compare the algebra (\ref{statealgebra}) with Virasoro algebra which arises 
in string theories. Virasoro algebra without central extension is homomorphic to $\suinfa$ 
algebra (\ref{statealgebra})~\cite{suninfvirasoro,suninfvirasoro0,suninftorus,suninfrep0}. 
Indeed, similar to (\ref{statealgebra}), Virasoro algebra is a connected subalgebra of area 
preserving diffeomorphism of torus $\mathbb{T}^{(2)}$~\cite{suninfvirasoro,suninfvirasoro0}. 
However, it is well known that for its application in quantization string theories, it must 
be extended by a central charge, such that quantum anomalies can be canceled out. In the 
case of \sqgr~such extension is not necessary, because quantization and construction of a 
Hilbert space are performed algebraically: there is no {\it background} spacetime, and no 
classical limit anomaly to be removed. In fact, because the diffeomorphism generated by the 
Poisson algebra (\ref{lapp}) must preserve the area, it cannot preserve distances and 
angles. Consequently, both conformal and Weyl (dilaton) symmetries are broken, and the 
central charge which is the conformal symmetry charge is necessarily zero.

\sqgr~is compared with both perturbative and non-perturbative approaches to string theory 
in~\cite{hourisqgrcomp}. Here we briefly remind their main differences. Although in both 
models a 2D space seems to have an essential role, its origin and physical interpretation is 
very different. The worldsheet in string theory is a real (1+1)D spacetime, which its 
deformation is interpreted as presenting quantum gravity of extended physical and quantum 
objects - strings - embedded in a multi-dimensional quantum space. In the perturbative 
approach, coordinates of this extended space are perceived locally as fields living on the 
(1+1)D worldsheet and their dynamics is ruled by a sigma model. At least 4 of these quantum 
fields should be real and non-compactified to realize a classical spacetime far from Planck 
scale. By contrast, in \sqgr~what we called diffeo-surface is a virtual surface, which its 
diffeomorphism algebra is homomorphic to $\suinf$. We use its coordinates to parameterize an 
abstract Hilbert space defined on the complex numbers $\mathbb{C}$. Therefore, \sqgr~and 
string theory are profoundly different. 

\section{Non-Abelian algebra as quantization }  \label{app:quant}
A system that satisfies axioms of quantum mechanics \`a la Dirac and Von Neumann (or their 
variants) can be called quantum and does not need {\it quantization}. Indeed, it is well 
known that canonical commutation relations between conjugate operators are meaningful only 
if at least one of them is unbounded and has a continuous spectrum (see e.g. the chapter on 
uncertainty relation in~\cite{qmmathbook}). The most direct evidence of this property is 
the fact that two finite dimensional hermitian operators $\hA$ and $\hB$ cannot satisfy the 
canonical quantization relation $[\hA, \hB] = -i\hbar \mathbbm {1}$, because the trace of 
the l.h.s. would be null, whereas that of the r.h.s would be a finite non-zero value. 
Therefore, canonical quantization is useful and applicable only for quantization of 
classical theories, which usually include unbounded quantities such as spacetime or fields. 
Nonetheless, the Planck constant $\hbar$ must be somehow involved in the construction of 
any quantum model, such that when $\hbar \rightarrow 0$ the model behave classically, that 
is conjugate observables, like coordinate and momentum commute with each other.

A typical example of quantum models without continuous degrees of freedom and usual 
uncertainty relations is quantum Heisenberg model of interacting spins on a lattice, which is 
widely used in condensed matters physics. The Hamiltonian of this model is 
$\hH_H \equiv -J \sum_{\{j\}} (\vec{S}_j \vec{S}_{j+1})$, where $J$ is a coupling constant and 
summation is over all sites of the lattice. The Hilbert space at each site represents $SU(2)$ 
symmetry and its generators $\hS_i,~ i=1,~2,~3$ can be normalized such that 
$[\hS_i, \hS_j] = -i\hbar \epsilon_{ijk} \hS_k$. 
%See also supplementary section \ref{app:sym} for more details about this model. 
It is evident that there is no pair of operators in this 
model that satisfies canonical quantization relation. Nonetheless, innumerable studies show 
that particles interacting according to Heisenberg model behave as predicted by axioms of 
quantum mechanics. In particular, as expected, the 3 components of spins or angular momentum 
vectors become simultaneously measurable when $\hbar \rightarrow 0$. 

The above example shows that non-commutativity of the algebra of observables does indeed 
replace the canonical quantization. Notably, this property is used in construction of 
several QGR models, such as: non-commutative geometry 
QGR~\cite{noncummut,noncummut0,qgrnoncommut,qgrmatrixnoncommut}, some string 
theories~\cite{noncommutstring}, and matrix 
models~\cite{qgrgaugedual,qgrmatrix,qgrmatrixrev,noncommutmatrix}. However, non-commutative 
spacetimes are stirringly constrained by 
observations~\cite{qgrtestnoncommut,qgrtestnoncommut0,qgrtestnoncommut1}.

\subsection {Dual conjugate of generators in spherical harmonic representation of $\suinfa$} \label{app:llmconj}
As we discussed in Sec. \ref{sec:quant}, generators $\hJ_{lm}$ of the dual Hilbert space 
$\hm_U^*$ defined in (\ref{lquantize}) are complex-valued functions. By using representation 
(\ref{lharminicexp}) for the generators of the $\suinfa$ algebra (\ref{statealgebra}) and 
applying two sides of (\ref{lquantize}) to an arbitrary vector $\psi \in \hm_U$, which is 
also a complex function, we find:
\be
\frac{\partial Y_{lm}}{\partial \eta} \frac{\partial \hJ_{lm}}{\partial \zeta} - 
\frac{\partial Y_{lm}}{\partial \zeta} \frac{\partial \hJ_{lm}}{\partial \eta} = i\hbar 
\label{jlmeq}
\ee
where here $\eta \equiv \cos \theta$ and $\zeta \equiv \phi$. Solutions of the first-order 
partial differential equation (\ref{jlmeq}) determines conjugate operators (functions) 
$\hJ_{lm}$. 

To solve equation (\ref{jlmeq}) we first remind the definition of spherical harmonic 
functions $Y_{lm}$:
\be
Y_{lm} (\eta, \zeta) \equiv (-1)^m \sqrt{\frac{(2l + 1)}{4\pi}~\frac{(l-m)!}{(l+m)!}} 
P_{lm} (\eta) e^{im\zeta} \label{sphereharmdef}
\ee
where $P_{lm}(\eta)$ is the associated Legendre polynomial for any $l$ and $|m| \leqslant l$. 
$(\eta, \zeta)$. A family of solutions for equation (\ref{jlmeq}) can be obtained by using 
Cauchy initial value method:
\be
%\hJ_{lm} (\eta, \zeta, \eta_0, \zeta_0, \hJ_{lm}(\eta_0, \zeta_0)) = \frac{\eta - \eta_0 + im \hJ_{lm}(\eta_0, \zeta_0) Y_{lm} (\eta_0, \zeta_0)}{im Y_{lm} (\eta, \zeta)} %Note: This solution is correct - up to a constant factor - only for $J_0 = \eta_0 = 0  
\hJ_{lm} (\eta, \zeta,) = - \frac{\hbar (\eta -\eta_0)}{m Y_{lm}} + J_0 \label{jlmsolgen}
\ee
where $\eta_0$, $\hJ_0$ are integration constants. Choosing $\eta_0 = J_0 = 0$ simplifies 
the expression of conjugate operator $\hJ_{lm} (\eta, \zeta)$ to:
\be
\hJ_{lm} (\eta, \zeta) = - \frac{\hbar\eta}{m Y_{lm} (\eta, \zeta)}  \label{jlmsol}
\ee

\subsection {Dual conjugate of continuous parameters in spherical harmonic representation of $\suinfa$} \label{app:positionop}
%######## Formula here are double checked. They are correct (January 22)
In the previous section we showed that conjugate generators $\hJ_{lm} \in \bm[\hm_U^*]$ are 
functions of of parameters $\eta$ and $\zeta$. Thus, $(\eta, \zeta) \in \bm[\hm_U^*]$ and 
thereby their conjugates $i\hbar ~ \frac{\partial}{\partial \eta}, i\hbar ~ \frac{\partial}{\partial \zeta} \in \bm[\hm_U]$, 
and they satisfy standard commutation relations of unbounded continuous operators. Here we 
use the representation (\ref{lharminicexp}) of the generators $\hL_{lm} (\eta, \zeta)$ of 
$\suinfa$, and properties of spherical harmonic functions $Y_{lm}$ and associated Legendre 
polynomials $P_{lm} (\eta)$, to relate $\hL_{lm}$ to the canonical quantization conjugates 
of parameters $(\eta, \zeta)$.

For $m = 0$, $P_{l0}(\eta) \equiv P_l (\eta)$ is the Legendre polynomial and generators 
$\hL_{l0}$ can be expended as:
\bea
\hL_{l0} & = & i\hbar \frac{\partial Y_{l0}}{\partial \eta} \frac{\partial}{\partial \zeta} = 
i\hbar ~ \sqrt{\frac{(2l + 1)}{4\pi}}~\frac{\partial P_l}{\partial \eta} ~
\frac{\partial}{\partial \zeta} = 
i\hbar ~ \sqrt{\frac{(2l +1)}{4\pi}}~\frac{l+1}{1-\eta^2} \biggl (\eta P_l (\eta) - 
P_{l+1} (\eta) \biggr ) \frac{\partial}{\partial \zeta} \nonumber \\
& = & \biggl (\frac{l+1}{1-\eta^2} \biggr ) \biggl (\eta Y_{l0} (\eta) - 
\sqrt{\frac{2l+1}{2l+3}} Y_{(l+1)0} (\eta) \biggr ) 
\biggl (i\hbar ~ \frac{\partial}{\partial \zeta} \biggr )  \label{zetaderiv}
\eea
Thus:
\be
i\hbar ~ \frac{\partial}{\partial \zeta} = \biggl (\frac{1-\eta^2}{l+1} \biggr ) 
\biggl (\eta Y_{l0} (\eta) - \sqrt{\frac{2l+1}{2l+3}} Y_{(l+1)0} (\eta) \biggr )^{-1} \hL_{l0}
\label{zetaderivsol}
\ee
Similarly, 
\be
(-1)^m e^{im\zeta} \hL_{l,-m} - e^{-im\zeta} \hL_{lm} = -2m\hbar ~ e^{-im\zeta} Y_{lm} 
\frac{\partial}{\partial \eta}  \label{etaderiv}
\ee
Using properties of spherical harmonic functions, namely $Y_{l,-m} = (-1)^m Y^*_{lm}$, it is 
straightforward to show that $\hL_{l,-m} = (-1)^{m+1} \hL^\dagger_{lm}$, and equation 
(\ref{etaderiv}) can be written as:
%Note: Y_l,-m = (-1)^m Y_lm^\dagger. The additional -1 is due to the presence of i\hbar in the definition \hL_lm.
\be
ie^{im\zeta} \hL^\dagger_{lm} + ie^{-im\zeta} \hL_{lm} = 2m ~ e^{-im\zeta} Y_{lm} 
\biggl (i\hbar ~ \frac{\partial}{\partial \eta} \biggr )  \label{etaderivp}
\ee
Finally:
\be
i\hbar ~ \frac{\partial}{\partial \eta} = i (2m ~ e^{-im\zeta} Y_{lm})^{-1} 
(e^{im\zeta} \hL^\dagger_{lm} + e^{-im\zeta} \hL_{lm})  \label{etaderivpsol}
\ee
%It is possible to invert (\ref{zetaderiv}) and (\ref{etaderivp}), and express the quantum conjugate of $\zeta$ and $\eta$, that is $i\hbar ~ \partial / \partial \zeta$ and $i\hbar ~ \partial / \partial \eta$ with respect to operators $\hL_{l0}$ and $\hL_{lm}$, respectively. However, 
Expressions (\ref{zetaderivsol}) and (\ref{etaderivpsol}) shows that the usual conjugates of 
$\eta$ and $\zeta$ are highly degenerate and can be obtained from many $\hL_{lm}$ operators. 
This observation demonstrates that quantization through the algebra (\ref{statealgebra}) is 
much richer than the canonical quantization of the finite dimensional parameter space - the 
spacetime, because the number generators $\hL{lm}$ and its conjugate $\hJ_{lm}$ is infinite, 
whereas 
$(i\hbar ~ \frac{\partial}{\partial \eta}, i\hbar ~ \frac{\partial}{\partial \zeta})$ 
and their conjugate $(\eta, zeta)$ constitute only a pair of operators. 
%Nonetheless, as we discussed in Sec. \ref{sec:quant}, the parameters and their canonical conjugate can be considered as representation of operators corresponding to these observables.

\subsubsection{About quantum uncertainty, nonlocality, and topological effects}  \label{app:uncertnonlocal}
The presence of usual uncertainty relations in a model 
that does not follow the standard quantization procedure is remarkable. For instance, if 
generators $\hL_{lm}$ depended on the second order 
derivatives of parameters, rather than the first order, there were no analytical expression 
for $i\hbar \partial / \partial \eta$ as a function of $\hL_{lm}$ generators. Indeed, it 
has been shown~\cite{hiesenbergnonlocal} that in any model with indeterminism and 
nonlocality, these properties and their quantification are related. On the other hand, 
nonlocality and contextuality of quantum mechanics are not maximal~\cite{qmmaxnonlocal}, 
in the sense that the CHSH inequalities~\cite{qmcontextshch} do not have the maximum 
possible deviations from prediction of a local and causal model. Therefore, a priori a 
model which does not explicitly follow the standard procedure for quantization may have 
commutation relations and nonlocalities different from what is predicted by quantum 
mechanics. 

Nonlocality from indeterminism reflects itself in entanglement. But, it does not explain 
global/topological quantum effects, such as Aharanov-Bohm effect~\cite{qmaharanovbohm} and 
quantum Hall phenomenon, see e.g.~\cite{qmhallrev} for a review. To explain these processes 
one has to virtually remove some points of the space to make it non-simply connected and 
topologically non-trivial. However, if the spacetime is a physical entity, removing or 
ignoring part of it to explain some processes is difficult to justify, because they can be 
detected by other observations. For instance, in the case of Aharanov-Bohm setup, the path 
of a neutral particle can pass through the singular magnetic flux without being disrupted. 
By contrast, if the spacetime is interpreted as a parameter space, absence of some point in 
some phenomena is not controversial. In Sec. \ref{sec:qmevol} we show that a symmetry 
invariant functional similar to Yang-Mills Lagrangian can be defined for \sqgr. Therefore, 
topological quantum phenomena observed in non-gravitational many-body systems may be 
relevant for quantum gravitational processes as well. 

\section{Subsystems}  \label{app:subsysdetail}
In this supplementary section we provide more details about how a quantum system may consider 
as an ensemble of subsystems, both logically and physically, and how the process of division 
of a system to subsystems affect the Hilbert space of ensemble, its symmetries and other 
properties.

\subsection{Criteria for identifying quantum subsystems} \label{app:subsyscriteria}
General conditions for the division of a quantum system with Hilbert space $\hm$ to a set 
of  quantum subsystem is studied in details in~\cite{sysdiv}. They can be summarized as the 
followings:
\setcounter{enumi}{0}
\renewcommand{\theenumi}{\alph{enumi}}
\begin{enumerate}
\item There must exist sets of operators $\{A_i\} \subset \bm [\hm]$ such that 
$\forall ~ \hat{a} \in \{A_i\}$ and $\forall ~ \hat{b} \in \{A_j\},$ 
and $i \neq j,~ [\hat{a}, \hat{b}] = 0$;  \label{subcommut}
\item Operators in each set $\{A_i\}$ must be local;   \label{sublocal}
\item $\{A_i\}$'s must be complementary, meaning 
$\otimes_i \{A_i\} \cong \text{End} ~ (\bm [\hm])$.  \label{subcomplement}
\end{enumerate}
In~\cite{sysdiv} the locality condition (\ref{sublocal}) refers to locality of operations 
regarding the background spacetime. In \sqgr~there is no {\it background} spacetime. 
Nonetheless, we can extend the concept of locality to the diffeo-surface of parameters, and 
thereby to the Hilbert space $\bm [\hm_U]$.

\subsection{Hilbert space, symmetries, and coherence of subsystems} \label{app:subsyssymm}
Division of $\bm[\hm]$ to subspaces $\{A_i\}$ as subsystems according to the criteria 
(\ref{subcommut} - \ref{subcomplement}) means that the symmetry $G$ represented by $\bm[\hm]$ 
breaks to a direct product $\bigotimes_i G_i \subseteq G$ such that 
$\bm[\hm] \supseteq \bigotimes_i \mathfrak{g}_i$ where $\mathfrak{g}_i$ is the algebra associated 
to group $G_i$ and $\hm \supseteq \bigotimes_i \hm_i$, where the Hilbert space $\hm_i$ represents 
$G_i$. Here, the order of factors in the tensor product is not important. Thus, through this 
work tensor products are assumed to be permutation symmetrized. The ensemble of subsystems with 
similar symmetries and representations present multiplicity of their type. 

\subsection{Entanglement of subsystems} \label{app:subsysentangle}
If the dimension of subspace $\{A_i\}$ is $N_{A_i}$, the minimum dimension of $\hm_i$ must 
be $N_i$ such that $N_i^2 - 1 \geq N_{A_i}$. The complementary condition (\ref{subcomplement}) 
for the division of a quantum system to subsystems dictates that $\sum_{i} N_i = N$, where 
$N$ is the dimension of Hilbert space of the system. Assuming a division to two subsystems, 
(\ref{sundecomp}) leads to $G \supseteq G_1 \times G_2$, where $G_1$ and $G_2$ are symmetry 
groups represented by Hilbert spaces $\hm_1$ and $\hm_2$ of the subspaces, respectively.  
Therefore, in general, division of a quantum system reduces the symmetry. 

Dimension $N_{12}$ of the tensor product $\hm_1 \times \hm_2$ of Hilbert spaces of subsystems 
is $N_{12} \equiv N_1 \times N_2 \geq N$. For $N_1 ,~ N_2 > 1$, $N \geqslant 2$ and 
$N,~N_1,~N_2 \leqslant \infty$. On the other hand, consistency imposes isometry between states 
in the initial Hilbert space $\hm$ and those of the ensemble of its components. Considering 
the relation between dimension of $\hm$ and $\hm_1 \times \hm_2$, the two components must be 
entangled to preserve the isometry. Iteration of this procedure on subsystems extends these 
properties to the case where $\hm$ is decomposed to an arbitrary subsystems. 

It is important to emphasize the crucial role of the assumed isometry between space of states 
of the system as a whole and that of the ensemble of its subsystems. Without this condition, 
considering a collection of quantum systems together does not impose entanglement on the 
components.  

This result can be considered as another demonstration of the Proposition \ref{propentang} in 
\sqgr~but without assuming a pure state for the Universe as a whole. However, as we discussed 
in Sec. \ref{sec:division}, in \sqgr~dimensions $N$, $N_1$ and $N_2$ are infinite. Therefore, 
according to (\ref{suinfmult}), the isomorphism condition is always satisfied and cannot be 
used to prove the Proposition \ref{propentang}. Nonetheless, as for $N$, $N_1$ and $N_2$ very 
large but finite the assumption of isomorphism leads to entanglement of components, it should 
be also true for the limit where these dimensions approach infinity. Indeed, in this limit 
(\ref{suinfmult}) is equivalent to the purity condition used in Sec. \ref{sec:entangle} for 
proving the Proposition \ref{propentang}. In~\cite{hourisqgrym} we quantify entanglement 
between subsystems and their environment. Another quantity related to both symmetry breaking 
and entanglement is quantum complexity~\cite{qmcommuncomlex,qmcommuncomlexrev}, which has been 
used in some QGR models such as ADS/CFT~\cite{qmcomplexity,qmcomplexity0}. However, its 
relation with entanglement is not straightforward and cannot be considered as entanglement 
measure.
%Note: Read 2311.04277 about complexity of entanglement (they use geometry in Hilbert space)
%Note: State symmetry is not the same as Hilbert space symmetry. For instance symmetry of GHZ$_n$ state is SU(2). But, symmetry of its Hilbert space is $SU(2)^n$.

\subsubsection{Consequences of the global entanglement}  \label {app:entanglecons}
The ever-existing quantum correlation/entanglement between every subsystem and the rest of the 
Universe reinforces decomposition (\ref{gdecomp}), because if there is a subsystem with only 
finite dimensional Hilbert space, its entanglement with the infinite degrees of freedom of the 
rest of the Universe makes its effective degrees of freedom, and thereby symmetry of its 
Hilbert space infinite dimensional. 

As $G$ symmetry emerges from clustering of states, it is expected that interaction/mixing 
between subsystems through $G$ symmetry be much stronger than via the global $\suinf$. This 
property is consistent with the observed universality of gravitational interaction and its 
weakness in comparison with internal interactions of subsystems/particles. It also provides an 
explanation for the weak gravity conjecture~\cite{weakgrconj}.

From (\ref{bmsuinftensor}) it is clear that in what concerns mathematical definition and 
action, there is no essential difference between finite rank internal symmetries of subsystems 
and $\suinf$ - {\it gravity} - symmetry. Nonetheless, the resemblance of gravitational 
interaction to a gauge field in \sqgr~is very different from what is called gauge-gravity 
duality models in the literature~\cite{qgrgauge,qgrgauge0,qgrgaugesep,qgrgaugesep0,qgrgaugesep1}, 
Notably, in \sqgr~the gauge symmetry of quantum gravity is not directly related to the 
classical Lorentz invariance. In \sqgr~Lorentz and Poincar\'e symmetries correspond to 
reparameterization invariance of the parameter space of the model. Hence, difficulties which 
arise in gauge-gravity proposals~\cite{ssymmtheorem,ssymmsusy,grgaugemix} are irrelevant. 
%We will describe other consequences of the global entanglement of subsystems through $\suinf$ symmetry along with other properties of \sqgr~in the following sections.

\subsubsection{Observability of the entanglement}  \label{app:entangleobs}
Finally, an important question is whether the global entanglement of subsystems is observable. 
The answer is: in principle yes, but in practice no ! For such observation we need to consider 
at least 3 subsystems: the target subsystem, a quantum clock as explained in sections 
\ref{sec:clock} and \ref{app:time}, and the rest of the Universe as a reference observer or 
environment. Assume that the observer performs a POVM on the target such that its state is 
projected to a pointer state defined in (\ref{subsysstate}). According to the proposition 
\ref{propentang} this operation projects the state of the rest of the Universe, that is the 
observer and the clock to a pointer state. How can the observer verify this expectation ? 
The only way is the measurement of the expected pointer state of the observer-clock system. 
However, this is impossible. In fact, it is even impossible to verify that the target 
subsystem is indeed in a projected state after the first operation. Such verification needs 
repetition of the same operation. However, a repetition implicitly means passage of time as 
a consequence of the change of clock state. But, if the clock state varies, the state of 
observer-clock subsystem changes, which in turn changes the state of the target subsystem 
according to the proposition \ref{propentang}. Neglecting this verification, the only way to 
verify the entanglement is either the target subsystem or the observer-clock themselves should 
perform a self-state verification test. However, none of these operations can be performed 
in isolation and without modifying some of the degrees of freedom of the target.

\section{Relative dynamics and its parameters}  \label{app:reldyn} %###### To do: Read 2405.15455 about relational QFT.
This section provides complementary discussions about parameter space of \sqgr~and how 
dynamics is induced.

\subsection {Relative time}  \label{app:time}
Introduction of relative time in quantum mechanics and quantum gravity is extensively 
studied in the literature~\cite{qmtimepage,qmtimepage0,qmtimepage1}, specially as a mean 
for solving the problem of Hamiltonian constraint and the absence of time in canonical 
quantization of gravity. Efficiency of this approach is criticized in~\cite{qgrtimecritic}, 
and equivalence of various definitions of a relative and dynamical time is demonstrated 
in~\cite{qmtimedef}. Many of complexities of quantum time and reference frame definition in 
QGR proposals arise because of seeking a diffeomorphism invariant quantized spacetime. Here 
we show that in \sqgr~spacetime is related to the parameter space of the model. Thus, it 
should not be quantized, and concerns about the validity of a relative 
time~\cite{qgrtimecritic} are irrelevant.

Once the Universe is divided to subsystems, it becomes possible to define dynamics as relative 
variation of their states. But, in \sqgr~there is no completely isolated subsystem to play the 
role of perfect clock. 
%In analogy to the division of the Universe to subsystems, our approach with the definition of a quantum clock and time parameter would be operational. 
Quantum entanglement between an imperfect clock and quantum systems, and its effects on the 
definition of a reference frame is demonstrated in~\cite{qmclockimperfect}. In \sqgr~similar 
effects arise due to the omnipresent entanglement of subsystems according to the Proposition 
\ref{propentang}. Consequently, successive applications of the same operator to a selected 
subsystem in general does not lead to the same outcome, and quantum 
contextuality~\cite{qmcontextual,qmcontextual0} becomes globally manifest. This leads to the 
emergence of an order - a time arrow - between operations. In particular, relative variation 
of states and observables of subsystems can be quantified with respect to those of the clock. 
Thus, the emergence of time and dynamics in \sqgr~is in spirit similar to the Page \& Wootters 
proposal~\cite{qmtimepage}. In particular, as we described in detail in 
Sec. \ref {sec:qmstatesub}, time is treated as one of the parameters characterizing states 
of subsystems\footnote{When a quantum system is constructed from quantization of a classical 
model, there are alternative ways to define dynamics, for instance by using relational 
observables \`a la Dirac~\cite{diraclecture}, or what is called Heisenberg symmetry reduction 
in the phase space. Equivalence of these methods is demonstrated in~\cite{qmtimedef}. As 
\sqgr~is not obtained from a classical model, the conditional approach of Page \& Wootters is 
more suitable. On the other hand, as a parameter characterizing states, a subsystem can be in 
its coherent superposition, which can be interpreted as a POVM measurement of time parameter.}.
%In the following subsections we explain in more details properties of the clock and 
%There is however no need for explicit conditional wave functions, because in contrast to the canonical quantization of QGR, there is no need for (3+1)D decomposition of formulation

\subsubsection {General properties of a quantum clock and time parameter}  \label{app:timeprop} 
Without considering specifics of a suitable quantum clock and how the change of its state 
and/or observables are quantified as a time parameter, we can conclude the following general 
properties about the clock and its observable used as time:
\begin{description}
\item{\it Irrelevance of initial state:} The initial state of the clock, that is the outcome 
of the first operation on it initializes relative dynamics of subsystems, but its value is 
irrelevant for further measurements. Only its variation in the next measurement is relevant.
\item{\it Arrow of time:} Selection of any subsystem as clock induces an ordering in 
operations on subsystems. This is a direct consequence of the proposition \ref{propentang}.
\item{\it Monotonous time:} Observables used for defining a time parameter are arbitrary, 
but, the time parameter must be a single-valued monotone or cyclic real or purely imaginary 
function of them. %For instance, a vector-valued observable is not suitable, unless its module is used as time.
\item{\it POVM time:} Measurement of time should be in general considered as POVM.
%Quantification of clock state variation must be a single-valued uniquely calculable function of parameters that characterize state of the clock. Nonetheless, its measurement does not need to be projective.
%\item{\it One real parameter:} The above conditions mean that without loss of generality, the  parameter called {\it time} $t$, which quantifies the change of clock's state can be chosen to be a real number, that is $t \in \mathbb{R}$.
\item{\it Time clicks for everyone:} Entanglement of the clock with the rest of the Universe 
means that after initialization of the clock, every other subsystem evolves along it. This is 
the origin of the {\it arrow of time}.
\end{description}

An extended and mathematically rigorous description of necessary conditions for a quantum 
clock can be found in~\cite{qmtimedef}. Nonetheless, simple properties described here are 
sufficient for our purpose of defining a relative dynamics, because we do not need to relate 
it to a classical model.

\subsection {Relation between reparameterization and diffeomorphism}  \label{app:reparam}
Diffeomorphism is a continuous and invertible map of a manifold $M \rightarrow M$.  
Parameterization is a continuous map $M \rightarrow N$, where in general $M \neq N$. Thus, 
diffeomorphism can be considered as a special case of parameterization when $N=M$. For this 
reason diffeomorphism and reparameterization are usually considered as the same operation. 
Moreover, diffeomorphism of $N$ induces a diffeomorphism transformation in $M$, when they 
are related under a reparameterization map. 

In \sqgr, for the whole Universe $M = \hm_U$ and $N = D_2 \cong \mathbb{R}^{(2)}$. A basis 
for $\hm_U$ consists of vectors 
$|x\rangle, ~ \forall x \equiv (\eta, \zeta) \in \mathbb{R}^{(2)}$. Reparameterization 
corresponds to a diffeomorphism in $\mathbb{R}^{(2)}$ such that: 
$x \rightarrow x' (x) ~ \forall x \in \mathbb{R}^{(2)}$. An infinitesimal $\suinf$ 
transformation of the basis $|x\rangle$ is 
$|x' (\epsilon)\rangle = (\mathbbm{1} + \epsilon \sum_{(l,m)} \hL_{l,m} (x)) |x\rangle$. 
A general $\suinf$ transformation 
$U = \exp (\int d^2 \Omega \sum_{(l,m)} \epsilon_{lm}\hL_{l,m} (x))$ for any set of constants 
$\epsilon_{lm}$ and $(|\psi \rangle \in \hm_U) \rightarrow |\psi'\rangle = U |\psi\rangle $. 
Considering the differential representation of $\hL_{l,m} (x)$ in (\ref{lharminicexp}), it 
is clear that reparameterization, that is a diffeomorphism in $D_2$ is equivalent to a 
$\suinf$ transformation of $\hm_U$. After division of the Universe to subsystems 
$N \cong \mathbb{R}^{(4)}$ and reparameterization corresponds to diffeomorphism in 
$\mathbb{R}^{(4)}$ and leads to $M = \hm_s \rightarrow \hm_s$ and 
$x \rightarrow x' (x) ~ \forall x \in \mathbb{R}^{(4)}$. Similar to the case of the whole 
Universe, this transformation can be described as application of $\suinf$ parameterized by 
only a pair of parameters. This is possible because of property (\ref{suinfmult}) of 
$\suinf$ and superposition of quantum states.

\subsection {Coordinate independent definition of curvatures} \label{app:curvature}
For any Riemannian or pseudo-Riemannian manifold $(\mm,g)$ of dimension $d \geqslant 2$ 
equipped with a Levi-Civita connection $\nabla$ the (1,3)-Riemann curvature tensor at point 
$p \in \mm$ is defined as:
\be
R_p (X, Y)Z = (\nabla_X \nabla_Y - \nabla_Y \nabla_X - \nabla_{[X,Y]}) Z  \label{riemanncurve13}
\ee
Vector fields $X,~Y,~Z \in T\mm_p$ and $T\mm_p$ is the tangent space at $p$. When $X,~Y,~Z$ 
are chosen to be $\partial_i \equiv \partial / \partial x^i$ basis of the tangent space for 
coordinates $x^i,~ i=0, \cdots d-1$, one recovers the usual coordinate dependent definition 
of the Riemann curvature tensor (we drop $p$ because it corresponds to the point with 
coordinates $x^i$):
\bea
&& R (\partial_i, \partial_j) \partial_k = R^{l}_{kij}, \label{riemanncurve13c} \\
&& R_{ijkl} \equiv R(\partial_i, \partial_j, \partial_k, \partial_l) \equiv \bigg 
\langle R (\partial_i, \partial_j) \partial_k, \partial_l \bigg \rangle = g_{ml} R^m_{kij} 
\label{riemanncurve04c}
\eea
Using the notation defined in (\ref{riemanncurve04c}) for (0,4)-Riemann curvature tensor,  
sectional curvature $K(\Pi) = K(X, Y)$ with respect to a 2D plane $\Pi \subset T\mm_p$ 
containing two vectors $X, Y \in T\mm_p$ at $p \in \mm$ is defined as:
\be
K(\Pi) \equiv K(X,Y) = \frac{R_p(X, Y, X, Y)}{\langle X, X\rangle \langle Y, Y\rangle - 
\langle X, Y\rangle^2}  \label{sectionalcurve}
\ee
Notice that $K(\Pi)$ is independent of the choice of $X$ and $Y$ and depends only on the 
plane passing through them. It can be shown that $\langle R_p (X, Y)Z, W)\rangle$ can be 
expanded with respect to sectional curvatures~\cite{seccurvproof}. Using relations between 
different curvature tensors of a Riemannian manifold, the scalar curvature at point 
$p \in \mm$ is defined as~\cite{seccurvproof0}:
\be
R(p) = \sum_{i \neq j} R_p (e_i, e_j, e_i, e_j) = \sum_{i \neq j} K_p (e_i, e_j) 
\label{riccicurvsec}
\ee
where $e_i, ~ i=0, \cdots, d-1$ is an orthonormal basis of $T\mm_p$. From (\ref{riccicurvsec}) 
we conclude that there is only one sectional curvature at each point of a 2D surface and it 
is equal to its scalar curvature.

\section{Effective geometry of the quantum state of subsystems}  \label{app:qslgeom}
In this section we first briefly review what is called {\it Quantum Speed Limit} (QSL) first 
discovered as the Mandelstam-Tamm inequality~\cite{qmspeed}. Then, we use QSL to define an 
average path in the parameter space $\Xi$ of the Hilbert space of subsystems, and show that 
it is related to their quantum state irrespective of quantum purity of the state, that is 
whether states of clock and reference and their back reaction on the rest of the Universe are 
taken into account, or subsystems are treated as being open. We also describe expansion of 
the density matrix and its variation with respect to the generators of $\suinf$. Finally, we 
discuss the physical interpretation of these results and their relationship with the 
perceived classical spacetime. 

\subsection{Review of QSL and emergence of effective spacetime in \sqgr} \label{app:qmspeed}
The origin of limited speed of information transfer in quantum mechanics is the Heisenberg 
uncertainty relation between time $t$ and energy $E$: 
\be
\Delta t \Delta E \geqslant \hbar/2  \label {timeeneruncertain}
\ee
where here $\Delta R$ means standard deviation of measurement outcomes for observable $R$. 

\subsubsection{Unitary evolution}  \label{sec:qslunitaryevol}
Standard deviations of any two statistical variables $R_1$ and $R_2$ satisfy the following 
inequality:
\be
\Delta S_1 \Delta S_2 \geqslant \frac{1}{2} \overline{(S_1S_2 - S_2S_1)}  \label{sdave}
\ee
where the bar means statistical average, Using the evolution relation for an observable 
$\hR$ of a unitarily evolving quantum system with Hamiltonian $\hH$ in a pure state 
$\hrho = |\psi (t)\rangle\langle \psi (t)|$:
\be
\frac{d\hR}{dt} = -\frac{i}{\hbar} [\hR , \hH]   \label{opevolpure}
\ee
Mandelstam and Tamm found the following uncertainty relation for any observable 
$\hR$~\cite{qmspeed}:
\be
\Delta \hR \Delta \hH \geqslant \biggl | \frac{i}{\hbar} \frac{d \bar{\hR}}{dt} 
\biggr |, \quad \quad \Delta \hX \equiv \tr (\hX^2 \hrho) - (\tr (\hX \hrho))^2 
\label{genuncertainty}
\ee
By choosing $\hR = \hrho(t) = |\psi (t)\rangle\langle \psi (t)|$ and integrating 
(\ref{genuncertainty}) for the evolution of state from $\hrho (t_0)$ to $\hrho(t)$ the 
following lower bound is obtained for the time $\Delta t = t - t_0$ necessary for such 
process~\cite{qmspeed,qmuncertainener,qmspeedformul,qmspeedgeom}:
\bea
\Delta t & \geqslant & \hbar \frac{\Delta \Theta}{\Delta \hH} = \hbar 
\frac{\arccos ~ ([\tr (\hrho (t_0) \hrho (t))]^{\frac{1}{2}})}{\Delta \hH}  \label{aqdef}  \\
\cos^2 (\Delta \Theta) & \equiv & \frac {\tr (\hrho (t_0) \hrho (t))}{\tr (\hrho^2 (t_0))} 
\equiv F_R \label{relativepurity}  \\ 
\Delta \hH & = & |-\frac{1}{2} \tr([\hrho, \hH]^2)|^{\frac{1}{2}} \label{aqdefdef}
\eea 
where $0 \leqslant F_R \leqslant 1$ is relative purity. For a pure initial state $\hrho (t_0)$ 
the denominator of $F_R$ is $\tr (\hrho^2 (t_0)) = 1$. From (\ref{relativepurity}) it is clear 
that for $|\psi (t_0) \rangle$ and $|\psi (t)\rangle$ to be orthogonal and completely 
distinguishable (linearly independent), $\cos (\Delta \Theta) = 0$. In this case 
$\Delta \Theta = \pi/2$ and the original Mandelstam-Tamm inequality~\cite{qmspeed} is obtained:
\be
\Delta t \geqslant \frac{\pi \hbar}{2 \Delta \hH}  \label{mtinequal}
\ee
This inequality clarifies the nature of measured time in the Heisenberg uncertainty relation 
(\ref{timeeneruncertain}) as the evolution time of the system's state~\cite{qmuncertainener}. 
For pure states it is independent of the path taken by the system in its Hilbert 
space~\cite{qmspeedgeom,qmspeedgeomopen}.

The lower bound on the evolution of state (\ref{aqdef}) is related to the geometry associated 
to the Hilbert space~\cite{qmspeedgeom,qmspeedgeomopen,qmspeedgeomopen0,qmspeedgeomgen0,qmspeedgeomaction,qmspeedgeomlocal,qmspeedgeommix} (and references therein, see also~\cite{qmspeedrev} 
for a review)\footnote{List of literature on the QSL is long and citations in this brief review 
are far from being exclusive.}. Indeed, in (\ref{aqdef}) the argument of $\arccos$ function is 
the Fubini-Study distance between $\hrho (t_0)$ and $\hrho (t)$~\cite{qmhilbertgeom}, and the 
denominator corresponds to Fubini-Study metric, see e.g.~\cite{qmspeedgeomaction}. They define 
a unique Riemannian geometry for the manifold of pure density matrices. Notably, according to 
the geometric interpretation of the Mandelstam-Tamm QSL~\cite{qmspeedgeom,qmspeedgeomopen}, 
the inequality (\ref{mtinequal}) can be generalized to:
\be
\Delta t \geqslant \frac{\dm (\psi (t_0), \psi (t))}{\llangle \sqrt{g_{tt}} \rrangle} 
\label{ineqgen}
\ee
where $\dm$ is a distance function between two pure or mixed density matrices and $g_{tt}$ is 
a metric associated to the manifold of density matrices. The double bracket means averaging 
over the time interval $\Delta t$ along the evolution path. In contrast to pure states, many 
metrics can be associated to the manifold of density matrices that also includes mixed 
states~\cite{densitymatgeom,densitymatgeom0,densitymatgeom1}. However, only for a handful of 
them geodesics, that is $\dm (\psi (t_0), \psi (t))$ in (\ref{ineqgen}), is 
known~\cite{qmspeedgeomgen0}. 

One of the most popular QSL for mixed states, called Uhlmann QSL, is based on the 
Wigner-Yanase skew information~\cite{wigneryanaseqminfo,wigneryanaseqminfogeo}\footnote{The 
Winger-Yanase skew information $I_{WY}(\hrho, \hA)$ for operator $\hA$ describes the limit 
on the amount of information that a quantum system can convoy through variation of its state 
by application of $\hA$.} as metric and Bures distance~\cite{buresmetric,qmspeedgeomopen}:
\bea
\Delta t & \geqslant & \frac{\hbar \cos^{-1} A (\hrho_0, \hrho_1)}{\sqrt{|}I_{WY}(\hrho, \hH)|}, 
\label{aqdefsqrt} \\
A (\hrho_0, \hrho_1) & \equiv & \tr (\sqrt {\hrho_0}\sqrt {\hrho_1}^\dagger) 
\label{affinity} \\
I_{WY}(\hrho, \hH) & \equiv & -\frac{1}{2} \tr([\sqrt{\hrho}, \hH]^2) = \tr (\hH^2 \hrho) - 
\tr(\sqrt{\hrho} \hH \sqrt{\hrho} \hH)  \label{wyinfo} 
\eea
For pure states $\sqrt{\hrho} = \hrho$ and (\ref{aqdef}) and (\ref{aqdefsqrt}) are equivalent 
to each other. In~\cite{houriqmsymmgr} we used this QSL in the framework of \sqgr~to find an 
effective classical spacetime with (\ref{tracemetricalpha}) as its metric\footnote{The 
expression for bound in (\ref{aqdefsqrt}) is different from that 
in~\cite{houriqmsymmgr,hourisqgrcomp} which was taken from~\cite{qmspeedgeomopen0} by a 
factor of $1/\sqrt{2}$.}. We also demonstrated that this metric has a negative signature and 
the effective geometry is pseudo-Riemannian. However, both (\ref{aqdef}) and (\ref{aqdefsqrt}) 
assume a unitary evolution for the state, that is $d\hrho / dt = -i/\hbar [\hrho, \hH]$. On the 
other hand, the state $\hrho$ of subsystems defined in (\ref{densitynog}) is mixed, because 
the contribution of local symmetry $G$ is traced out. In addition, according to the 
Proposition~\ref{propentang} subsystems are entangled to clock and reference. A priori it is 
possible to take into account the backreaction of these subsystems on each others and find a 
Hamiltonian describing the evolution of $\hrho$ unitarily. But, in practice it is very difficult 
to accomplish such task. Therefore, both clock-reference ensemble and other subsystems should 
be treated as open systems, which may or may not be Markovian. Here we only consider the 
Markovian case.

\subsubsection{Markovian evolution}  \label{secqslmarkov}
The general evolution equation of a density matrix $\hrho$ is 
$d \hrho/dt = \mathsf{\hL} (\hrho)$. For Markovian systems this equation takes a linear form 
with a Lindbladian operator~\cite{qmopenbook} such that $d \hrho/dt = \mathsf{\hL}\hrho$. 
Unlike unitary evolution, many QSL-like inequality can be written for open quantum systems. 
Moreover, they are not always strict and/or attainable limits, because in contrast to the 
unitary case they depend on the path of the state in the space of density matrices of the 
system, see e.g.~\cite{qmspeedgeomlocal,qmspeedoperat} for examples. This can be understood 
from the fact that purification of mixed states of open systems used in some 
approaches~\cite{qmspeedgeomopen,qmspeedgeomfisher} is not unique. 

Considering geometric description (\ref{ineqgen}) of SQL's, several distance functions and 
metrics are used to obtain Mandelstam-Tamm-like QSL inequalities for 
Markovian~\cite{qmspeedgeomaction,qmspeedopenunif} and general evolution of open 
systems~\cite{qmspeednonmarkov,qmspeedgeomgen0,qmspeedgeomlocal}. They use various distance 
definitions such as trace distance~\cite{qmspeedgeomaction}, quantum Fisher 
information~\cite{qmspeedgeomfisher,qmspeedgeomlocal,qmspeedopenunif}, and path length in the 
parameter space characterizing the state~\cite{qmspeedgeomgen0,qmspeedgeomaction}. In the 
latter case, due to the absence of a description for geodesics in a general geometry of state 
space, QSL's can be explicitly calculated only for a handful of cases. Attainability and 
tightness of some of these QSL's are discussed in~\cite{qmspeedgeomlocal,qmspeedhamiltonian}. 
The QSL's can be also obtained without explicitly referring to the geometry of states. 
In~\cite{qmspeednonmarkov} a single QSL applicable to both unitary and non-unitary evolution 
is obtained. Another approach is using relative purity $F_R$ defined in (\ref{relativepurity}) 
for finding a QSL relation for any type of state evolution of open systems~\cite{qmspeedopen}. 
These QSL's are explicit and simpler to use, specially in the context of \sqgr. Here we use 
this approach for Markovian evolution of subsystems of the \sqgr~Universe. 

Using the definition of relative purity (\ref{relativepurity}) and its evolution with time 
$dF_R /dt$, the following QSL for infinitesimal evolution of the state $\hrho (t_0)$ is 
found~\cite{qmspeedopen}:
\be
\Delta t \geqslant \frac{|\cos (\Theta) - 1| \tr(\hrho^2 (t_0)}{\sqrt{\tr 
[(\mathsf{\hL}\hrho (t))^2]}}, \quad \quad \Theta \equiv \arccos (F_R)  \label{qslrelpurity} 
\ee
We can then proceed in the same manner as in~\cite{houriqmsymmgr}. Consider the state of 
subsystems:
\be
\hrho = \tr_G \hrho_G  \label{densitynog}
\ee
where $\hrho_G$ is the state of subsystems as defined in Sec. \ref{sec:fullpath}. Assume that 
there is a clock with time parameter $t$ and a Lindbladian super-operator 
$\mathsf{\hL} : \bm[\hm_s] \rightarrow \bm[\hm_s]$ describing the evolution of $\hrho$ such 
that in (\ref{qslrelpurity}) the equality is achieved:
\be
g_{tt} dt^2 = (\tr(\hrho \delta \hrho))^2 \equiv ds_R^2, \quad \quad g_{tt} \equiv 
\tr[(\mathsf{\hL} \hrho)^2]  \label{metriclindblad}
\ee
where $\delta \hrho$ is variation $\hrho$ during $dt$. For another choice of clock and time 
parameter $t'$ the affine separation $ds_R$ does not change, because it is a scalar function 
of the parameter space $\Xi$ of subsystem and independent of its reparameterization. 
By contrast, in general the equality (\ref{metriclindblad}) changes to an inequality, 
that is $g_{tt} dt^2 \geqslant ds_R^2$. The equality can be established by adding a negative 
term to the l.h.s of this equality. Considering the differential form of the l.h.s., the 
general form of an equality similar to (\ref{metriclindblad}) is $g_{\mu\nu} dx^\mu dx^\nu$, 
where $g_{\mu\nu}$ a symmetric 2-tensor with negative signature, leading to the Lorentzian 
effective metric (\ref{tracemetricalpha})\footnote{In this notation $g_{00}$ is equivalent to 
$g_{tt}$ in (\ref{ineqgen}).}. We remind that in contrast to this demonstration of Lorentzian 
effective geometry, in special and general relativity the negative signature of metric is an 
axiom dictated by the observation of constant speed of light in classical vacuum.

For comparison, in case of the unitary evolution of a pure or mixed state, when the equality 
is attained in the inequality (\ref{aqdefsqrt}), we obtain:
\be
g_{tt}^{(WY)} dt^2 = -\frac{1}{\hbar^2} \tr [\sqrt{\hrho_1}, \hH]^2 dt^2 = 
\tr (\sqrt{\delta\hrho_1} \sqrt{\delta\hrho_1}^\dagger) \equiv ds_{WY}^2 \label{separation}
\ee
As expected, $ds_R \neq ds_{WY}$. Notably, for pure states (\ref{qslrelpurity}) becomes 
trivial, that is $\Delta t \geqslant 0$, because $F_R = 1$ for pure states, and consequently. 
$\Delta F_R = 0$. However, it is straightforward to prove that the order of $ds_R$ and 
$ds_{WY}$, that is which one is larger depends on the details of $\hrho$ and $\delta \hrho$. 
%If we neglect this consistency condition and compare $ds^2_R$ in (\ref{metriclindblad}) with $ds^2_{WY}$ in (\ref{tracemetricalpha}), we find $0 \leqslant ds^2_R \leqslant ds^2_{WY}$, which is consistent with the trivial constraint on $\Delta t$. 

There is another lower bound on the evolution time between two orthogonal states called 
Margolus-Levitin QSL~\cite{qmspeedlevitin,qmspeedlimitequi}, which states that 
$\Delta t \geqslant \pi \hbar / 2 E$ where $E$ is the expectation value of energy difference 
with the ground state of a system along its evolution path, rather than energy dispersion 
as in (\ref{mtinequal}). Although, according to this bound and (\ref{mtinequal}) for 
$E ~ \text{or} ~ \Delta E \rightarrow \infty$ apparently the evolution time can approach 
zero, Lieb-Robinson bound~\cite{liebrobinsonbound} (see e.g.~\cite{liebrobinsonboundrev} for 
a review) rules out such possibility. The ultimate lower bound on $\Delta t$ is the maximum of 
the above bounds~\cite{qmspeednonmarkov}, and may provide a natural UV cutoff. We leave the 
exploration of this possibility and application of geometric description of QSL to \sqgr~to 
future works. For an observer who can only measure the effective metric and affine separation 
deviation from maximum speed appears as an {\it effective mass} induced by interactions that 
prevent the system to have the maximum speed.

Finally, the close relationship between quantum uncertainty relations, signature of the 
effective metric, and its Lorentz symmetry in the framework of \sqgr~shows that modifying 
Heisenberg uncertainty relation proposed in some QGR 
models~\cite{qmmaxnonlocal,hiesenbergnonlocal}, may violate Lorentz invariance and constant 
speed of light, which are stringently constrained~\cite{grestgrb090510a,grbphotonmass,grlorentz}. 
% Note: In~\cite{qmspeedlevitin} $h$ is Planck constant not $\hbar = h/2\pi$. Therefore, the limit would be $h/4E$. 

\subsection{Expansion of density matrix of subsystems and its variation}  \label{app:rhoexpad}
As we discussed in Sec. \ref{sec:subsysalgebra} diffeo-surface of representations of 
$\suinf$ by subsystems is embedded in the parameter space $\Xi$ and pairs of parameters 
$(\eta, \zeta) \in \Xi$ can be assumed t describe the diffeo-surfaces. Thus, the space of 
linear operators $\bm [\hm_s]$ can be generated by operators $\hL_{lm} (\eta, \zeta)$, which 
satisfy the algebra (\ref{algebra4d}). The $\suinf$ generators $\hL_{lm} (\eta, \zeta)$ 
defined in (\ref{lharminicexp}) have the structure of a 2-form with respect to 
$(\eta, \zeta)$ and to make this property explicit, we may write:
\be
\hL_{lm} (\eta, \zeta) \equiv \hL_{lm}^{\mu\nu} (\eta, \zeta), \quad \quad 
\hL_{lm}^{\mu\nu} = - \hL_{lm}^{\nu\mu}  \label{rhoformdef}
\ee
where indices $\mu, \nu$ correspond to those of coordinates $(\eta, \zeta)$ of $\Xi$. More 
generally, Here we concentrate on $\suinf$ symmetry and ignore $G$ symmetry representation by 
$\hm_s$. 

The density matrix $\hrho$ defined in (\ref{densitynog}) can be expanded as:
\bea
&& \hrho = \frac{1}{2} \int dx^4 \sqrt{|\upeta|} ~ \sum_{l,m} \rho^{lm}_{\mu\nu} (x) 
\hL_{lm}^{\mu\nu}, \quad \quad \rho^{lm}_{\nu\mu} = - \rho^{lm}_{\mu\nu} \label{rhoexpan}  \\
&& \frac{1}{2} \int dx^4 \sqrt{|\upeta|} \sum_{l,m; \nu, \nu} \epsilon^{\mu\nu} 
\rho^{lm}_{\mu\nu} (x) = 1, \label{rhocomnorm}
\eea
The range of $(l,m)$ are defined in (\ref{lharminicexp}). The constant 2-form 
$\epsilon_{\mu\nu}$ in the integrand of (\ref{rhoexpan}) is necessary to make it a scalar of 
the parameter space $\Xi$. The expression 
$\rho^{lm} (x) \epsilon_{\mu\nu} \equiv \rho^{lm}_{\mu\nu}$ is a 2-form field on the parameter 
space $\Xi$. Although it has the same geometrical structure as a gauge field 
strength, its physical interpretation is very different. Specifically, components 
$\rho^{lm}$ are positive probability densities and must satisfy the constraint 
(\ref{rhocomnorm}). By contrast, the gauge field strength presents the number of quanta of 
the gauge boson.

Variation of the density matrix $\delta \hrho \in \bm[\hm_s]$ can be expanded in the same 
manner:
\be
\delta\hrho = \frac{1}{2} \int dx^4 \sqrt{|\upeta|} ~ 
\sum_{l,m} \delta\rho_{\mu\nu}^{lm} (x) \hL_{lm}^{\mu\nu}  \label{varhrho}
\ee
%Note: \epsilon_\mu\nu..... is a constant pseudo tensor because in odd dimensions it gets a minus sign under reflection. For this reason in 4D \epsilon_ijk is not Lorentz invarient, but \epsilon_\mu\nu is - we can assume it as \epsilon (Levi-Civitta symbol) in 2-surfaces defined by x^\mu, x^\nu.
These expansions can be applied to (\ref{distcase}) to determine affine separation as 
a function of quantum state and its variation. In principle, one can write a similar 
expansion for $\sqrt{\delta\hrho}$, which is usually difficult to calculate 
directly~\cite{densitysqrt}:
\be
\sqrt{\delta\hrho} \equiv \frac{1}{2} \int dx^4 \sqrt{|\upeta|} ~ \sum_{l,m} 
\delta \rho_{\mu\nu}^{'lm} (x) \hL_{lm}^{\mu\nu} \label{sqrtvarhrho}
\ee
Using $\delta\hrho = \sqrt{\delta\hrho} \sqrt{\delta\hrho}$, expansions (\ref{varhrho}) and 
(\ref{sqrtvarhrho}) and algebra of generators $\hL_{lm}^{\mu\nu}$ (\ref{algebra4d}), it is 
straightforward to see that $\delta \rho_{\mu\nu}^{'lm}$ modes depend on the mixture of all 
$\delta \rho_{\mu\nu}^{lm}$ modes, but remain local with respect to parameters $x$.

\subsection{Spacetime as a property and its consequences}  \label{app:effgeometry}
The emergence of the effective metric (\ref{tracemetricalpha}) and its interpretation as 
the perceived classical spacetime indicates that in \sqgr~spacetime is not a physical entity, 
but a property - a sort of order parameter enveloping quantum state of the contents of the 
Universe and its evolution. Such change of nature has various consequences. For example, 
the wave function of subsystems can be extended to classically singular points of the 
effective metric inside a black hole without any concern, because the metric is related to 
the quantum state, and thereby to a probability distribution, which can be singular in a 
measure zero subspace of parameters - {\it coordinates} in the metric 
(\ref{tracemetricalpha}) - without violating unitarity. We leave the investigation of black 
holes in the framework of \sqgr~for future works. In the rest of this section we discuss the 
concepts of locality, causality, light-cone, etc. arise or are defined in this new 
interpretation of spacetime.

\subsubsection{Locality and causality}  \label{sec:causal}
In both classical and quantum models on a background spacetime locality is usually defined 
as decreasing of interactions and correlations with distance. In classical physics such 
behaviour is guaranteed by the limited speed of light. However, due to superposition, 
coherence, and entanglement quantum systems do not follow the strict locality and causality 
definition of classical physics. A definition more suitable for quantum systems is based on 
the tensor product structure (TPS), that is a decomposition similar to (\ref{hilbertsum}) 
such that interactions are bounded to the TPS factors~\cite{localitydef}. This definition is 
dynamical and only depends on the resolution available to observers.

In \sqgr~all subsystems interact through $\suinf$ symmetry - the gravity - and locality 
based on the TPS is only an approximation. Nonetheless, using the relation between the 
effective metric (\ref{tracemetricalpha}) and the variation of quantum state, we can define - 
in analogy with general relativity - a {\it lightcone} in the Hilbert space $\hm_s$ for each 
state $\hrho$. We call states $\hrho' \neq \hrho$ causally related if there is a path 
connecting them such that at each point of the path $ds^2 \geqslant 0$. In analogy with 
general relativity, the necessary and sufficient condition for existence of such path is the 
intersection of lightcones of $\hrho'$ and $\hrho$. This definition is global and 
independent of clock, observers, and resolution of their experiments. According to 
(\ref{metriclindblad}) causally connected states are attainable from each other.
 
%Note: Recalling Einstein's opinion about the nature of spacetime described in the Introduction, he thought that only the affine separation has a proper physical meaning~\cite{etherhistory} (and references therein). However, if the spacetime is not a physical entity - otherwise there had to be the infamous ether - the association of the metric tensor to a geometry seems meaningless. The modern view of metric/gravity as a spin-2 quantum particle apparently solves this puzzle. But, it does not explain why this field determines the geometry of the spacetime, which according to Einstein does not have a physical reality of its own. \sqgr~provides straightforward explanation for these puzzles. 

\subsubsection {Remnant coherence in the state of the Universe}  \label{app:coherdisnting}
%{Relation with coherence and distinguishability}
The Wigner-Yanase skew information $I_{WY}$ defined in (\ref{wyinfo}) satisfies conditions 
for a coherence measure according to~\cite{qminfocohere} and can be used as a measure of 
coherence~\cite{skewinfocohere}. Giving the important role of $I_{WY}$ in the definition of a 
Riemann geometry and QSL for unitary systems, we conclude a relationship between coherence, QSL, 
and effective spacetime in \sqgr. Specifically, we observe that for a completely incoherent 
state $\hrho_{CI} \equiv \sum_i \rho_i |i\rangle \langle i|$, with 
$\sum_i \rho_i = 1, ~ \hH |i\rangle = \lambda_i |i\rangle$, $Q(\hrho_{CI}, \hH) = 0$ and 
equation (\ref{aqdef}) leads to $\Delta t \rightarrow \infty$. This means that the state of the 
system is stable and does not change, unless it is non-unitarily modified by application of an 
operator $\hH'$, such that $[\hH , \hH'] \neq 0$. In \sqgr, this corresponds to a state of the 
Universe where in the energy eigen basis, subsystems are decohered to a simple statistical 
distribution with no quantum superposition. A Universe with static state is very far from 
what we are observing. Therefore, our Universe cannot be completely decohered and it should be 
eventually possible to detect the signature of quantumness in cosmological observations. See 
e.g.~\cite{qmclasscorr,qmdecohere,infcohermode0} for attempts to quantify such a signal.

\section{Candidate processes in \sqgr~for dark energy}  \label{app:sqgrcc}
Any candidate process for dark energy should answer the following questions:
\setcounter{enumi}{0}
\renewcommand{\theenumi}{\alph{enumi}}
\begin{enumerate}
\item Can it leave a remnant that survive the apparent expansion of the Universe ? 
\label{decondb}
\item Is the amplitude of the remnant and its variation across time and space sufficient 
to explain cosmological observations? \label{deconda}
\item Can it solve the coincidence problem ? In other words, can it justify, without 
fine-tuning, the dominance of dark energy well after the formation of galaxies ? 
\label{decondc}
\end{enumerate}
Here for each of the following candidate phenomena we address briefly and qualitatively these 
questions and leave their quantitative assessments to future works.

\subsection {Static $\suinf$ field} \label{sec:staticf} 
Although the Lagrangians (\ref{yminvar}) and (\ref{yminvarsub}) do not explicitly include 
a constant term, in analogy with static electromagnetic, the field strength $F^{\mu\nu}$ can 
include a zero mode generated by all subsystems. In particular, due to the non-commutative 
nature of $\suinf$, Yang-Mills graviton source themselves and in analogy with parton swarms 
in high energy QCD, can condensate. In Einstein gravity gravitons self-interact, but its 
quantized version is non-renormalizable. By contrast, as a quantum Yang-Mills theory \sqgr~is 
renormalizable and its self-interaction of spin-1 gravitons is controlled and does not lead 
to instability. Such condensation is also suggested in other QGR models~\cite{grcondens1} and 
in models for quantum black holes~\cite{grcondens,grcondens0}.

In this framework, in analogy with the QCD glueball, the \sqgr~Universe may be considered as 
a {\it gravity-ocean} in which other fields are immersed. It is certain that in a classical 
view of the spacetime, gravitons are everywhere, and considering their huge number, in a 
quantum view they form a condensate. In dark energy models which propose what they call 
{\it vacuum} of QCD or other Yang-Mills fields as the origin of dark energy, it is the 
condensation of gauge bosons (and/or fermions) that plays the role of dark energy. 
Consequently, IR modes rather than UV ones are able to preserve their quantum coherence and 
energy density of the condensate $\rho_\Lambda$ is expected to be approximately proportional 
to $\rho_\Lambda \sim \mathcal{O}(V^{-4})$ rather than usual QGR (UV) estimation of 
$\rho_\Lambda \sim \mathcal{O}(M_P^{4})$. Thus, the answer to the question (\ref{deconda}) 
is yes. The answers to questions (\ref{decondb}) and (\ref{decondc}) are more subtle and need 
quantification.

\subsection{${\mathbf \suinf ~\Theta}-$vacuum}  \label{sec:vacuum}
The cosmological constant term in the classical Einstein equation has been famously 
interpreted as the vacuum energy~\cite{devacuum,devacuum0}. Non-Abelian Yang-Mills models can 
have non-trivial vacuums, corresponding to a vanishing field strength $F^{\mu\nu} = 0$, but 
non-zero gauge field $A^\mu \equiv A^\mu_a \hL^a$ (for $\suinf$ symmetry). We remind that 
$A^\mu = 0$ is not gauge invariant and changes under local application of gauge symmetry 
according to $A^\mu \rightarrow U A^\mu U^{-1} + i/\lambda (D^\mu U) U^{-1}$, where 
$U(x) \in \bm[\hm_s] \cong \suinf$. Moreover, for large gauge transformations the instanton 
field configurations, that is $A^\mu = i/\lambda (D^\mu U) U^{-1}$ can be topologically 
nontrivial and are classified by their winding number. It is equal to (\ref{thetaact}) with 
$\Theta = 1$. Specifically, as we mentioned earlier, the integrand of (\ref{thetaact}) is a 
total derivative $D_\mu K^\mu$, where the Bardeen current $K^\mu$ is:
\be
K^\mu = \frac{1}{16\pi^2} \epsilon^{\mu\nu\rho\sigma} \tr (A_\nu F_{\rho\sigma} - A_\mu A_\rho A_\sigma) 
\label{bardeencurrent}
\ee
The value of (\ref{thetaact}) with $\Theta = 1$ depends only on the configuration of the 
field $A^\mu$ on the boundary. This means that $\lm_{\Theta = 1}$ has the same value as a vacuum, 
even if $F^{\mu\nu} \neq 0$ except for $|x| \rightarrow \infty$. Different classes of field 
configurations represent homotopy group $\pi_3$ of $SU(2)$, the global rotation symmetry of 
the effective classical 3D space, which persists even after fixing the 
gauge~\cite{instantonrev,instantonrev0}. As $\pi_3 (SU(2) = \mathbb{Z}$, the value of 
$\lm_{\Theta = 1} \in \mathbb{Z}$ and a general vacuum state $|\Theta \rangle$ 
can be written as:
\be
|\Theta \rangle = \sum_n e^{-in \theta_g} |n \rangle  \label{thetavac}
\ee
where $|n\rangle$ are instanton configurations with winding number $n$ and 
$0 \leqslant \theta_g \leqslant \pi$ is a phase angle. If in the early Universe, when QGR 
processes were important, the Yang-Mills graviton field had a $\Theta$-vacuum configuration 
of the form (\ref{thetavac}), local decreasing of its strength $F_{\rho\sigma}$ could not change 
the configuration of zero modes and the field could contribute to the observed dark energy. 
Specifically, in the framework of Schwinger-Dyson Closed Time Path-integral (CTP), the 
1-Particle Irreducible (1PI) path integral of subsystems/fields can be formally written as:
\be
\mathcal{Z} [J_a;\hrho] \equiv e^{i W[J_a]} = \int \mathcal{D} \Phi^a \mathcal{D} 
\Phi^b \exp \biggl [i S(\Phi^a) + i\int d^4x \sqrt {-g} J_a(x) \Phi^a(x) \biggr ] 
\langle \Phi^a | \hrho (| \Phi^b \rangle \label{genpathint}
\ee
where $\Phi$ generically presents all the fields and parameter space (spacetime) and gauge 
symmetry indices are implicit. Indices $a,b \in \{+,-\}$ indicate fields on two opposite 
time branches. They are contracted by the diagonal tensor $c_{ab} \equiv \text{diag} (1, -1)$ 
and satisfy the condition $\langle\Phi^+(x)\rangle = \langle\Phi^-(x)\rangle$. The last 
factor in (\ref{genpathint}) presents the projection of density matrix on the fields eigen 
vectors at initial time $t_0$. Usually $t_0 \rightarrow -\infty$ is used. As in CTP every 
local observable is determined by integration of the generating functional (\ref{genpathint}) 
over the interval $(-\infty, +\infty)$ and back, the past boundary condition affect 
observables at all time. In general, the result has the form of a functional, which without 
loss of generality can be written as:
\be
\langle \Phi^a | \hat{\varrho} | \Phi^b \rangle = \exp (iF[\Phi^a,\Phi^b] \delta(t-t_0)) 
\label{densitymatrix}
\ee 
The functional $iF[\Phi^a,\Phi^b]$ can be added to the action $iS(\Phi^a)$. Therefore, the 
effect of a non-perturbative configuration on the $t_0 \rightarrow -\infty$ boundary, such as 
an instanton state (\ref{thetaact}), appears as a constant term in (\ref{genpathint}). 
In principle it can be separated from other terms and treated as a constant phase. However, 
in \sqgr~according to the Proposition \ref{propentang} other modes, both gravitational and 
$G$ symmetry related, remain entangled to the zero modes. Phenomenologically, this means that 
the effect of $\suinf$ instanton - the gravity $\Theta-$vacuum - on the subsystems can be 
parameterized by an amplitude $\phi_g(x)$, which its effective contribution to the action 
(\ref{genpathint}) would be similar to an axion, but without kinetic and mass terms of a 
true field. It is perceivable that at $|x| \ll \infty$, the effect of instanton settles to 
a roughly constant and behaves similar to a cosmological constant. In contrast to UV 
explanation of cosmological constant, very roughly $|\phi_g|^2 \sim \mathcal{O}((M_P H_0)^2)$ 
is expected. Evidently, these claims need detailed investigation. Meanwhile, in QCD an 
analogous phenomenological model for the $\Theta$-vacuum called {\it topological liquid of 
instanton and anti-instanton} offers a compelling non-perturbative description of the model 
in strong coupling regime, see e.g.~\cite{qcdinstantonliquid} for a review. 

The idea of cosmological and gravitational instantons and their contribution in dark energy 
is also studied both for inflation in de Sitter background~\cite{infinstanton} and in the 
framework of QGR models of spin-2 gravitons~\cite{bfgrvity,desittergrinstanton}. Notably, to 
associate cosmological constant to instanton of a spin-2 field as QGR, significant deviation 
from Einstein gravity is necessary~\cite{bfgrvity}.

\subsection{Manifestation of global entanglement}
In Sec. \ref{sec:vacuum} the Proposition \ref{propentang} about entanglement of subsystems 
was crucial for our arguments in favour of a $\suinf$ topological vacuum as the origin or 
contributor to the observed dark energy. Even in the case of graviton condensation proposed 
in Sec. \ref{sec:staticf} this proposal is indirectly involved, because in absence of a 
global entanglement between subsystems expansion of the Universe may reduce the amplitude 
of the quantum condensate to a negligibly small value, see e.g. simulations of a scalar 
field condensation in an expanding Universe in~\cite{houriqmcond}. Therefore, it would be 
interesting to consider whether in \sqgr~the global entanglement alone can explain the 
roughly constant density of dark energy. 

The contents of the Universe behave in large extent classically. Quantum effects such as 
coherence and entanglement persist locally for a short time, before being decohered through 
interaction with the infinite environment formed by other subsystems/particles. In general 
decoherence increases entropy and releases energy. On the other hand, preservation of global 
entanglement of subsystems in \sqgr~requires creation and/or transfer of entanglement, which in 
turn needs energy~\cite{qmentangleener} (and references therein). Specifically, the mutual 
quantum information provides a lower bound on the energy $E_{entang}$ necessary for 
correlating/entangling two subsystems $A$ and $B$ (see e.g.~\cite{qminfothermo} for a review):
\be
E_{entang} \geqslant k_B T I(A:B)   \label{entangener}
\ee 
where $I(A:B)$ is the mutual information, $k_B$ is the Boltzmann constant, and $T$ the temperature 
of subsystems. In~\cite{hourisqgrym} we calculate the mutual information between gravitational 
and internal degrees of freedom of subsystems of the Universe and show that it is not zero. 
On the other hand, the process of cosmological decoherence and entanglement should arrive to 
an equilibrium, otherwise, either the minimum global entanglement will be lost, which is in 
contradiction with the Proposition \ref{propentang}, or the entanglement grows and reduces 
entropy, which is inconsistent with the second law of thermodynamics. The observed dark energy 
may be the result of this permanent decoherence and recoherence.


\begin{thebibliography}{99}
\bibitem {kaluza} \Name{Th.}{Kaluza}, ~\emph{Zum Unitätsproblem der Physik}, \Journal{\em Sitz. Preuss. Akad. Wiss. Phys. Math.}{K1}{1921}{966}. 
\bibitem {klein} \Name{O.}{Klein}, ~\emph{Quantentheori und f\"unfdimensionale Relativist\"atstheori}, \Journal {\em Zeits Phys.}{37}{1926}{895}.
\bibitem {stringrev} \Name {M.B.}{Green}, \Name {J.H.}{Schwarz}, \Name {E.}{Witten}, ~\emph{Superstring Theory I \& II}, Cambridge University Press, Cambridge, UK, (1987).
\bibitem {stringrev0} \Name {J.}{Polchinski}, ~\emph{String Theory I \& II}, Cambridge University Press, Cambridge, UK, (2005).
\bibitem {qgrmatrixm} \Name{T.}{Banks}, \Name{W.}{Fischler}, \Name{S.H.}{Shenker}, \Name{L.}{Susskind}, ~\emph{M Theory As A Matrix Model: A Conjecture}, \href{https://journals.aps.org/prd/abstract/10.1103/PhysRevD.55.5112}{\Journal {\PRD}{55}{1997}{5112}}, [\href{http://arxiv.org/abs/hep-th/9610043}{arXiv:hep-th/9610043}].
\bibitem {qgrmatrix0} \Name{R.}{Dijkgraaf}, \Name{E.}{Verlinde}, \Name{H.}{Verlinde}, ~\emph{Matrix String Theory}, \href{https://www.sciencedirect.com/science/article/abs/pii/S055032139700326X}{\Journal{\NPB}{500}{1997}{43}}, [\href{http://arxiv.org/abs/hep-th/9703030}{arXiv:hep-th/9703030}]. 
\bibitem {qgrtensornet} \Name{F.}{Piazza}, ~\emph{Glimmers of a pre-geometric perspective}. \href{https://link.springer.com/article/10.1007/s10701-009-9387-5}{\Journal{\FOP}{40}{2010}{239}}, [\href{http://arxiv.org/abs/hep-th/0506124}{arXiv:hep-th/0506124}].
\bibitem {qgrtensornet0} \Name{A.}{Eichhorn}, \Name{T.}{Koslowski}, \Name{J.}{Lumma}, \Name{A.D.}{Pereira}, ~\emph{Towards background independent quantum gravity with tensor models}, \href{https://iopscience.iop.org/article/10.1088/1361-6382/ab2545}{\Journal{\CQG}{36}{2019}{15}}, [\href{http://arxiv.org/abs/1811.00814}{arXiv:1811.00814}].
\bibitem {lqgrev} \Name{C.}{Rovelli}, ~\emph{Quantum Gravity}, Cambridge University Press,  Cambridge, UK, 2004.
\bibitem {lqgrev0} \Name{A.}{Ashtekar}, \Name{J.}{Lewandowski}, ~\emph{Background Independent Quantum Gravity: A Status Report}, \href{https://iopscience.iop.org/article/10.1088/0264-9381/21/15/R01}{\Journal{\CQG}{21}{2004}{R53}}, [\href{http://arxiv.org/abs/gr-qc/0404018}{arXiv:gr-qc/0404018}].
\bibitem {qgrgroup} \Name{D.}{Oriti}, ~\emph{The group field theory approach to quantum gravity}, (2006) [\href{http://arxiv.org/abs/gr-qc/0607032}{arXiv:gr-qc/0607032}].
\bibitem {qgrgroup0} \Name{D.}{Oriti}, ~\emph{The microscopic dynamics of quantum space as a group field theory, in Foundations of Space and Time: Reflections on Quantum Gravity}, edited by G. Ellis, J. Murugan, A. Weltman, published by Cambridge University Press (2012), [\href{http://arxiv.org/abs/1110.5606}{arXiv:1110.5606}]. 
\bibitem {grinconsist} \Name{K.}{Eppley} ; \Name{E.}{Hanna}, ~\emph{The Necessity of Quantizing the Gravitational Field}, \href{https://link.springer.com/article/10.1007/BF00715241}{\Journal{\FOP}{7}{1977}{51}}.
\bibitem {grinconsist0} \Name{D.N.}{Page}, \Name{C.D.}{Geilker}, ~\emph{Indirect Evidence for Quantum Gravity}, \href{https://journals.aps.org/prl/abstract/10.1103/PhysRevLett.47.979}{\Journal {\PRL}{47}{1981}{979}}. 
\bibitem {grinconsist1} \Name{D.R.}{Terno}, ~\emph{Inconsistency of quantum classical dynamics and what it implies}, \href{https://link.springer.com/article/10.1007/s10701-005-9007-y}{\Journal {\FOP}{36}{2006}{102}}, [\href{http://arxiv.org/abs/quant-ph/0402092}{arXiv:/quant-ph/0402092}].
\bibitem {houriqgr} \Name{H.}{Ziaeepour}, ~\emph{And what if gravity is intrinsically quantic ?}, \href{https://iopscience.iop.org/article/10.1088/1742-6596/174/1/012027}{\Journal{\JPC}{174}{2009}{012027}}, [\href{http://arxiv.org/abs/0901.4634}{arXiv:0901.4634}].
\bibitem {grclassic} \Name{J.}{Oppenheim}, ~\emph{A post-quantum theory of classical gravity?}, (2018) [\href{http://arxiv.org/abs/1811.03116}{arXiv:1811.03116}]. 
\bibitem {grclassic0} \Name{J.}{Oppenheim}, \Name{C.}{Sparaciari}, \Name{B.}{Soda}, \Name{Z.}{Weller-Davies}, ~\emph{Gravitationally induced decoherence vs space-time diffusion: testing the quantum nature of gravity}, (2022) [\href{http://arxiv.org/abs/2203.01982}{arXiv:2203.01982}].
\bibitem {statesymmbreak} \Name{W.H.}{Zurek}, ~\emph{Quantum Darwinism, Classical Reality, and the Randomness of Quantum Jumps}, \href{https://physicstoday.scitation.org/doi/10.1063/PT.3.2550}{\Journal {\PTO}{67}{2014}{44}}, [\href{https://arxiv.org/abs/1412.5206}{arXiv:1412.5206}]. 
\bibitem {infcohermode} \Name{D.}{Campo}, \Name{R.}{Parentani}, ~\emph{Space-time correlations in inflationary spectra, a wave packet analysis}, \href{https://journals.aps.org/prd/abstract/10.1103/PhysRevD.70.105020}{\Journal {\PRD}{70}{2004}{105020}}, [\href{http://arxiv.org/abs/gr-qc/0312055}{arXiv:gr-qc/0312055}].
\bibitem {infcohermode0} \Name{C.P.}{Burgess}, \Name{R.}{Holman}, \Name{G.}{Kaplanek}, \Name{J.}{Martin}, \Name{V.}{Vennin}, ~\emph{Minimal decoherence from inflation}, (2022) [\href{http://arxiv.org/abs/2211.11046}{arXiv:2211.11046}].
\bibitem {infdecohere} \Name{C.}{Kiefer}, \Name{D.}{Polarski}, \Name{A.A.}{Starobinsky}, ~\emph{Quantum-to-classical transition for fluctuations in the early Universe}, \href{https://www.worldscientific.com/doi/abs/10.1142/S0218271898000292}{\Journal {\IMD}{7}{1998}{455}}, [\href{https://arxiv.org/abs/gr-qc/9802003}{arXiv:gr-qc/9802003}].
\bibitem {infdecohere0} \Name{E.A.}{Calzetta}, \Name{B.L.}{Hu}, \Name{F.D.}{Mazzitelli}, ~\emph{Coarse-Grained Effective Action and Renormalization Group Theory in Semiclassical Gravity and Cosmology}, \href{https://www.sciencedirect.com/science/article/abs/pii/S0370157301000436}{\Journal {\\PRE}{352}{2001}{459}}, [\href{https://arxiv.org/abs/hep-th/0102199}{arXiv:hep-th/0102199}].
\bibitem {qbitdef} \Name{L.}{Viola}, \Name{E.}{Knill}, \Name{R.}{Laflamme}, ~\emph{Constructing Qubits in Physical Systems}, \href{https://iopscience.iop.org/article/10.1088/0305-4470/34/35/331}{\Journal {\JPA}{Math. Gen. 34}{2001}{7067}}, [\href{https://arxiv.org/abs/quant-ph/0101090}{arXiv:quant-ph/0101090}]. 
\bibitem {sysdiv} \Name{P.}{Zanardi}, \Name{D.}{Lidar}, \Name{S.}{Lloyd}, ~\emph{Quantum tensor product structures are observable-induced}, \href{https://journals.aps.org/prl/abstract/10.1103/PhysRevLett.92.060402}{\Journal{\PRL}{92}{2004}{060402}}, [\href{http://arxiv.org/abs/quant-ph/0308043}{arXiv:quant-ph/0308043}].
\bibitem {houriqmsymm} \Name{H.}{Ziaeepour}, ~\emph{Symmetry as a foundational concept in Quantum Mechanics}, \href{https://iopscience.iop.org/article/10.1088/1742-6596/626/1/012074/meta}{\Journal{\JPC}{626}{2015}{012074}}, [\href{http://arxiv.org/abs/1502.05339}{arXiv:1502.05339}].
\bibitem {qmref} \Name{S.D.}{Bartlett}, \Name{T.}{Rudolph}, \Name{R.W.}{Spekkens}, ~\emph{Reference frames, superselection, and quantum information}, \href{https://journals.aps.org/rmp/abstract/10.1103/RevModPhys.79.555}{\Journal{\RMP}{79}{2007}{555}}, [\href{https://arxiv.org/abs/quant-ph/0610030}{arXiv:quant-ph/0610030}].
\bibitem {qmrefchangedecoher} \Name{M.C.}{Palmer}, \Name{F.}{Girelli}, \Name{S.D.}{Bartlett}, ~\emph{Changing quantum reference frames}, \href{https://journals.aps.org/pra/abstract/10.1103/PhysRevA.89.052121}{\Journal {\PRA}{89}{2014}{052121}}, [\href{http://arxiv.org/abs/1307.6597}{arXiv:1307.6597}].
\bibitem {qmrefdecohere} \Name{F.}{Giacomini}, \Name{E.}{Castro-Ruiz}, \Name{C.}{Brukner}, ~\emph{Quantum mechanics and the covariance of physical laws in quantum reference frames}, \href{https://www.nature.com/articles/s41467-018-08155-0}{\Journal {\NAC}{10}{2019}{494}}, [\href{http://arxiv.org/abs/1712.07207}{arXiv:1712.07207}]. 
\bibitem {qmrefsubsys} \Name{S.A.}{Ahmad}, \Name{T.D.}{Galley}, \Name{P.A.}{Hoehn}, \Name{M.P.E.}{Lock}, \Name{A.R.H.}{Smith}, ~\emph{Quantum Relativity of Subsystems}, \href{https://journals.aps.org/prl/abstract/10.1103/PhysRevLett.128.170401}{\Journal {\PRL}{128}{2022}{170401}}, [\href{https://arxiv.org/abs/2103.01232}{arXiv:2103.01232}].
\bibitem {qmtimepage} \Name{D.N.}{Page}, \Name{W.K.}{Wootters}, ~\emph{Evolution without evolution: Dynamics described by stationary observables}, \href{https://journals.aps.org/prd/abstract/10.1103/PhysRevD.27.2885}{\Journal{\PRD}{27}{1983}{2885}}.
\bibitem {qmtimedef} \Name{P.A.}{Hoehn}, \Name{A.R.H.}{Smith}, \Name{M.P.E.}{Lock}, ~\emph{The Trinity of Relational Quantum Dynamics}, (2019), [\href{http://arxiv.org/abs/1912.00033}{arXiv:1912.00033}].
\bibitem {qgrhistory} \Name{J.B.}{Hartle}, ~\emph{Generalizing Quantum Mechanics for Quantum Spacetime}, In ~\emph{The Quantum Structure of Space and Time},  Gross, D., Henneaux, M., Sevrin, A., Eds., World Scientific: Singapore, (2007) [\href{https://arxiv.org/abs/gr-qc/0602013}{arXiv:gr-qc/0602013}].
\bibitem {qgrlocalqm} \Name{S.B.}{Giddings}, ~\emph{Hilbert space structure in quantum gravity: An algebraic perspective}, \href{https://link.springer.com/article/10.1007/JHEP12(2015)099}{\Journal {\JHE}{2015}{2015}{1}}, [\href{https://arxiv.org/abs/1503.08207}{arXiv:1503.08207}].
\bibitem {hologprin} \Name{J.D.}{Bekenstein}, ~\emph{Statistical black-hole thermodynamics}, \href{https://journals.aps.org/prd/abstract/10.1103/PhysRevD.12.3077}{\Journal {\PRD}{12}{1975}{3077}}.
\bibitem {hologprin0} \Name{G.}{t'Hooft}, ~\emph{Dimensional Reduction in Quantum Gravity}, (1993), [\href{https://arxiv.org/abs/gr-qc/9310026}{arXiv:gr-qc/9310026}].
\bibitem {hologprin1} \Name{L.}{Susskin}, ~\emph{The World as a Hologram}, \href{https://aip.scitation.org/doi/10.1063/1.531249}{\Journal{\JMP}{36}{1995}{6377}}, [\href{https://arxiv.org/abs/hep-th/9409089}{arXiv:hep-th/9409089}].
\bibitem {adscftrev} \Name{J.}{Maldacena}, ~\emph{The gauge/gravity duality}, in ~\emph{Black Holes in Higher Dimensions}, G. Horowitz, Ed., Cambridge University Press: Cambridge, UK, (2012), [\href{https://arxiv.org/abs/1106.6073}{arXiv:1106.6073}].
\bibitem {hourisqgrcomp} \Name{H.}{Ziaeepour}, ~\emph{Comparing Quantum Gravity Models: Loop Quantum Gravity, Entanglement and AdS/CFT versus $SU(\infty)$-QGR}, \href{http://dx.doi.org/10.3390/sym14010058}\href{https://www.mdpi.com/2073-8994/14/1/58}{\Journal{\SYM}{14}{2022}{58}}, [\href{https://arxiv.org/abs/2109.05757}{arXiv:2109.05757}].
\bibitem {houriqmsymmgr} \Name{H.}{Ziaeepour}, ~\emph{Making a Quantum Universe: Symmetry and Gravity}, \href{https://www.mdpi.com/2218-1997/6/11/194}{\Journal{\MDU}{6(11)}{2020}{194}}, [\href{https://arxiv.org/abs/2009.03428}{arXiv:2009.03428}].
\bibitem {etherhistory} \Name{D.}{Meschini}, \Name{M.}{Lehto}, ~\emph{Is empty spacetime a physical thing?}, \href{https://link.springer.com/article/10.1007/s10701-006-9058-8}{\Journal {\FOP}{36}{2006}{1193}}, [\href{http://arxiv.org/abs/gr-qc/0506068}{arXiv:gr-qc/0506068}].
\bibitem {qgrentangle} \Name{M.}{Van Raamsdonk}, ~\emph{Building up spacetime with quantum entanglement}, \href{https://link.springer.com/article/10.1007/s10714-010-1034-0}{\Journal{\GRG}{42}{2010}{2323}}, ~\emph{\href{https://www.worldscientific.com/doi/abs/10.1142/S0218271810018529}{\Journal {\IMD}{19}{2010}{2429}}, [\href{http://arxiv.org/abs/1005.3035}{arXiv:1005.3035}]. 
\bibitem {qgrentangle1} \Name{C.}{Cao}, \Name{S.M.}{Carroll}, \Name{S.}{Michalakis}, ~\emph{Space from Hilbert Space: Recovering Geometry from Bulk Entanglement}, \href{https://journals.aps.org/prd/abstract/10.1103/PhysRevD.95.024031}{\Journal{\PRD}{95}{2017}{024031}}, [\href{http://arxiv.org/abs/1606.08444}{arXiv:1606.08444}].
\bibitem {qgrentangle2} \Name{C.}{Cao}, \Name{S.M.}{Carroll} , Bulk Entanglement Gravity without a Boundary: Towards Finding Einstein's Equation in Hilbert Space}, \href{https://journals.aps.org/prd/abstract/10.1103/PhysRevD.97.086003}{\Journal{\PRD}{97}{2018}{086003}}, [\href{https://arxiv.org/abs/1712.02803}{arXiv:1712.02803}].
\bibitem {gwmemory} \Name{Ya.B.}{Zel’dovich}, \Name{A.G.}{Polnarev}, ~\emph{Radiation of gravitational waves by a cluster of superdense stars}, {\em Astron. Zh. 51, (1974) 30 [Sov. Astron. 18 17(1974)]}.
\bibitem {gwmemory0} \Name{V.B.}{Braginsky}, \Name{L.P.}{Grishchuk}, ~\emph{Kinematic resonance and the memory effect in free mass gravitational antennas}, {\em Zh. Eksp. Teor. Fiz. 89 744-750 (1985) [Sov.Phys. JETP 62, 427 (1985)]}.
\bibitem {gwmemory1} \Name{D.}{Christodoulou}, ~\emph{Nonlinear nature of gravitation and gravitational-wave experiments}, \href{https://journals.aps.org/prl/abstract/10.1103/PhysRevLett.67.1486}{\Journal {\PRL}{67}{1991}{1486}}.
\bibitem {gwmemory2} \Name{M.}{Favata}, ~\emph{The gravitational-wave memory effect}, \href{https://iopscience.iop.org/article/10.1088/0264-9381/27/8/084036}{\Journal {\CQG}{27}{2010}{084036}}, [\href{http://arxiv.org/abs/1003.3486}{arXiv:1003.3486}].
\bibitem {hourisqgrym} \Name{H.}{Ziaeepour}, ~\emph{Quantum state of fields in $\mathbf{\suinf}$ Quantum Gravity}, [\href{https://arxiv.org/abs/2402.18237}{arXiv:2402.18237}].
\bibitem {suninfhoppthesis} \Name{J.}{Hoppe}, ~\emph{Quantum Theory of a Massless Relativistic Surface and a Two-dimensional Bound State Problem}, \href{https://dspace.mit.edu/handle/1721.1/15717}{Ph.D. Thesis, MIT}, Cambridge, MA, USA, (1982).
\bibitem {suninfym} \Name{E.G.}{Floratos}, \Name{J.}{Iliopoulos}, \Name{G.}{Tiktopoulos}, ~\emph{A note on $SU(\infty)$ classical Yang-Mills theories}, \href{https://www.sciencedirect.com/science/article/abs/pii/0370269389908678}{\Journal{\PLB}{217}{1989}{285}}. 
\bibitem {suninftorus} \Name{J.}{Hoppe}, ~\emph{Diffeomorphism Groups, Quantization, and $SU(\infty)$}, \href{https://www.worldscientific.com/doi/abs/10.1142/S0217751X89002235}{\Journal {\IMA}{4}{1989}{5235}}.
\bibitem {suninfrep} \Name{J.}{Hoppe}, \Name{P.}{Schaller}, ~\emph{Infinitely Many Versions of $SU(\infty)$}, \href{https://www.sciencedirect.com/science/article/abs/pii/037026939091197J}{\Journal {\PLB}{237}{1990}{407}}.
\bibitem {suninfrep0} \Name{Y.}{Zunger}, ~\emph{Why Matrix theory works for oddly shaped membranes}, \href{https://journals.aps.org/prd/abstract/10.1103/PhysRevD.64.086003}{\Journal{\PRD}{64}{2001}{086003}}, [\href{http://arxiv.org/abs/hep-th/0106030}{arXiv:hep-th/0106030}].
\bibitem {cartandecomp} \Name{Z-Y.}{Su}, ~\emph{A Scheme of Cartan Decomposition for su(N)}, (2006), [\href{http://arxiv.org/abs/quant-ph/0603190}{arXiv:quant-ph/0603190}].
\bibitem {qmmathbook} \Name{B.C.}{Hall}, ~\emph{Quantum Theory for Mathematicians}, Springer, (2013). % IHP code 513 89 a
\bibitem {weylquant} \Name{H.}{Weyl}, ~\emph{Quantenmechanik und Gruppentheorie}, \Journal {\em Zeitschrift f\"ur Physik}{46}{1927}{1}, reprint: ~\emph{The Theory of Groups and Quantum Mechanics, Dover Publications}, (1950).
\bibitem {weylquantiz} \Name{S.}{Chaturvedi}, \Name{E.}{Ercolessi}, \Name{G.}{Marmo}, \Name{G.}{Morandi}, \Name{N.}{Mukunda}, \Name{R.}{Simon}, ~\emph{Wigner–Weyl correspondence in quantum mechanics for continuous and discrete systems—a Dirac-inspired view}, \href{https://iopscience.iop.org/article/10.1088/0305-4470/39/6/014}{\Journal {\JPA}{Math. Gen. 39}{2006}{1405}}.
\bibitem {noncummut} \Name{H.S.}{Snyder}, ~\emph{Quantized Space-Time}, \Journal{\PRV}{71}{1947}{38}.
\bibitem {noncummut0} \Name{C.N.}{Yang}, ~\emph{On quantized space-time}, \Journal{\PRV}{72}{1947}{874}.
\bibitem {qgrnoncommut} \Name{A.}{Connes}, ~\emph{Gravity coupled with matter and foundation of non-commutative geometry}, \href{https://link.springer.com/article/10.1007/BF02506388}{\Journal {\CMP}{182}{1996}{155}}, [\href{https://arxiv.org/abs/hep-th/9603053}{arXiv:hep-th/9603053}].
\bibitem {qgrmatrixnoncommut} \Name{A.}{Connes}, \Name{M.R.}{Douglas}, \Name{A.}{Schwarz}, ~\emph{Noncommutative Geometry and Matrix Theory: Compactification on Tori}, \href{https://iopscience.iop.org/article/10.1088/1126-6708/1998/02/003}{\Journal{\JHE}{02}{1998}{003}}, [\href{http://arxiv.org/abs/hep-th/9711162}{arXiv:hep-th/9711162}].
\bibitem {noncommutstring} \Name{N.}{Seiberg}, \Name{E.}{Witten}, ~\emph{String Theory and Noncommutative Geometry}, \href{https://iopscience.iop.org/article/10.1088/1126-6708/1999/09/032}{\Journal{\JHE}{9909}{1999}{032}}, [\href{http://arxiv.org/abs/hep-th/9908142}{arXiv:hep-th/9908142}].
\bibitem {qgrgaugedual} \Name{T.}{Banks}, \Name{W.}{Fischler}, \Name{S.H.}{Shenker}, \Name{L.}{Susskind}, ~\emph{M Theory As A Matrix Model: A Conjecture}, \href{https://journals.aps.org/prd/abstract/10.1103/PhysRevD.55.5112}{\Journal{\PRD}{55}{1997}{5112}}, [\href{https://arxiv.org/abs/hep-th/9610043}{arXiv:hep-th/9610043}].
\bibitem {qgrmatrix} \Name{N.}{Ishibashi}, \Name{H.}{Kawai}, \Name{Y.}{Kitazawa}, \Name{A.}{Tsuchiya}, ~\emph{A Large-N Reduced Model as Superstring}, \href{https://www.sciencedirect.com/science/article/abs/pii/S0550321397002903}{\Journal{\NPB}{498}{1997}{467}}, [\href{http://arxiv.org/abs/hep-th/9612115}{arXiv:hep-th/9612115}].
\bibitem {qndecohersubsys} \Name{D.A.}{Lidar}, ~\emph{Review of Decoherence Free Subspaces, Noiseless Subsystems, and Dynamical Decoupling}, \href{https://onlinelibrary.wiley.com/doi/book/10.1002/9781118742631}{\Journal {\em Adv. Chem. Phys.}{154}{2014}{295}}, [\href{http://arxiv.org/abs/1208.5791}{arXiv:1208.5791}].
\bibitem {qmframeperspect} \Name{A.}{Vanrietvelde}, \Name{Ph.A.}{Hoehn}, \Name{F.}{Giacomini}, \Name{E.}{Castro-Ruiz}, ~\emph{A change of perspective: switching quantum reference frames via a perspective-neutral framework}, \href{https://quantum-journal.org/papers/q-2020-01-27-225/}{\Journal {\QNM}{4}{2020}{225}}, [\href{http://arxiv.org/abs/1809.00556}{arXiv:1809.00556}]. 
\bibitem {qmframeperspect0} \Name{A.}{Vanrietvelde}, \Name{Ph.A.}{Hoehn}, \Name{F.}{Giacomini}, ~\emph{Switching quantum reference frames in the N-body problem and the absence of global relational perspectives}, (2018), [\href{http://arxiv.org/abs/1809.05093}{arXiv:1809.05093}.
\bibitem {qmclock} \Name{V.}{Giovannetti}, \Name{S.}{Lloyd}, \Name{L.}{Maccone}, \Name{S.M.}{Shahriar}, ~\emph{Limits to clock synchronization induced by completely dephasing communication channels}, \href{https://journals.aps.org/pra/abstract/10.1103/PhysRevA.65.062319}{\Journal {\PRA}{65}{2002}{062319}}, [\href{http://arxiv.org/abs/quant-ph/0110156}{arXiv:quant-ph/0110156}].
\bibitem {qmclock0} \Name{A.}{Frenkel}, ~\emph{Dependence of the time-reading process of the Salecker--Wigner quantum clock on the size of the clock}, \href{https://link.springer.com/article/10.1007/s10701-015-9938-x}{\Journal {\FOP}{45}{2015}{1561}}, [\href{http://arxiv.org/abs/1501.00840}{arXiv:1501.00840}].
\bibitem {qmclockevent} \Name{E.O.}{Dias}, ~\emph{A quantum formalism for events and how time can emerge from its foundations}, \href{https://journals.aps.org/pra/abstract/10.1103/PhysRevA.103.012219}{\Journal {\PRA}{103}{2021}{012219}}, [\href{http://arxiv.org/abs/2007.00513}{2007.00513}].
\bibitem {admgr} \Name{R.}{Arnowitt}, \Name{S.}{Deser}, \Name{C.}{Misner}, ~\emph{Dynamical Structure and Definition of Energy in General Relativity}, \href{https://journals.aps.org/pr/abstract/10.1103/PhysRev.116.1322}{\Journal{\PRV}{116}{1959}{1322}}, [\href{https://arxiv.org/abs/gr-qc/0405109}{arXiv:gr-qc/0405109}] (reprint).
\bibitem {qgrether} \Name{C.H.}{Brans}, ~\emph{Absolute spacetime: the twentieth century ether}, \href{https://link.springer.com/article/10.1023/A:1026632709502}{\Journal {\GRG}{31}{1999}{597}}, [\href{http://arxiv.org/abs/gr-qc/9801029}{arXiv:gr-qc/9801029}].
\bibitem {compacttime} \Name{J.}{de Haro}, \Name{S.}{Nojiri}, \Name{S.D.}{Odintsov}, \Name{V.K.}{Oikonomou}, \Name{S.}{Pan}, ~\emph{Finite-time Cosmological Singularities and the Possible Fate of the Universe}, \href{https://www.sciencedirect.com/science/article/abs/pii/S0370157323002910?via\%3Dihub}{\Journal{\PRE}{1034}{2023}{1}}, [\href{https://arxiv.org/abs/2309.07465}{arXiv:2309.07465}].
\bibitem {comosclosed} \Name{J.D.}{Barrow}, \Name{J.}{Levin}, ~\emph{The Copernican Principle in Compact Spacetimes}, \href{https://academic.oup.com/mnras/article/346/2/615/1476046?login=false}{\Journal{\MRA}{346}{2003}{615}}, [\href{https://arxiv.org/abs/gr-qc/0304038}{gr-qc/0304038}]. 
\bibitem {comosclosedcmb} \Name{A.N.}{Lasenby}, \Name{C.}{Doran}, ~\emph{Conformal models of de Sitter space, initial conditions for inflation and the CMB}, \href{https://pubs.aip.org/aip/acp/article-abstract/736/1/53/675176/Conformal-Models-of-de-Sitter-Space-Initial?redirectedFrom=fulltext}{\Journal{\AIP}{736}{2004}{53}}, [\href{https://arxiv.org/abs/astro-ph/0411579}{astro-ph/0411579}].
\bibitem {comosclosedcmb0} \Name{J.}{Alberto V\'azquez}, \Name{A.N.}{Lasenby}, \Name{M.}{Bridges}, \Name{M.P.}{Hobson}, ~\emph{A Bayesian study of the primordial power spectrum from a novel closed universe model}, \href{https://academic.oup.com/mnras/article/422/3/1948/1041187?login=false}{\Journal{\MRA}{442}{2012}{1948}}, [\href{https://arxiv.org/abs/1103.4619}{arXiv:1103.4619}].
\bibitem {lssclose} \Name{M.J.}{Mortonson}, ~\emph{Testing flatness of the universe with probes of cosmic distances and growth}, \href{https://journals.aps.org/prd/abstract/10.1103/PhysRevD.80.123504}{\Journal{\PRD}{80}{2009}{123504}}, [\href{https://arxiv.org/abs/0908.0346}{arXiv:0908.0346}]. 
\bibitem {lssclose0} \Name{W.}{Yang}, \Name{W.}{Giar\`e}, \Name{S.}{Pan}, \Name{E.}{Di Valentino}, \Name{A.}{Melchiorri}, \Name{J.}{Silk}, ~\emph{Revealing the effects of curvature on the cosmological models}, \href{https://journals.aps.org/prd/abstract/10.1103/PhysRevD.107.063509}{\Journal{\PRD}{107}{2023}{063509}}, [\href{https://arxiv.org/abs/2210.09865}{arXiv:2210.09865}].
\bibitem {qmmbl} \Name{P.W.}{Anderson}, ~\emph{Absence of Diffusion in Certain Random Lattices}, \href{https://journals.aps.org/pr/abstract/10.1103/PhysRev.109.1492}{\Journal{\PRV}{109}{1958}{1492}}.
\bibitem {inttablebook} \Name{I.S.}{Gradshteyn}, \Name{I.M.}{Ryzhik}, ~\emph{Table of Integral, Series and Products}, $7^{th}$ Edition, Ed. \Name{A.}{Jeffery} and \Name{D.}{Zwillinger}, Elsevier, (2007).
\bibitem{qmheisenuncert} \Name{W.}{Heisenberg} -\emph{\"Uber den anschaulichen Inhalt der quantentheoretischen Kinematik und Mechanik}, \Journal{\em Zeitschrift f\"ur Physik}{43}{1927}{172}. 
\bibitem {qmspeed} \Name{L.}{Mandelstam}, \Name{I.}{Tamm}, ~\emph{The Uncertainty Relation Between Energy and Time in Non-relativistic Quantum Mechanics}, \href{https://link.springer.com/chapter/10.1007/978-3-642-74626-0_8}{\Journal{\JPU}{9}{1945}{249}}.
\bibitem {smlorentz} \Name{B.}{Audren}, \Name{D.}{Blas}, \Name{M.M.}{Ivanov}, \Name{J.}{Lesgourgues}, \Name{S.}{Sibiryakov}, ~\emph{Cosmological constraints on deviations from Lorentz invariance in gravity and dark matter}, \href{https://iopscience.iop.org/article/10.1088/1475-7516/2015/03/016}{\Journal {\JCA}{1503}{2015}{016}}, [\href{http://arxiv.org/abs/1410.6514}{arXiv:1410.6514}].
\bibitem {grlorentz} The LIGO Scientific Collaboration, the Virgo Collaboration, the KAGRA Collaboration, ~\emph{Tests of General Relativity with GWTC-3}, \href{https://journals.aps.org/prd/accepted/17075Qf4Z7b11729787e85f1c18faca230d51e013}{\Journal {\PRD}{}{2022}{Accepted}} [\href{http://arxiv.org/abs/2112.06861}{arXiv:2112.06861}].
\bibitem {leptobaryo} \Name{W.}{Buchmuller}, \Name{R.D.}{Peccei}, \Name{T.}{Yanagida}, ~\emph{Leptogenesis as the origin of matter}, \href{https://www.annualreviews.org/doi/10.1146/annurev.nucl.55.090704.151558}{\Journal {\em Ann.Rev.Nucl.Part.Sci.}{55}{2005}{311}}, [\href{http://arxiv.org/abs/hep-ph/0502169}{arXiv:hep-ph/0502169}].
\bibitem {curvatureinvar} \Name{Sh-T.}{Yau}, ~\emph{Curvature Preserving Diffeomorphisms}, \href{https://www.jstor.org/stable/1970843?origin=crossref}{\Journal {\ANM}{100}{1974}{121}}.
\bibitem {qgrgaugesep} \Name{F.}{Wilczek}, ~\emph{Riemann-Einstein Structure from Volume and Gauge Symmetry}, \href{https://journals.aps.org/prl/abstract/10.1103/PhysRevLett.80.4851}{\Journal {\PRL}{80}{1998}{4851}}, [\href{http://arxiv.org/abs/hep-th/9801184}{arXiv:hep-th/9801184}].
\bibitem {qgrgaugesep0} \Name{A.}{Torres-Gomez}, \Name{K.}{Krasnov}, ~\emph{Gravity-Yang-Mills-Higgs unification by enlarging the gauge group}, \href{https://journals.aps.org/prd/abstract/10.1103/PhysRevD.81.085003}{\Journal{\PRD}{81}{2010}{085003}}, [\href{http://arxiv.org/abs/0911.3793}{arXiv:0911.3793}]. %and references therein gravity from extension of gauge group on a Minkowski background
\bibitem {qgrgaugesep1} \Name{J.W.}{Barrett}, \Name{S.}{Kerr}, ~\emph{Gauge gravity and discrete quantum models}, (2013), [\href{http://arxiv.org/abs/1309.1660}{arXiv:1309.1660}]. %gauge gravity QM and references 2 papers from Sciama and Kibble in ref.
\bibitem {gaugestringcorr} \Name {S.S.}{Gubser}, \Name {I.R.}{Klebanov}, \Name {A.M.}{Polyakov}, ~\emph{Gauge Theory Correlators from Non-Critical String Theory}, \href{https://www.sciencedirect.com/science/article/abs/pii/S0370269398003773}{\Journal{\PLB}{428}{1998}{105}}, [\href{https://arxiv.org/abs/hep-th/9802109}{arXiv:hep-th/9802109}].
\bibitem {stringgauge0} \Name{E.}{Witten}, ~\emph{Anti De Sitter Space And Holography}, \href{https://www.intlpress.com/site/pub/pages/journals/items/atmp/content/vols/0002/0002/a002/}{\Journal{\ATM}{2}{1998}{253}}, [\href{https://arxiv.org/abs/hep-th/9802150}{arXiv:hep-th/9802150}].
\bibitem {stringgauge1} \Name{O.}{Aharony}, \Name{S.S.}{Gubser}, \Name{J.}{Maldacena}, \Name{H.}{Ooguri}, \Name{Y.}{Oz}, ~\emph{Large N Field Theories, String Theory and Gravity}, \href{https://www.sciencedirect.com/science/article/abs/pii/S0370157399000836}{\Journal{\PRE}{323}{2000}{183}}, [\href{https://arxiv.org/abs/hep-th/9905111}{arXiv:hep-th/9905111}].
\bibitem {bhfeline} \Name{A.C.}{Fabian}, \Name{K.}{Iwasawa}, \Name{C.S.}{Reynolds}, \Name{A.J.}{Young}, ~\emph{Broad iron lines in Active Galactic Nuclei}, \href{https://iopscience.iop.org/article/10.1086/316610}{\Journal {\PAP}{112}{2000}{1145}}, [\href{http://arxiv.org/abs/astro-ph/0004366}{arXiv:astro-ph/0004366}].
\bibitem {bhmirage} \Name{D.}{Psaltis}, \Name{L.}{Medeiros}, \Name{P.}{Christian}, \Name{F.}{Ozel}, \Name{K.}{Akiyama}, \Name{A.}{Alberdi}, \Name{W.}{Alef}, \Name{K.}{Asada}, ~\emph{\etal, Gravitational Test Beyond the First Post-Newtonian Order with the Shadow of the M87 Black Hole}, \href{https://journals.aps.org/prl/abstract/10.1103/PhysRevLett.125.141104}{\Journal {\PRL}{125}{2020}{141104}}, [\href{http://arxiv.org/abs/2010.01055}{arXiv:2010.01055}].
\bibitem {qftmetricsignature} \Name{S.}{Mukohyama}, \Name{J-Ph.}{Uzan}, ~\emph{From configuration to dynamics -- Emergence of Lorentz signature in classical field theory}, \href{https://journals.aps.org/prd/abstract/10.1103/PhysRevD.87.065020}{\Journal {\PRD}{87}{2013}{065020}}, [\href{http://arxiv.org/abs/1301.1361}{arXiv:1301.1361}].
\bibitem {qmspeedgeom} \Name {J.}{Anandan}, \Name{Y.}{Aharonov}, ~\emph{Geometry of quantum evolution}, \href{https://journals.aps.org/prl/abstract/10.1103/PhysRevLett.65.1697}{\Journal{\PRL}{65}{1990}{1697}}.
\bibitem {qmspeedgeomopen} \Name{A}{Uhlmann}, -\emph{An energy dispersion estimate}, \href{https://www.sciencedirect.com/science/article/abs/pii/037596019290555Z}{\Journal{\PLA}{161}{1992}{329}}.  %#### Bures metric
\bibitem {qmspeedgeomgen0}\Name{D.}{Paiva Pires}, \Name{M.}{Cianciaruso}, \Name{L.C.}{Céleri}, \Name{G.}{Adesso}, \Name{D.O.}{Soares-Pinto}, ~\emph{Generalized Geometric Quantum Speed Limits}, \href{https://journals.aps.org/prx/abstract/10.1103/PhysRevX.6.021031}{\Journal {\PRX}{6}{2016}{021031}}, [\href{http://arxiv.org/abs/1507.05848}{arXiv:1507.05848}].
\bibitem {qmspeedgeomlocal} \Name{E.}{O'Connor}, \Name{G.}{Guarnieri}, \Name{S.}{Campbell}, ~\emph{Action quantum speed limits}, \href{https://journals.aps.org/pra/abstract/10.1103/PhysRevA.103.022210}{\Journal {\PRA}{103}{2021}{022210}}, [\href{http://arxiv.org/abs/2011.05232}{arXiv:2011.05232}]. % QSL for 3 different metric and distance, path dependence for mixed states, comp with non-attainable geodesic.
\bibitem {devacuum0} \Name{S.}{Weinberg}, ~\emph{Anthropic Bound on the Cosmological Constant}, \href{https://journals.aps.org/prl/abstract/10.1103/PhysRevLett.59.2607}{\Journal{PRL}{59}{1987}{2607}}.
\bibitem {devacuum1} \Name{C.P.}{Burgess}, ~\emph{The Cosmological Constant Problem: Why it's hard to get Dark Energy from Micro-physics}, (2013),[\href{http://arxiv.org/abs/1309.4133}{arXiv:1309.4133}].
\bibitem {cosmobook} \Name{S.}{Weinberg}, ~\emph{Gravity and Cosmology}, John Wiley \& Sons, Inc. (1972).
\bibitem {spinor2d0} \Name{C.}{Koke}, \Name{C.}{Noh}, \Name{D.G.}{Angelakis}, ~\emph{Dirac equation in 2-dimensional curved spacetime, particle creation, and coupled waveguide arrays}, \href{https://www.sciencedirect.com/science/article/abs/pii/S0003491616301609}{\Journal {\APH}{375}{2016}{162}}, [\href{http://arxiv.org/abs/1607.04821}{arXiv:1607.04821}].
\bibitem {spinor2d} \Name{P.}{Yip}, ~\emph{Spinors in two dimensions}, \href{https://aip.scitation.org/doi/10.1063/1.525798}{\Journal {\JMP}{24}{1983}{1206}}.
\bibitem {grymselfint} \Name{S.}{Deser}, ~\emph{Gravity from self-interaction in a curved background}, \href{https://iopscience.iop.org/article/10.1088/0264-9381/4/4/006}{\Journal{\CQG}{4}{1987}{L99}}.
\bibitem {qftbook} \Name{S.}{Weinberg}, ~\emph{The Quantum Theory of Fields II}, Cambridge University Press, (1996).
\bibitem {qcdaxionrev} \Name{R.D.}{Peccei}, ~\emph{The Strong CP Problem and Axions}, \href{https://link.springer.com/chapter/10.1007/978-3-540-73518-2_1}{\Journal {\LNP}{741}{2008}{3}}, [\href{http://arxiv.org/abs/hep-ph/0607268}{arXiv:hep-ph/0607268}].
\bibitem {instantonrev} \Name{A.I.}{Vainshtein}, \Name{V.I}{Zakharov}, \Name{V.A.}{Novikov} \Name{M.A.}{Shifman}, ~\emph{ABC of Instanton}, \href{https://iopscience.iop.org/article/10.1070/PU1982v025n04ABEH004533/meta?casa_token=G_pzfYtI5JIAAAAA:HUD3j33Kth9z_WEkjSH4esC1CX9QeWpjXMWvhhncgHKGLVWmx_9CMzoAu92dbdQPZ7y970MS3FA}{\Journal{\em Sov.Phys.Usp}{25}{1982}{195}}.
\bibitem {instantonrev0} \Name{S.}{Vandoren}, \Name{P.}{van Nieuwenhuizen}, ~\emph{Lectures on instantons}, (2008), [\href{http://arxiv.org/abs/0802.1862}{arXiv:0802.1862}].
\bibitem {leptosphaleron} \Name{V.A.}{Rubakov}, \Name{M.E.}{Shaposhnikov}, ~\emph{Electroweak Baryon Number Non-Conservation in the Early Universe and in High Energy Collisions}, \href{https://iopscience.iop.org/issue/1063-7869/39/5}{\Journal{\em Phys.Usp.}{39}{1996}{461}}, [\href{http://arxiv.org/abs/hep-ph/9603208}{arXiv:hep-ph/9603208}].
\bibitem {leptosphaleron0} \Name{D.}{Bodeker}, \Name{W.}{Buchmuller}, ~\emph{Baryogenesis from the weak scale to the grand unification scale}, \href{https://journals.aps.org/rmp/abstract/10.1103/RevModPhys.93.035004}{\Journal{\RMP}{93}{2021}{035004}}, [\href{http://arxiv.org/abs/2009.07294}{arXiv:2009.07294}].
\bibitem {grparityviol} \Name{N.}{Yunes}, \Name{R.}{O'Shaughnessy}, \Name{B.J.}{Owen}, \Name{S.}{Alexander}, ~\emph{Testing gravitational parity violation with coincident gravitational waves and short gamma-ray bursts}, \href{https://journals.aps.org/prd/abstract/10.1103/PhysRevD.82.064017}{\Journal {\PRD}{82}{2010}{064017}}, [\href{http://arxiv.org/abs/1005.3310}{arXiv:1005.3310}].
\bibitem {grparityviol0} \Name{K.W.}{Masui}, \Name{Ue-Li}{Pen}, \Name{N.}{Turok}, ~\emph{Two- and Three-Dimensional Probes of Parity in Primordial Gravity Waves}, \href{https://journals.aps.org/prl/abstract/10.1103/PhysRevLett.118.221301}{\Journal {\PRL}{118}{2017}{221301}}, [\href{http://arxiv.org/abs/1702.06552}{arXiv:1702.06552}].
\bibitem {grparityviol1} \Name{S.H.}{Alexander}, \Name{N.}{Yunes}, ~\emph{Gravitational Waves Probes of Parity Violation in Compact Binary Coalescence}, \href{https://journals.aps.org/prd/abstract/10.1103/PhysRevD.97.064033}{\Journal {\PRD}{97}{2018}{064033}}, [\href{http://arxiv.org/abs/1712.01853}{arXiv:1712.01853}].
\bibitem {curvaturfunc} \Name{A.L.}{Besse}, ~\emph{Einstein manifolds}, in ~\emph{Results in Mathematics and Related Areas (3)}, Springer-Verlag, Berlin, (1987).
\bibitem {ymgwsimul} \Name{G.}{Travaglini}, \Name{A.}{Brandhuber}, \Name{P.}{Dorey}, \Name{T.}{McLoughlin}, \Name{S.}{Abreu}, \Name{Z.}{Bern}, \Name{N.E.J.}{Bjerrum-Bohr}, \Name{J.}{Blümlein}, \etal, ~\emph{The SAGEX Review on Scattering Amplitudes}, \href{https://iopscience.iop.org/article/10.1088/1751-8121/ac8380}{\Journal{\JPA}{55}{2022}{443001}}, \href{http://arxiv.org/abs/2203.13011}{arXiv:2203.13011}].
\bibitem {ymgwsimul0} \Name{Z.}{Bern}, \Name{J.}{Joseph Carrasco}, \Name{M.}{Chiodaroli}, \Name{H.}{Johansson}, \Name{R.}{Roiban}, ~\emph{The SAGEX Review on Scattering Amplitudes, Chapter 2: An Invitation to Color-Kinematics Duality and the Double Copy}, [\href{http://arxiv.org/abs/2203.13013}{arXiv:2203.13013}].
\bibitem {bimetricgr} \Name{N.}{Rosen}, ~\emph{General Relativity and Flat Space. I}, \href{https://journals.aps.org/pr/abstract/10.1103/PhysRev.57.147}{\Journal {\PRV}{57}{1940}{147}}.
\bibitem {bimetricgr0} \Name{N.}{Rosen}, ~\emph{General Relativity and Flat Space. II}, \href{https://journals.aps.org/pr/abstract/10.1103/PhysRev.57.158}{\Journal {\PRV}{57}{1940}{150}}.
\bibitem {bimetricgr1} \Name{N.}{Rosen}, ~\emph{A bi-metric Theory of Gravitation}, \href{https://link.springer.com/article/10.1007/BF01215403}{\Journal {\GRG}{4}{1973}{435}}.
\bibitem {bimetricmassivegr} \Name{C.}{de Rham}, \Name{G.}{Gabadadze}, \Name{A.J.}{Tolley}, ~\emph{Resummation of Massive Gravity}, \href{https://journals.aps.org/prl/abstract/10.1103/PhysRevLett.106.231101}{\Journal {\PRL}{106}{2011}{231101}}, [\href{http://arxiv.org/abs/1011.1232}{arXiv:1011.1232}].
\bibitem {grspeedvar} \Name{J.}{Magueijo}, ~\emph{New varying speed of light theories}, \href{https://iopscience.iop.org/article/10.1088/0034-4885/66/11/R04}{\Journal {\RPP}{66}{2003}{2025}}, [\href{http://arxiv.org/abs/astro-ph/0305457}{arXiv:astro-ph/0305457}].
\bibitem {hubbletension} \Name{E.}{Abdalla}, \Name{G.F.}{Abellán}, \Name{A.}{Aboubrahim}, \Name{A.}{Agnello}, \Name{O.}{Akarsu}, \Name{Y.}{Akrami}, \Name{G.}{Alestas}, \Name{D.}{Aloni}, \etal, ~\emph{Cosmology Intertwined: A Review of the Particle Physics, Astrophysics, and Cosmology Associated with the Cosmological Tensions and Anomalies}, \href{https://www.sciencedirect.com/science/article/pii/S2214404822000179}{\Journal {\JHA}{34}{2022}{49}}, [\href{http://arxiv.org/abs/2203.06142}{arXiv:2203.06142}].
\bibitem {hubbletensionmod} \Name{H.}{Ziaeepour}, ~\emph{Nonparametric determination of redshift evolution index of Dark Energy}, \href{https://www.worldscientific.com/doi/10.1142/S0217732307023845}{\Journal{\MPL}{22}{2007}{1569}}, [\href{http://arxiv.org/abs/astro-ph/0702519}{astro-ph/0702519}].
\bibitem {hubbletensionmod0} \Name{S.}{Vagnozzi}, ~\emph{New physics in light of the H0 tension: an alternative view}, \href{https://journals.aps.org/prd/abstract/10.1103/PhysRevD.102.023518}{\Journal{\PRD}{102}{2020}{023518}}, [\href{http://arxiv.org/abs/1907.07569}{ arXiv:1907.07569}].
\bibitem {hubbletensionmodrev} \Name{N.}{Schöneberg}, \Name{G.F.}{Abellán}, \Name{A.}{P\'erez S\`anchez}, \Name{S.J.}{Witte}, \Name{V.}{Poulin}, \Name{J.}{Lesgourgues}, ~\emph{The $H_0$ Olympics: A fair ranking of proposed models}, \href{https://www.sciencedirect.com/science/article/abs/pii/S0370157322002538}{\Journal {\PRE}{984}{2022}{1}}, [\href{http://arxiv.org/abs/2107.10291}{arXiv:2107.10291}].
\bibitem {qmsymmfound} \Name{A.}{Bohr}, \Name{O.}{Ulfbeck}, ~\emph{Primary manifestation of symmetry. Origin of quantal indeterminacy}, \href{https://journals.aps.org/rmp/abstract/10.1103/RevModPhys.67.1}{\Journal{\RMP}{67}{1995}{1}}.
\bibitem{qmsymmfound0} \Name{I.S.}{Helland}, ~\emph{Quantum Mechanics from Focusing and Symmetry}, \href{https://link.springer.com/article/10.1007/s10701-008-9239-8}{\Journal{\FOP}{38}{2008}{818}}, [\href{http://arxiv.org/abs/0801.2026}{arXiv:0801.2026}].
\bibitem {qmdirac} \Name{P.A.M.}{Dirac}, ~\emph{The Principles of Quantum Mechanics}, Oxford University Press (1958).
\bibitem {qmvonneumann} \Name{J.}{Von Neumann}, ~\emph{Mathematical Foundation of Quantum Theory}, Princeton University Press, (1955).
\bibitem {qminfocohere} \Name{T.}{Baumgratz}, \Name{M.}{Cramer}, \Name{M.B.}{Plenio}, ~\emph{Quantifying Coherence}, \href{https://journals.aps.org/prl/abstract/10.1103/PhysRevLett.113.140401}{\Journal {\PRL}{113}{2014}{140401}}, [\href{http://arxiv.org/abs/1311.0275}{arXiv:1311.0275}]. 
\bibitem {qmspeedcohere} \Name{I.}{Marvian}, \Name{R.W.}{Spekkens}, \Name{P.}{Zanardi}, ~\emph{Quantum speed limits, coherence and asymmetry}, \href{https://journals.aps.org/pra/abstract/10.1103/PhysRevA.93.052331}{\Journal {\PRA}{93}{2016}{052331}}, [\href{http://arxiv.org/abs/1510.06474}{arXiv:1510.06474}].  % QSL relation with coherence
\bibitem {qgrlocalqm1} \Name{S.B.}{Giddings}, ~\emph{Quantum-first gravity}, \href{https://link.springer.com/article/10.1007/s10701-019-00239-1}{\Journal {\FOP}{49}{2019}{177}}, [\href{http://arxiv.org/abs/1803.04973}{arXiv:1803.04973}].
\bibitem{qmcategory} \Name{T.}{Leinster}, ~\emph{Basic Category Theory, Cambridge Studies in Advanced Mathematics}, Vol. 143, Cambridge University Press, Cambridge (2014), [\href{https://arxiv.org/abs/1612.09375B}{arXiv:1612.09375B}].
\bibitem {qgrlocalqm0} \Name{W.}{Donnelly}, \Name{S.B.}{Giddings}, ~\emph{Diffeomorphism-invariant observables and their nonlocal algebra}, \href{https://journals.aps.org/prd/abstract/10.1103/PhysRevD.93.024030}{\Journal {\PRD}{93}{2016}{024030}}, [\href{http://arxiv.org/abs/1507.07921}{arXiv:1507.07921}].
\bibitem {hawkingrad} \Name{S.}{Hawking}, ~\emph{Breakdown of predictability in gravitational collapse}, \href{https://journals.aps.org/prd/abstract/10.1103/PhysRevD.14.2460}{\Journal{\PRD}{14}{1976}{2460}}.
\bibitem {bhentropy0} \Name{W.H.}{Zurek}, ~\emph{Entropy Evaporated by a Black Hole}, \href{https://journals.aps.org/prl/abstract/10.1103/PhysRevLett.49.1683}{\Journal{\PRL}{49}{1982}{1683}}.
\bibitem {hourivacuum} \Name{H.}{Ziaeepour}, ~\emph{Issues with vacuum energy as the origin of dark energy}, \href{https://www.worldscientific.com/doi/abs/10.1142/S0217732312501544}{\Journal{\MPL}{27}{2012}{1250154}}, [\href{http://arxiv.org/abs/1205.3304}{arXiv:1205.3304}]. 
\bibitem {coherglauber} \Name{R.J.}{Glauber}, ~\emph{Coherent and Incoherent States of the Radiation Field}, \href{https://journals.aps.org/pr/abstract/10.1103/PhysRev.131.2766}{\Journal {\PRV}{131}{1963}{2766}}.
\bibitem {greos} \Name{T.}{Jacobson}, ~\emph{Thermodynamics of Spacetime: The Einstein Equation of State} \href{https://journals.aps.org/prl/abstract/10.1103/PhysRevLett.75.1260}{\Journal{\PRL}{75}{1995}{1260}}, [\href{http://arxiv.org/abs/gr-qc/9504004}{arXiv:gr-qc/9504004}]. 
\bibitem {suninfvirasoro} \Name{E.G.}{Floratos}, \Name{I}{Iliopoulos}, ~\emph{A Note on the Classical Symmetries of the Closed Bosonic Membranes}, \href{https://www.sciencedirect.com/science/article/abs/pii/0370269388902201}{\Journal {\PLB}{201}{1988}{237}}.
\bibitem {suninfvirasoro0} \Name{I.}{Antoniadis}, \Name{P.}{Ditsas}, \Name{E.F.}{Floratos}, \Name{J.}{Iliopoulos}, ~\emph{New Realizations of the Virasoro Algebra as Membrane Symmetries}, \href{https://www.sciencedirect.com/science/article/abs/pii/0550321388906128}{\Journal {\NPB}{300}{1988}{549}}.
\bibitem {qgrmatrixrev} \Name{A.}{Konechny}, \Name{A.}{Schwarz}, ~\emph{Introduction to M(atrix) theory and noncommutative geometry}, \href{https://www.sciencedirect.com/science/article/abs/pii/S0370157301000965}{\Journal{\PRE}{360}{2002}{353}}, [\href{http://arxiv.org/abs/hep-th/0012145}{arXiv:hep-th/0012145}].
\bibitem {noncommutmatrix} \Name{H.}{Steinacker}, ~\emph{Emergent Gravity from Noncommutative Gauge Theory}, \href{https://iopscience.iop.org/article/10.1088/1126-6708/2007/12/049}{\Journal{\JHE}{0712}{2007}{049}}, [\href{http://arxiv.org/abs/0708.2426}{arXiv:0708.2426}].
\bibitem {qgrtestnoncommut} \Name{M.}{Moumni}, \Name{A.}{BenSlama}, \Name{S.}{Zaim}, ~\emph{A New Limit for the Non-Commutative Space-Time Parameter}, \href{https://www.sciencedirect.com/science/article/pii/S0393044010001774}{\Journal {\JGP}{61}{2011}{151}}, [\href{http://arxiv.org/abs/0907.1904}{arXiv:0907.1904}].
\bibitem {qgrtestnoncommut0} \Name{A.}{Kobakhidze}, \Name{C.}{Lagger}, \Name{A.}{Manning}, ~\emph{Constraining noncommutative spacetime from GW150914}, \href{https://journals.aps.org/prd/abstract/10.1103/PhysRevD.94.064033}{\Journal {\PRD 94}{2016}{064033}}, [\href{http://arxiv.org/abs/1607.03776}{arXiv:1607.03776}].
\bibitem {qgrtestnoncommut1} \Name{K.}{Piscicchia}, \Name{A.}{Addazi}, \Name{A.}{Marciano}, \Name{M.}{Bazzi}, \Name{M.}{Cargnelli}, \Name{A.}{Clozza}, \Name{L.}{De Paolis}, \Name{R.}{Del Grande}, ~\etal, ~\emph{Strongest atomic physics bounds on Non-Commutative Quantum Gravity Models}, \href{https://journals.aps.org/prl/abstract/10.1103/PhysRevLett.129.131301}{\Journal {\PRL}{129}{2022}{131301}}, [\href{http://arxiv.org/abs/2209.00074}{arXiv:2209.00074}].
\bibitem {hiesenbergnonlocal} \Name{J.}{Oppenheim}, \Name{S.}{Wehner}, ~\emph{The Uncertainty Principle Determines Nonlocality of Quantum Mechanics}, \href{https://www.science.org/doi/10.1126/science.1192065}{\Journal {\SCI}{330}{2010}{1072}}, [\href{http://arxiv.org/abs/1004.2507}{arXiv:1004.2507}].
\bibitem {qmmaxnonlocal} \Name{S.}{Popescu}, \Name{D.}{Rohrlich} -\emph{Quantum Nonlocality as an Axiom}, \href{https://link.springer.com/article/10.1007/BF02058098}{\Journal{\FOP}{24}{1994}{379}}, [\href{http://arxiv.org/abs/quant-ph/9508009}{arXiv:quant-ph/9508009}].
\bibitem {qmcontextshch} \Name{J.F.}{Clauser}, \Name{M.A.}{Horne}, \Name{A.}{Shimony}, \Name{R.A.}{Holt}, ~\emph{Proposed Experiment to Test Local Hidden-Variable Theories}, \href{https://journals.aps.org/prl/abstract/10.1103/PhysRevLett.23.880}{\Journal{\PRL}{23}{1969}{880}}.
\bibitem {qmaharanovbohm} \Name{Y.}{Aharanov}, \Name{D.}{Bohm}, -\emph{Significance of Electromagnetic Potentials in the Quantum Theory}, \href{https://journals.aps.org/pr/abstract/10.1103/PhysRev.115.485}{\Journal {\PRV}{115}{1959}{485}}.
\bibitem {qmhallrev} \Name{D.}{Tong}, -\emph{Lectures on the Quantum Hall Effect}, (2016), [\href{http://arxiv.org/abs/1606.06687}{arXiv:1606.06687}].
\bibitem {qminfopreskill} \Name{j.}{Preskill}, ~\emph{Lecture Notes for Physics 229:Quantum Information and Computation}, CreateSpace Independent Publishing Platform (2015).
\bibitem {qmcommuncomlex} \Name{H.}{Buhrman}, \Name{R.}{Cleve}, \Name{W.}{van Dam}, ~\emph{Quantum Entanglement and Communication Complexity}, \href{https://epubs.siam.org/doi/10.1137/S0097539797324886}{\Journal{\em SIAM J.Comput.}{30}{2001}{1829}}, [\href{http://arxiv.org/abs/quant-ph/9705033}{quant-ph/9705033}].
\bibitem {qmcommuncomlexrev} \Name{J.}{Watrous}, ~\emph{Quantum Computational Complexity}, [\href{http://arxiv.org/abs/0804.3401}{arXiv:0804.3401}]. 
\bibitem {qmcomplexity} \Name{A.R.}{Brown}, \Name{L.}{Susskind}, ~\emph{The Second Law of Quantum Complexity}, \href{https://journals.aps.org/prd/abstract/10.1103/PhysRevD.97.086015}{\Journal {\PRD}{97}{2018}{086015}}, [\href{https://arxiv.org/abs/1701.01107}{arXiv:1701.01107}]. %Note:qm complexity is a general concept and does not need a reference.But it is used as a resource in this paper and 2110.11371
\bibitem {qmcomplexity0} \Name{N.}{Yunger Halpern}, \Name{N.B.T.}{Kothakonda}, \Name{J.}{Haferkamp}, \Name{A.}{Munson}, \Name{J.}{Eisert}, \Name{Ph.}{Faist}, ~\emph{Resource theory of quantum uncomplexity}, \href{https://journals.aps.org/pra/abstract/10.1103/PhysRevA.106.062417}{\Journal {\PRA}{106}{2021}{062417}}, [\href{https://arxiv.org/abs/2110.11371}{arXiv:2110.11371}].
\bibitem {weakgrconj} \Name{N.}{Arkani-Hamed}, \Name{L.}{Motl}, \Name{A.}{Nicolis}, \Name{C.}{Vafa}, ~\emph{The String Landscape, Black Holes and Gravity as the Weakest Force}, \href{https://iopscience.iop.org/article/10.1088/1126-6708/2007/06/060}{\Journal {\JHE}{06}{2007}{060}}, [\href{http://arxiv.org/abs/hep-th/0601001}{arXiv:hep-th/0601001}].
\bibitem {qgrgauge} \Name{R.}{Utiyama}, ~\emph{Invariant theoretical interpretation of interaction}, \href{https://journals.aps.org/pr/abstract/10.1103/PhysRev.101.1597}{\Journal{\PRV}{101}{1956}{1597}}.
\bibitem {qgrgauge0} \Name{T.W.B.}{Kibble}, ~\emph{Lorentz invariance and the gravitational field}, \href{https://aip.scitation.org/doi/10.1063/1.1703702}{\Journal{\JMP}{2}{1961}{212}}.
\bibitem {ssymmtheorem} \Name {S.}{Coleman}, \Name{J.}{Mandula}, ~\emph{All possible symmetries of the $S$ matrix}, \href{https://journals.aps.org/pr/abstract/10.1103/PhysRev.159.1251}{\Journal{\PRV}{159}{1967}{1251}}.
\bibitem {ssymmsusy} \Name{R.}{Haag}, \Name{J.T.}{Lopuszanski}, \Name{M.}{Sohnius}, ~\emph{All possible generators of supersymmetries of the $S$ matrix}, \href{https://www.sciencedirect.com/science/article/abs/pii/0550321375902795}{\Journal{\NPB}{88}{1975}{257}}. 
\bibitem {grgaugemix} \Name{R.}{Percacci}, ~\emph{Mixing internal and spacetime transformations: Some examples and counterexamples}, \href{https://iopscience.iop.org/article/10.1088/1751-8113/41/33/335403}{\Journal{\JPA}{41}{2008}{335403}}, [\href{http://arxiv.org/abs/0803.0303}{arXiv:0803.0303}].
\bibitem {qmtimepage0} \Name{L.}{Maccone}, \Name{K.}{Sacha}, ~\emph{Quantum measurements of time}, \href{https://journals.aps.org/prl/abstract/10.1103/PhysRevLett.124.110402}{\Journal {\PRL}{124}{2020}{110402}}, [\href{http://arxiv.org/abs/1810.12869}{arXiv:1810.12869}]. 
\bibitem {qmtimepage1} \Name{V.}{Baumann}, \Name{M.}{Krumm}, \Name{Ph.}{Allard Gu\'erin}, \Name{C.}{Brukner}, ~\emph{Noncausal Page-Wootters circuits}, \href{https://journals.aps.org/prresearch/abstract/10.1103/PhysRevResearch.4.013180}{\Journal {\PRR}{4}{2022}{013180}}, [\href{http://arxiv.org/abs/2105.02304}{arXiv:2105.02304}].
\bibitem {qgrtimecritic} \Name{K.V.}{Kuchar}, ~\emph{Time and interpretations of quantum gravity}, \href{https://www.worldscientific.com/doi/abs/10.1142/S0218271811019347?casa_token=KFBxWqgglJoAAAAA:C6ZFAuGLe4y-kAAiiTmd8nc_XgBiNqbxcW3jMF-5N8Y695f-JDrsdi7GGqZmouV6jbP5nn3QmkXA}{\Journal{\IMD}{20}{2011}{3}}.
\bibitem {qmclockimperfect} \Name{A-C.}{de la Hamette}, \Name{S.L.}{Ludescher}, \Name{M.P.}{Mueller}, ~\emph{Entanglement-asymmetry correspondence for internal quantum reference frames}, \href{https://journals.aps.org/prl/abstract/10.1103/PhysRevLett.129.260404}{\Journal {\PRL}{129}{2022}{260404}}, [\href{http://arxiv.org/abs/2112.00046}{arXiv:2112.00046}].
\bibitem{qmcontextual} \Name{N.}{Bohr}, ~\emph{Can Quantum-Mechanical Description of Physical Reality be Considered Complete?}, \href{https://journals.aps.org/pr/abstract/10.1103/PhysRev.48.696}{\Journal {\PRV}{48}{1935}{696}}. %first about QM complementarity/contextuality
\bibitem{qmcontextual0} \Name{S.}{Kochen}, \Name{E.P.}{Specker}, -\emph{The problem of hidden variables in quantum mechanics}, \href{https://link.springer.com/chapter/10.1007/978-94-010-1795-4_17}{\Journal{\JMM}{17}{1967}{59}}.
\bibitem {diraclecture} \Name{P.A.M.}{Dirac}, ~\emph{Lectures on Quantum Mechanics}, Yeshiva Univ. New York, (1964).
\bibitem {seccurvproof} \Name{W.}{K\"uhnel}, ~\emph{Differential Geometry}, AMS, Third Edition, Rhode Island, (2010). %Wolfgang
\bibitem {seccurvproof0} \Name{J}{Gallier}, ~\emph{Differential Geometry and Lie Groups}, Vol, I, Springer, (2020).
\bibitem {qmuncertainener} \Name{M.}{Bauer}, \Name{P.A.}{Mello}, ~\emph{The time-energy uncertainty relation}, \href{https://www.sciencedirect.com/science/article/abs/pii/0003491678902233}{\Journal {\APH}{111}{1978}{38}}.
\bibitem {qmspeedformul} \Name {K.}{Bhattacharyya}, -\emph{Quantum decay and Mandelstam-Tamm energy inequality}, \href{https://iopscience.iop.org/article/10.1088/0305-4470/16/13/021}{\Journal{\JPA: {\em Math. Gen}}{16}{1983}{2993}}.
\bibitem {qmspeedgeomopen0} \Name{D.}{Mondal}, \Name{C.}{Datta}, \Name{S.}{Sazim}, ~\emph{Quantum Coherence Sets The Quantum Speed Limit For Mixed States}, \href{https://www.sciencedirect.com/science/article/abs/pii/S0375960115010518}{\Journal{\PLA}{380}{2016}{689}}, [\href{http://arxiv.org/abs/1506.03199}{arXiv:1506.03199}].
\bibitem {qmspeedgeomaction} \Name{K.}{Funo}, \Name{N.}{Shiraishi}, \Name{K.}{Saito}, ~\emph{Speed limit for open quantum systems}, \href{https://iopscience.iop.org/article/10.1088/1367-2630/aaf9f5}{\Journal{\NJP}{21}{2019}{013006}}, [\href{https://arxiv.org/abs/1810.03011}{arXiv:1810.03011}].
\bibitem {qmspeedgeommix} \Name{N.}{H\"ornedal}, \Name{D.}{Allan}, \Name{O.}{S\"onnerborn}, ~\emph{Extensions of the Mandelstam-Tamm quantum speed limit to systems in mixed states}, \href{https://iopscience.iop.org/article/10.1088/1367-2630/ac688a}{\Journal {\NJP}{24}{2022}{055004}}, [\href{http://arxiv.org/abs/2112.08017}{arXiv:2112.08017}].
\bibitem {qmspeedrev} \Name{S.}{Deffner}, \Name{S.}{Campbell}, ~\emph{Quantum speed limits: from Heisenberg's uncertainty principle to optimal quantum control}, \href{https://iopscience.iop.org/article/10.1088/1751-8121/aa86c6}{\Journal{\JPA}{50}{2017}{453001}}, [\href{https://arxiv.org/abs/1705.08023}{arXiv:1705.08023}].
\bibitem {qmhilbertgeom} \Name{G.W.}{Gibbons}, ~\emph{Typical states and density matrices}, \href{https://www.sciencedirect.com/science/article/abs/pii/0393044092900464}{\Journal{\em J. Geometry \& Phys.}{8}{1992}{147}}.
\bibitem {densitymatgeom} \Name{E.A.}{Morozova}, \Name{N.N.}{Chentsov}, ~\emph{Markov invariant geometry on manifolds of states}, \href{https://link.springer.com/article/10.1007/BF01095975}{\Journal{\em Journal of Soviet Math.}{56}{1991}{2648}}.
\bibitem {densitymatgeom0} \Name{D.}{Petz}, \Name{H.}{Hasegawa}, ~\emph{Metric of $\alpha$-Entropies of Density Matrices}, \href{https://link.springer.com/article/10.1007/BF00398324}{\Journal{\em Lett. Math. Phys.}{38}{1996}{221}}. % proves I_WY as distance for stochastic maping of density matrices
\bibitem {densitymatgeom1} \Name{D.}{Petz}, ~\emph{Monoton Matrices on Matrix spaces}, \href{}{\Journal{\em Linear Algebra Appl.}{244}{1996}{81}}.
\bibitem {wigneryanaseqminfo} \Name{E.P.}{Wigner}, \Name{M.M.}{Yanase}, ~\emph{Information Content of Distributions}, \href{https://www.pnas.org/doi/abs/10.1073/pnas.49.6.910}{\Journal{\em Proc. Nation. Acad. Sci. USA}{49}{1963}{910}}.
\bibitem {wigneryanaseqminfogeo} \Name{P.}{Gibilisco}, \Name{T.}{Isola}, ~\emph{Wigner-Yanase information on quantum state space:the geometric approach}, \href{https://pubs.aip.org/aip/jmp/article-abstract/44/9/3752/440570/Wigner-Yanase-information-on-quantum-state-space?redirectedFrom=fulltext}{\Journal{\JMP}{44}{2003}{3752}}, [\href{https://arxiv.org/abs/math/0304170}{math/0304170}].  % geodesics for WY metric (skew info)
\bibitem {buresmetric} \Name{D.}{Bures}, ~\emph{An Extension of Kakutant's Theorem on Infinite Product Measures to the Tensor Product of Semifinite $w^*$-Algebras}, \href{https://www.jstor.org/stable/1995012?casa_token=BCgMgaFHRY4AAAAA\%3AbFS9omYbO_tgJ3KEu8x8WBX3W-ej4KPdwhZh7ysWB6YPhVqwgpyH326sZxcStnwp0bdugN00Rs1Y-OzpJbQcOv_MzzzD2G3Fby_bP9FqgoeNM4aADF0}{\Journal{\em Trans.Am.Math.Soc}{135}{1969}{199}}. 
\bibitem {qmopenbook} \Name{H.P.}{Breuer}, \Name{F.}{Petruccione}, ~\emph{The theory of open quantum systems}, Clarendon Press, Oxford, (2010).
\bibitem {qmspeedoperat} \Name{M.}{Zhang}, \Name{H-M.}{Yu}, \Name{J.}{Liu}, ~\emph{Quantum speed limit for complex dynamics}, \href{https://www.nature.com/articles/s41534-023-00768-8}{\Journal{\em npj Quantum Inf} {9}{2023}{97} (2023)}, [\href{https://arxiv.org/abs/2301.00566}{arXiv:2301.00566}].
\bibitem {qmspeedgeomfisher} \Name{M.M.}{Taddei}, \Name{B.M.}{Escher}, \Name{L.}{Davidovich}, \Name{R.L.}{de Matos Filho}, -\emph{Quantum speed limit for physical processes}, \href{https://journals.aps.org/prl/abstract/10.1103/PhysRevLett.110.050402}{\Journal{\PRL}{110}{2013}{050402}}, [\href{https://arxiv.org/abs/1209.0362}{arXiv:1209.0362}]. % uses fidelity diatance and Fisher info.
\bibitem {qmspeedopenunif} \Name{L.P.}{García-Pintos}, \Name{S.}{Nicholson}, \Name{J.R.}{Green}, \Name{A.}{del Campo}, \Name{A.V.}{Gorshkov}, ~\emph{Unifying Quantum and Classical Speed Limits on Observables}, \href{https://journals.aps.org/prx/abstract/10.1103/PhysRevX.12.011038}{PhysRevX.12.2022.011038}, [\href{https://arxiv.org/abs/2108.04261}{arXiv:2108.04261}].  %open sys, decomposition to coherent and incoherent QSL, by conditions attainability for observables 
%\bibitem {qmspeeddyn} New J. Phys. 19 (2017) 103018 1704.03357 % It is in Wignerspace, i.e. phase space {(x,p)}
\bibitem {qmspeednonmarkov} \Name{S.}{Deffner}, \Name{E.}{Lutz}, ~\emph{Quantum speed limit for non-Markovian dynamics}, \href{https://journals.aps.org/pre/abstract/10.1103/PhysRevE.77.021128}{\Journal{\PRL}{111}{2013}{010402}}, \href{https://arxiv.org/abs/1302.5069}{[arXiv:1302.5069]} % It is Levnin (energy) type QSL
\bibitem {qmspeedhamiltonian} \Name{A.}{Mostafazadeh}, ~\emph{On Hamiltonians Generating Optimal-Speed Evolutions}, \href{https://journals.aps.org/pra/abstract/10.1103/PhysRevA.79.014101}{\Journal {\PRA}{79}{2009}{014101}}, [\href{http://arxiv.org/abs/0804.4755}{arXiv:0804.4755}]. %all invariant metrics of density matrices
\bibitem {qmspeedopen} \Name{A.}{del Campo}, \Name{I.L.}{Egusquiza}, \Name{M.B.}{Plenio}, \Name{S.F.}{Huelga}, ~\emph{Quantum speed limits in open system dynamics}, \href{https://journals.aps.org/prl/abstract/10.1103/PhysRevLett.110.050403}{\Journal{\PRL}{110}{2013}{050403}}, [\href{https://arxiv.org/abs/1209.1737}{arXiv:1209.1737}]. % This is not about geometry
\bibitem {qmspeedlevitin} \Name{N.}{Margolus}, \Name{L.B.}{Levitin}, ~\emph{The maximum speed of dynamical evolution}, \href{}{\Journal{\PYA}{120}{1998}{188}}, \href{https://arxiv.org/abs/quant-ph/9710043}{quant-ph/9710043}.
\bibitem {qmspeedlimitequi} \Name{L.B.}{Levitin}, \Name{T.}{Toffoli}, ~\emph{The fundamental limit on the rate of quantum dynamics: the unified bound is tight}, \href{https://journals.aps.org/prl/abstract/10.1103/PhysRevLett.103.160502}{\Journal {\PRL}{103}{2009}{160502}}, [\href{http://arxiv.org/abs/0905.3417}{arXiv:0905.3417}].
\bibitem {liebrobinsonbound} \Name{E.H.}{Lieb}, \Name{D.W.}{Robinson}, ~\emph{The finite group velocity of quantum spin systems}, \href{https://link.springer.com/article/10.1007/BF01645779}{\Journal{\CMP}{28}{1972}{251}}.
\bibitem {liebrobinsonboundrev} \Name{C-F.}{Chen}, \Name{A.}{Lucas}, \Name{C.}{Yin}, ~\emph{Speed limits and locality in many-body quantum dynamics}, \href{https://iopscience.iop.org/article/10.1088/1361-6633/acfaae}{\Journal{\RPP}{86}{2023}{116001}}, [\href{https://arxiv.org/abs/2303.07386}{arXiv:2303.07386}].
\bibitem {grestgrb090510a} Fermi GBM/LAT Collaborations:\Name{A.A.}{Abdo}, \Name{M.}{Ackermann}, \Name{M.}{Ajello}, \Name{K.}{Asano}, \Name{W.B.}{Atwood}, \Name{M.}{Axelsson}, \Name{L.}{Baldini}, \Name{J.}{Ballet}, \Name{G.}{Barbiellini}, \Name{M.G.}{Baring}, \etal, ~\emph{A limit on the variation of the speed of light arising from quantum gravity effects}, \href{https://www.nature.com/articles/nature08574}{\Journal {\NAT}{462}{2009}{331}}, [\href{http://arxiv.org/abs/0908.1832}{arXiv:0908.1832}].
\bibitem {grbphotonmass} \Name{D.J.}{Bartlett}, \Name{H.}{Desmond}, \Name{P.G.}{Ferreira}, \Name{J.}{Jasche}, ~\emph{Constraints on quantum gravity and the photon mass from gamma ray bursts}, \href{https://journals.aps.org/prd/abstract/10.1103/PhysRevD.104.103516}{\Journal {\PRD}{104}{2021}{103516}}, [\href{http://arxiv.org/abs/2109.07850}{arXiv:2109.07850}]. 
\bibitem {densitysqrt} \Name{A.}{Yahalom}, \Name{R.}{Englman}, ~\emph{A ``Square-root'' Method for the Density Matrix and its Applications to Lindblad Operators}, \href{https://www.sciencedirect.com/science/article/abs/pii/S0378437106003748?via\%3Dihub}{\Journal{\PYA}{371}{2006}{368}}, [\href{https://arxiv.org/abs/cond-mat/0512474}{cond-mat/0512474}].
\bibitem {localitydef} \Name{J.S.}{Cotler}, \Name{G.R.}{Penington}, \Name{D.H.}{Ranard}, ~\emph{Locality from the Spectrum}, \href{https://link.springer.com/article/10.1007/s00220-019-03376-w}{\Journal {\CMP}{368}{2019}{1267}}, [\href{http://arxiv.org/abs/1702.06142}{arXiv:1702.06142}].
\bibitem {skewinfocohere} \Name{D.}{Girolami}, ~\emph{Observable measure of quantum coherence in finite dimensional systems}, \href{https://journals.aps.org/prl/abstract/10.1103/PhysRevLett.113.170401}{\Journal{\PRL}{113}{2014}{170401}}, [\href{http://arxiv.org/abs/1403.2446}{arXiv:1403.2446}].
%\binitem \Name{A.}{Wehrl}, ~\emph{General properties of entropy}, \href{https://journals.aps.org/rmp/abstract/10.1103/RevModPhys.50.221}{\Journal{\RMP}{50}{1978}{221}}. %##### It is not sure that it is relevant for coherence measure
\bibitem {qmclasscorr} \Name{L.}{Henderson}, \Name{V.}{Vedral}, ~\emph{Classical, quantum and total correlations}, \href{https://iopscience.iop.org/article/10.1088/0305-4470/34/35/315}{\Journal {\JPA}{Math. Gen. 34}{2001}{6899}}, [\href{http://arxiv.org/abs/quant-ph/0105028}{arXiv:quant-ph/0105028}].
\bibitem {qmdecohere} \Name{J.}{Martin}, \Name{A.}{Micheli}, \Name{V.}{Vennin}, ~\emph{Comparing quantumness criteria}, (2022), [\href{http://arxiv.org/abs/2211.10114}{arXiv:2211.10114}].
\bibitem {grcondens1} \Name{G.}{Dvali}, \Name{R.}{Venugopalan}, ~\emph{Classicalization and unitarization of wee partons in QCD and Gravity: The CGC-Black Hole correspondence}, \href{https://journals.aps.org/prd/abstract/10.1103/PhysRevD.105.056026}{\Journal {\PRD}{105}{2022}{056026}}, [\href{http://arxiv.org/abs/2106.11989}{arXiv:2106.11989}].
\bibitem {grcondens} \Name{G.}{Dvali}, \Name{C.}{Gomez}, ~\emph{Black Hole's Quantum N-Portrait}, (2011), [\href{http://arxiv.org/abs/1112.3359}{arXiv:1112.3359}].
\bibitem {grcondens0} \Name{G.}{Dvali}, \Name{C.}{Gomez}, ~\emph{Black Holes as Critical Point of Quantum Phase Transition}, (2012), [\href{http://arxiv.org/abs/1207.4059}{arXiv:1207.4059}].
\bibitem {devacuum} Letter from Lema\^itre to Einstein on 3 Oct. 1947, (see e.g. J.P. Luminet, ``Essais de cosmologie'', Seuil, Paris, (1997).
\bibitem {qcdinstantonliquid} \Name{I.}{Zahed}, ~\emph{Mass sum rule of hadrons in the QCD instanton vacuum}, \href{https://journals.aps.org/prd/abstract/10.1103/PhysRevD.104.054031}{\Journal {\PRD}{104}{2021}{054031}}, [\href{http://arxiv.org/abs/2102.08191}{arXiv:2102.08191}].
\bibitem {infinstanton} \Name{A.}{Kaya}, ~\emph{Stationary Phase Approximation and Instanton-like States for Cosmological In-In Path Integrals}, \href{https://journals.aps.org/prd/abstract/10.1103/PhysRevD.86.123511}{\Journal {PRD}{86}{2012}{123511}}, [\href{http://arxiv.org/abs/1209.4694}{arXiv:1209.4694}].
\bibitem {bfgrvity} \Name{S.H.}{Alexander}, ~\emph{A Quantum Gravitational Relaxation of The Cosmological Constant}, \href{https://www.sciencedirect.com/science/article/abs/pii/S0370269305013389}{\Journal {\PLB}{629}{2005}{53}}, [\href{http://arxiv.org/abs/hep-th/0503146}{arXiv:hep-th/0503146}].
\bibitem {desittergrinstanton} \Name{B.}{Julia}, \Name{J.}{Levie}, \Name{S.}{Ray}, ~\emph{Gravitational duality near de Sitter space}, \href{https://iopscience.iop.org/article/10.1088/1126-6708/2005/11/025}{\Journal{\JHE}{0511}{2005}{025}}, [\href{http://arxiv.org/abs/hep-th/0507262}{hep-th/0507262}].
\bibitem {houriqmcond} \Name{H.}{Ziaeepour}, ~\emph{Non-equilibrium evolution of quantum fields during inflation and late accelerating expansion}, in \href{https://novapublishers.com/product-tag/dark-energy/}{\emph{A Closer Look at Dark Energy}}, Nova Science Inc. New York (2019), [\href{http://arxiv.org/abs/1711.01925}{arXiv:1711.01925}].
\bibitem {qmentangleener} \Name{C.}{Beny}, \Name{Ch.T.}{Chubb}, \Name{T.}{Farrelly}, \Name{T.J.}{Osborne}, ~\emph{Energy cost of entanglement extraction in complex quantum systems}, \href{https://www.nature.com/articles/s41467-018-06153-w}{\Journal {\NAC}{9}{2018}{3792}}, [\href{http://arxiv.org/abs/1711.06658}{arXiv:1711.06658}].
\bibitem {qminfothermo} \Name{J.}{Goold}, \Name{F.}{Plastina}, \Name{A.}{Gambassi}, \Name{A.}{Silva}, ~\emph{The role of quantum information in thermodynamics --- a topical review}, \href{https://iopscience.iop.org/article/10.1088/1751-8113/49/14/143001}{\Journal{\JPA}{49}{2016}{143001}}, [\href{https://arxiv.org/abs/1505.07835}{arXiv:1505.07835}].
\end{thebibliography}
\end{document}